\documentclass[12pt, right]{iopart}

\providecommand{\docversion}{fi} 

\usepackage{ifthen}
\usepackage[normalem]{ulem}   
\usepackage{xcolor}
\usepackage{comment}
\usepackage[most]{tcolorbox}
\usepackage{tikz}
\usepackage{currfile}         
\usepackage{zref-abspage}     
\usepackage{lineno}           

\ifthenelse{\equal{\docversion}{mp}}{\linenumbers}{}

\newwrite\changelog
\immediate\openout\changelog=\jobname-changes.csv
\immediate\write\changelog{category,action,content,file,line,abspage}
\AtEndDocument{\immediate\closeout\changelog}

\makeatletter
\newcommand{\LogChange}[3]{
  \begingroup
    \zref@labelbyprops{chg@\the\year @\the\time @\the\inputlineno}{abspage}%
    \edef\thispage{\zref@extractdefault{chg@\the\year @\the\time @\the\inputlineno}{abspage}{0}}%
    \edef\thisfile{\currfilename}%
    \edef\safecontent{\detokenize{#3}}%
    \immediate\write\changelog{#1,#2," \safecontent ",\thisfile,\the\inputlineno,\thispage}%
  \endgroup
}
\makeatother

\DeclareRobustCommand{\removed}[1]{%
  \ifthenelse{\equal{\docversion}{mp}}{%
    \textcolor{red}{\sout{#1}}%
    \LogChange{text}{remove}{#1}%
  }{%
    \ifthenelse{\equal{\docversion}{og}}{%
      #1\unskip 
    }{%
      \ignorespaces 
    }%
  }%
}

\DeclareRobustCommand{\new}[1]{%
  \ifthenelse{\equal{\docversion}{mp}}{%
    \textcolor{blue}{#1}%
    \LogChange{text}{add}{#1}%
  }{%
    \ifthenelse{\equal{\docversion}{fi}}{%
      #1\unskip 
    }{%
      \ignorespaces 
    }%
  }%
}

\newenvironment{addedfigure}[1][tbp]{%
  \begin{figure}[#1]%
  \ifthenelse{\equal{\docversion}{og}}{%
  }{%
      \end{minipage}%
      \ifthenelse{\equal{\docversion}{mp}}{\end{tcolorbox}}{}%
      \endgroup
  }%
  \end{figure}%
}

\newenvironment{addedtable}[1][tbp]{%
  \begin{table}[#1]%
  \ifthenelse{\equal{\docversion}{og}}{%
  }{%
      \end{minipage}%
      \ifthenelse{\equal{\docversion}{mp}}{\end{tcolorbox}\end{nolinenumbers}}{}%
      \endgroup
  }%
  \end{table}%
}

\NewEnviron{removedtable}[1][tbp]{%
  \ifthenelse{\equal{\docversion}{fi}}%
    {}
    {%
      \begin{table}[#1]
        \centering
        \begin{tikzpicture}
          \node[inner sep=0,outer sep=0] (B) {%
            \begin{minipage}{\linewidth}
              \BODY
            \end{minipage}
          };
          \draw[red,line width=1.25pt] (B.north west) rectangle (B.south east);
          \draw[red,line width=1.25pt] (B.north west) -- (B.south east);
          \draw[red,line width=1.25pt] (B.north east) -- (B.south west);
        \end{tikzpicture}
      \end{table}%
    }%
}

\renewcommand{\docversion}{fi} 

\usepackage{subcaption} 
\usepackage[numbers]{natbib}
\usepackage{tabularx}
\usepackage{multirow}
\newcommand{\multrow}[1]{\begin{tabular}{@{}c@{}} #1 \end{tabular}}
\usepackage{adjustbox}
\usepackage{calc}       
\usepackage{float}

\usepackage[T1]{fontenc}
\usepackage{amsmath,amssymb}
\usepackage{nameref,hyperref}

\usepackage{ifthen}

\usepackage{longtable,booktabs, makecell,array, siunitx}
\renewcommand{\arraystretch}{1.15}
\newcolumntype{Y}{>{\raggedright\arraybackslash}X}

\newboolean{blind}
\ifthenelse{\equal{\docversion}{mp}}{%
  \setboolean{blind}{true}%
}{%
  \setboolean{blind}{false}%
}

\begin{document}

\title[Detecting Cognitive Impairment and Psychological Well-being among Older Adults]{Feasibility of Detecting Cognitive Impairment and Psychological Well-being among Older Adults Using Facial, Acoustic, Linguistic, and Cardiovascular Patterns Derived from Remote Conversations}

\ifthenelse{\boolean{blind}}{
}{
\author{
Xiaofan Mu\textsuperscript{1\dag},
Merna Bibars\textsuperscript{2,3\dag}, 
Salman Seyedi\textsuperscript{4\dag}, 
Iris Zheng\textsuperscript{1},
Zifan Jiang\textsuperscript{4}, 
Liu Chen\textsuperscript{5}, 
Bolaji Omofojoye\textsuperscript{4}, 
Rachel Hershenberg\textsuperscript{6}, 
Allan I. Levey\textsuperscript{7}, 
Gari D. Clifford\textsuperscript{2,4}, 
Hiroko H. Dodge\textsuperscript{5\ddag}, 
Hyeokhyen Kwon\textsuperscript{2,4\ddag} 
}
\address{
\textbf{1} Department of Computer Science, Emory University, Atlanta, GA, 30322, USA
\\
\textbf{2} The Wallace H. Coulter Department of Biomedical Engineering, Emory University and Georgia Institute of Technology, Atlanta, GA, 30332, USA
\\
\textbf{3} Department of Systems and Biomedical Engineering, Cairo University, Giza, Egypt
\\
\textbf{4} Department of Biomedical Informatics, Emory University, Atlanta, GA, 30322, USA
\\
\textbf{5} Department of Neurology, Massachusetts General Hospital, Harvard Medical School, Boston, Massachusetts, USA
\\
\textbf{6} Department of Psychiatry and Behavioral Sciences, Emory University, Atlanta, GA, 30322, USA
\\
\textbf{7} Department of Neurology, Emory University, Atlanta, GA, 30322, USA
}
\ead{hyeokhyen.kwon@emory.edu}

\vspace{10pt}
\begin{indented}
\item[] \dag Equally contributing authors.
\item[] \ddag Joint senior authors.
\end{indented}
}


\begin{abstract}
The aging society urgently requires scalable methods to monitor cognitive decline and identify social and psychological factors indicative of dementia risk in older adults.
Our machine learning (ML) models captured facial, acoustic, linguistic, and cardiovascular features from 39 \removed{individuals} \new{older adults} with normal cognition or Mild Cognitive Impairment \new{(MCI)}\new{,} derived from remote video conversations and \removed{classified} \new{quantified their} cognitive status, social isolation, neuroticism,  and psychological well-being. 
Our model could distinguish Clinical Dementia Rating Scale (CDR) of 0.5 (vs. 0) with \removed{0.78} \new{0.77} area under the receiver operating characteristic curve (AUC), social isolation with \removed{0.75} \new{0.74} AUC, \new{social satisfaction with 0.75 AUC}\removed{neuroticism with 0.71 AUC}, \new{psychological well-being with 0.72 AUC,} and negative affect \removed{scales} with \removed{0.79} \new{0.74} AUC.
\removed{Recent advances in machine learning offer new opportunities to remotely detect cognitive impairment and assess associated factors, such as neuroticism and psychological well-being.} Our \removed{experiment} \new{feature importance analysis} showed that speech and language patterns were useful for quantifying cognitive impairment, whereas facial expression\new{s} and cardiovascular patterns \removed{using photoplethysmography (PPG)} were useful for quantifying social and psychological well-being.
\new{Our bias analysis showed that the best-performing models for quantifying psychological well-being and cognitive states in older adults exhibited significant biases concerning their age, sex, disease condition, and education levels.}
\new{Our comprehensive analysis shows the feasibility of monitoring the cognitive and psychological health of older adults, as well as the need for collecting large-scale interview datasets of older adults to benefit from the latest advances in deep learning technologies to develop generalizable models across older adults with diverse demographic backgrounds and disease conditions.}

\end{abstract}

%
%
%
%
%

\section{Introduction}\label{sec:intro}

The number of older adults living with Alzheimer's disease and related dementias (ADRD) is expected to rise by nearly 13 million in the U.S. by the year 2060 \cite{better2023alzheimer,kinsella2005globa}.
Mild Cognitive Impairment (MCI) often precedes ADRD. It is characterized by cognitive decline that is greater than expected for an individual's age and education level, while the person remains capable of performing daily activities independently. 
Along with cognition, emotional well-being, such as anxiety, loneliness, or depressive symptoms, may have bidirectional relationships with cognitive impairments through shared neurobiological and behavioral mechanisms~\cite{mildcognitiveimpairment2006,hikichi2017social,fernandez2024depression}.
Additionally, mood disorders, social isolation, and negative emotions often co-occur with or even precede MCI and can further accelerate cognitive decline \cite{gardener2021social,steenland2012late,mourao2016depressive}.
Cognitive impairment and psychological well-being significantly impact the lives of elderly individuals, highlighting the importance of early detection, monitoring, and intervention of these symptoms to maintain their quality of life.

Clinical tools, such as the Montreal Cognitive Assessment (MoCA) and Clinical Dementia Rating (CDR), have become widely accepted as standardized methods for evaluating and monitoring cognitive impairment, which assess a range of cognitive functions~\cite{nasreddine2005montreal,Morris1993}.
However, traditional cognitive assessments are not sensitive enough to identify MCI or monitor the progression during the early MCI stage~\cite{mora2012mild,pinto2019montreal,chehrehnegar2020early}.
In standard geriatric care, subtle behavioral changes related to psychological well-being, such as reduced social engagement and declining mental health, are often overlooked.
These factors, however, may signal early dementia risk and could offer opportunities for timely intervention~\cite{zhaoyang2024features,hikichi2017social}, underscoring the need for innovative approaches to monitor them~\cite{Baumeister1995,john2023associations}.
The expected shortage of care services further exacerbates this situation.
It is expected that over 1 million additional direct care workers will be needed by 2031, and the U.S. must nearly triple its geriatricians by 2050~\cite{better2023alzheimer}.
The current global shortage of mental health professionals and geriatricians further complicates disparities in care, especially in underserved regions~\cite{cummings2023challenges,detectionratesofmildcognitive}.
Even developed countries are facing such challenges, with twenty states in the U.S. identified as ``dementia neurology deserts'' \cite{dementianeurologydeserts}, let alone developing countries.

The recent widespread adoption of video conferencing platforms for telehealth~\cite{worster2023increasing} presents an opportunity to use these tools to prescreen cognitive impairment and to assess associated factors, such as social isolation and psychological well-being, in older adults remotely~\cite{griffiths2006telemedicine,garcia2018mental}.
Recent advancements in artificial intelligence (AI) have spurred research into using telehealth platforms to quantify \removed{psychological-relevant} \new{psychologically-relevant} behaviors in individuals with MCI through facial, audio, and text analysis \cite{developoingamachinelearningmodelfordetecting, amultimodaldialogapproach,jiang2020classifying}. 
Despite \removed{the} active research in computerized interview analysis \cite{d2020ai,graham2019artificial,garcia2018mental}, there are still relatively few publications that focus on quantifying cognitive abilities, social isolation, and psychological well-being in older adults automatically through remote interviews. 
Furthermore, with the recent emergence of generative AI models, so-called foundation AI models trained on vast amounts of video, audio, and text data from internet sources, it is imperative to explore whether these innovations can be effectively leveraged to quantify behaviors in older adults \cite{oquab2023dinov2, chen2022wavlm, touvron2023llama}. 
Advances in \removed{machine learning}ML in video analysis also enable non-contact assessments of cardiovascular features, such as remote Photoplethysmography (rPPG) from facial videos \cite{10227326}, which could serve as another valuable modality for assessing cognitive impairment and psychological well-being in older adults. For example, rPPG allows the extraction of heart rate variability (HRV) features that are found to be linked with anxiety and other negative emotions~\cite{jung2024changes,chalmers2014anxiety}.

This work aims to investigate the feasibility of quantifying the psychological well-being, social network, and cognitive ability of individuals living with normal cognition or MCI, utilizing digital markers extracted from facial, acoustic, linguistic, and cardiovascular patterns detected by foundation AI models pretrained from large-scale datasets available from the internet. 
By objectively quantifying various modalities associated with cognitive decline through a scalable, remote, and automated assessment system, we expect this work to provide a step toward enhancing \removed{the} accessibility, reducing \removed{the} disparities \removed{of}\new{in} mental health and dementia services~\cite{viers2015efficiency}, and promoting evidence-based therapeutics~\cite{evidencebasedpsychotherapy}.
\new{Our systematic analysis of multimodal AI systems on bias, feature importances, and performances of various models, including the latest advances in deep learning techniques, is also expected to provide important insights on future research directions in developing generalizable ML systems for monitoring cognitive impairment and psychological well-being in older adults with normal cognition or MCI from various demographic groups and disease conditions.}

\section{Methods and Materials}\label{sec:methods}

\subsection{Ethics Statement}

\ifthenelse{\boolean{blind}}{
This study has received approval from the Institutional Review Board (IRB) at the institutions of the authors. The details are not provided for the double-blind review process. Full details will be disclosed when accepted.
}{
The Internet-Based Conversational Engagement
Clinical Trial (I-CONECT) study received approval from the Institutional Review Board (IRB) at Oregon Health \& Science University (OHSU) (IRB STUDY00015937) through a single IRB process.
The \removed{patients/}participants provided their written informed consent to participate in this study, which also allowed for secondary data analysis.
The I-CONECT trial began recruiting participants and collecting data in July 2018. In the second year of recruitment, the COVID-19 pandemic struck the USA. The trial ran for approximately 19 months before the pandemic, from July 2018 to March 2020.
The secondary analysis of the dataset from the I-CONECT study was approved by \new{the} IRB at Emory University (IRB STUDY00007231), as the dataset included identifiable information, such as videos and audio. 
The IRB was approved on January 8th, 2024, when the study team at Emory University started accessing the dataset for the proposed work.
}

\subsection{Participants and Interview Protocol}
\label{protocol}

\begin{table*}[t]
\centering
\small
\caption{\textbf{Demographics and Clinical Characteristics of the Participants.} 
 For sex, ``M'' stands for male, and ``F'' is for female. For race, ``C'' denotes Caucasians, ``A'' denotes African American\new{s}, and ``O'' denotes other. The years of education reflect the total number of academic years completed in formal education. The CDR categories include no (impairment) (CDR=0)  and questionable dementia (CDR=0.5).}
\renewcommand{\arraystretch}{1.3}
\begin{tabularx}{\textwidth}{|>{\raggedright\arraybackslash}X|>{\raggedright\arraybackslash}X|>{\raggedright\arraybackslash}X|>{\raggedright\arraybackslash}X|>{\raggedright\arraybackslash}X|}
\hline
\textbf{Type} & \textbf{Clinical Diagnosis} & \textbf{Normal Cognition (NC)} & \textbf{MCI } & \textbf{Combined (NC+MCI)} \\
\hline
\multirow{4}{*}{Demographics} & \textbf{Counts} & 17 & 22 & 39\\
\cline{2-5}
 & \textbf{Sex (F/M)} & 15/2 & 14/8 & 29/10\\
\cline{2-5}
 & \textbf{Race (C/A/O)} & 16/1/0 & 18/3/1 & 34/4/1\\
\cline{2-5}
 & \textbf{Age} & 79.85 $\pm$ 4.39 & 81.53 $\pm$ 4.81 & {80.69 $\pm$ 4.6}\\
\cline{2-5}
 & \textbf{Education (years)} & 15.47 $\pm$ 2.12 & 15.41 $\pm$ 2.54 & {15.44 $\pm$ 2.34}\\
\hline
\multirow{2}{*}{\multrow{Cognitive\\Ability}} & \textbf{MoCA} & 26.58 $\pm$ 2.51 & 22.96 $\pm$ 3.57 & {24.54 $\pm$ 3.61}\\
\cline{2-5}
 & \textbf{CDR (no / questionable dementia)} & 14/3 & 6/16 & 20/19\\
\hline
\multirow{4}{*}{\multrow{Social Network,\\Personality,\\Psychological\\Well-being}} & \textbf{LSNS-6} & 15.12 $\pm$ 6.08 & 13.36 $\pm$ 5.71 & {14.13 $\pm$ 5.86}\\
\cline{2-5}
 & \textbf{Neuroticism} & 15.35 $\pm$ 9.16 & 17.27 $\pm$ 7.89 & {16.44 $\pm$ 8.41}\\
\cline{2-5}
 & \textbf{Negative affect} & 46.06 $\pm$ 8.29 & 51.08 $\pm$ 12.82 & {48.57 $\pm$ 10.93}\\
\cline{2-5}
 & \textbf{Social satisfaction} & 49.64 $\pm$ 10.70 & 48.35 $\pm$ 13.12 & {49.02 $\pm$ 11.77}\\
\cline{2-5}
 & \textbf{Psychological well-being} & 54.66 $\pm$ 7.17 & 47.08 $\pm$ 11.12 & {50.87 $\pm$ 9.99}\\
\hline
\end{tabularx}
\label{table:demo}
\end{table*}

The data source for this project was the I-CONECT (NCT02871921) study \citep{Yu2021,dodge2024internet}. 
This behavioral intervention \new{study} aimed to enhance cognitive functions by providing social interactions (conversational interactions) to older subjects living in social isolation.  The study was based on the accumulating evidence that social isolation is a risk factor for dementia~\citep{livingston2024dementia}. The study recruited older adults (> 75 years old) with MCI or normal cognition from \removed{2}\new{two} sites: Portland, Oregon, which focused on Caucasian participants, and Detroit, Michigan, which focused on African American participants. The experimental group participated in 30-minute video chats with trained conversational specialists four times a week, along with weekly 10-minute phone check-ins for \removed{6}\new{six} months. In contrast, the control group only received weekly 10-minute phone check-ins. Conversations \removed{are}\new{were} semi-structured with predetermined themes each day, ranging from historical events to leisure activities, \new{and} using pictures to promote conversations. The participants with severe depressive symptoms (GDS-15 $\ge$7)~\citep{yesavage1982development} were excluded. 
Exclusion criteria included a clinical diagnosis of dementia. Clinical diagnoses were made through a consensus process involving neurologists and neuropsychologists, using the National Alzheimer's Coordinating Center (NACC) Uniform Data Set Version 3 (UDS-3) \removed{.}\citep{Weintraub2018,dodge2020differentiating}. Inclusion criteria required participants to be socially isolated according to at least one of the following: (1) a score of 12 or less on the 6-item Lubben Social Network Scale (LSNS-6) \citep{Lubben2006}, (2) engaging in conversations lasting 30 minutes or more\new{,} no more than twice a week, based on self-report, or (3) responding “often” to at least one question on the 3-item UCLA Loneliness Scale \citep{Hughes2004}. The intervention results showed that \removed{the} global cognitive functions improved significantly among the intervention group (i.e., \new{the} video-chats engaged group) compared with the control group after \removed{6}\new{six} months of intervention\new{,} with \new{a} Cohen's d of 0.73. The topline results of this behavioral intervention were published earlier~\citep{dodge2024internet,wu2024benefited}.

Cognitive tests were administered by telephone to participants engaging in the study during the COVID-19 pandemic. MoCA was changed to Telephone MoCA during this period.   
Out of 94 subjects randomized into the intervention group, 52 subjects had in-person MoCA (as opposed to Telephone MoCA).  Among them, 39 participants with all the available data, including transcribed data, personality, and   NIH-Toolbox emotional battery assessment (discussed later), were used in \removed{the current study} \new{this work}. The demographic characteristics of these participants are shown in \autoref{table:demo}.

\new{For this feasibility study, we opted to use the data from the first week of interviews to capture the participants’ baseline state of cognitive function and psychological well-being at the time of the gold-standard cognitive and psychological assessments. The I-CONECT intervention study indicated that participants experienced improvements in both cognitive functions and emotional well-being over the six months of study participation~\citep{dodge2024internet,wu2024benefited}. Consequently, the behaviors observed after the first week were likely influenced by the intervention, which is why we focused on the recordings from the enrollment week for studying the feasibility of quantifying baseline conditions of cognitive impairment and psychological well-being of older adults, including both those with normal cognition and MCI.}

\subsection{Outcomes and Clinical Assessment}\label{assessmentScoreExplaination}

In this study, we aimed to classify participants with cognitive assessments derived from various scales. \new{Following previous work for quantifying MCI \citep{chaitra_2025_mci_camera},} the MoCA scores were dichotomized into `high' and `low' categories using a cutoff of 24, which \removed{is}\new{was} based on the median score of our participants. A high score indicate\removed{s}\new{d} better cognitive function. The Normal Cognition (vs. MCI) assessment is the binary encoding of clinician evaluations (NACC UDS V3, Form D1: Clinical Diagnosis Section 1). This encoding assign\removed{s}\new{ed} a value of 1 to indicate normal cognition and 0 to indicate MCI.   
Regarding the CDR, participants had scores of either 0, indicating no cognitive impairment, or 0.5, indicating questionable or very mild dementia. These scores were dichotomized accordingly.

Social network and \removed{P}\new{p}sychological well-being assessments included LSNS-6 \citep{Lubben2006} for the amount of social interaction, neuroticism from the NEO Five-Factor Inventory (Neuro) \citep{McCrae2005}, and NIH Toolbox Emotional Battery (NIHTB-EB) \citep{NIHTBEB}.   
The \removed{latter}\new{NIHTB-EB} has three composite scores:  negative affect, social satisfaction, and psychological well-being \citep{Babakhanyan2018}.
For the LSNS-6\removed{ items version, which is used here}, the cutoff score of 12 is the suggested threshold to define social isolation~\citep{Lubben1984lsns}.
For neuroticism, our participants' median score of 16 \removed{is}\new{was} used to dichotomize our participants into groups that have higher or lower negative emotional reactivity to stressful stimuli.
From NIHTB-EB, negative affect, social satisfaction, and psychological well-being composite scores \citep{Babakhanyan2018} were dichotomized with medians of 44.10, 48.66, and 53.70, respectively, to group our participants into high and low-score groups. \new{The distribution following dichotomization was approximately balanced across all tasks, as shown in~\autoref{fig:grid}.}

\subsection{Predictors and Multimodal Analysis System for Remote Interview}
\begin{figure*}[t]
    \centering
    \includegraphics[width=\textwidth]{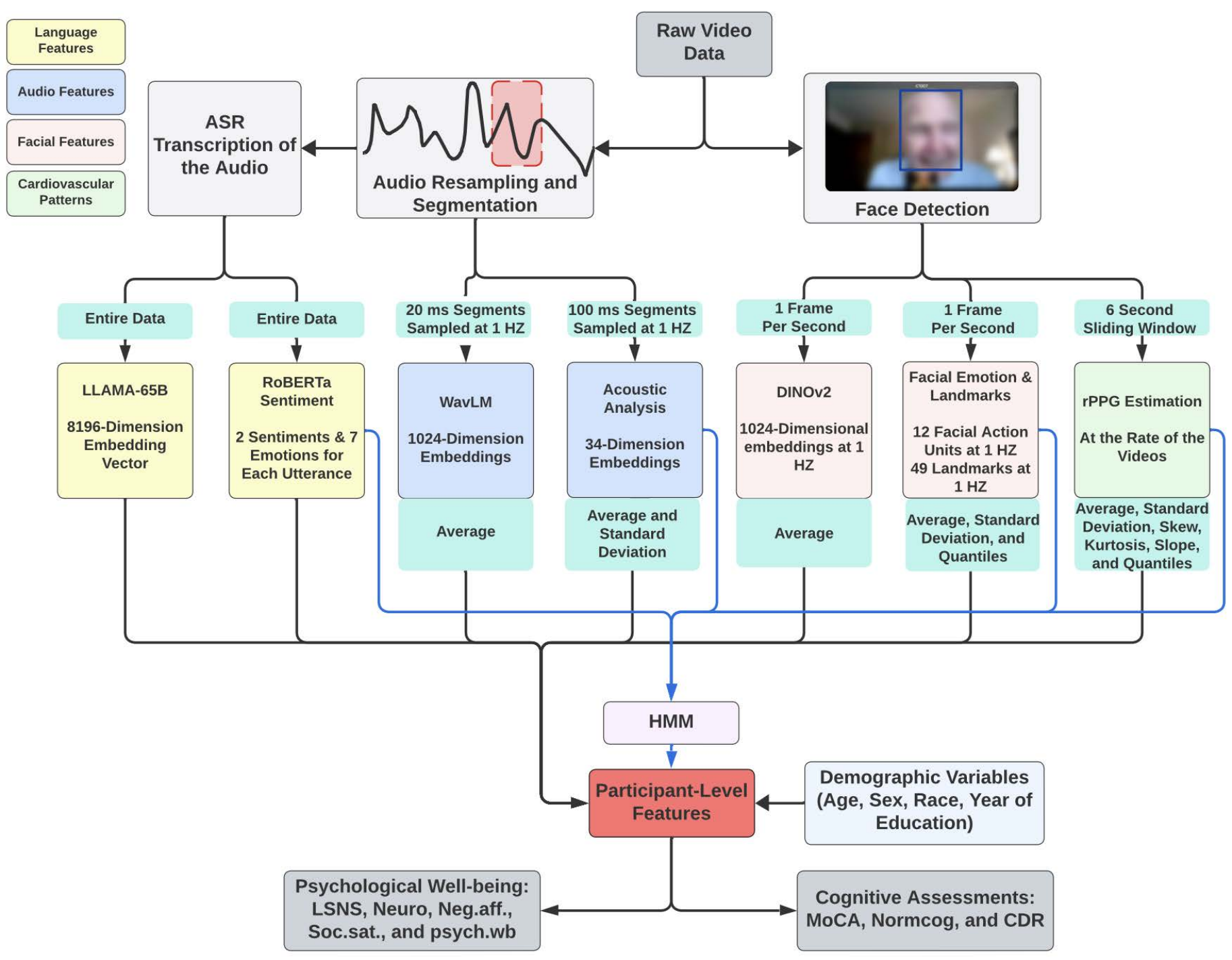}
    \caption{Overview of the processing pipeline illustrating the extraction and analysis of language, audio, facial, and cardiovascular patterns from conversation\new{al} video data. 
    The pipeline integrate\removed{s}\new{d} state-of-the-art feature representation models for each modality, including LLaMA-65B\new{~\citep{touvron2023llama}} for language feature embedding, RoBERTa~\citep{liu2019roberta} and WavLM~\citep{chen2022wavlm} for sentiment and audio processing, DINOv2\new{~\citep{oquab2023dinov2}} for facial feature extraction, and rPPG\new{~\citep{boccignone2022pyvhr}} for cardiovascular signal estimation. 
    }
    \label{fig:pipeline}
\end{figure*}

\paragraph{Overall Pipeline}
Our proposed multimodal analysis framework \removed{uses}\new{used} facial, acoustic, linguistic, and cardiovascular patterns to quantify the cognitive function and psychological well-being of the participants during remote interviews (i.e., semi-structured conversations).\removed{The participant segment of the interview recordings for the facial, vocal, linguistic, and cardiovascular patterns are extracted.} ~\new{All multimodal patterns were systematically extracted from the participant segment of the video, each represented as a time series. These multimodal patterns were then analyzed using binary shallow ML and deep learning classifiers to predict the }
\removed{The extracted time-series multimodal features are aggregated over the video using temporal pooling or Hidden Markov Model (HMM).
The video-level features are processed with binary classifiers, logistic regression, and/or gradient-boosting classifiers to classify the} dichotomized (high or low) rating \removed{scales} of the cognitive, social network, personality, and psychological well-being assessments. The overall pipeline is shown in \autoref{fig:pipeline}.

\paragraph{Preprocessing}
Our video data \removed{records}\new{recorded} both the participant's and moderator's activities during the remote interview, which require\removed{s}\new{d} segmenting the participant portion of the recording for analysis.
Our video recording show\removed{s}\new{ed} the indicator of the speaker on the screen, either as a moderator or participant ID (starting with ``C'' followed by 4 digits).
We used optical character recognition (OCR), namely EasyOCR \citep{easyocr,baek2019STRcomparisons,shi2016end}, and only the frames indicating participant ID \removed{are}\new{were} kept for further analysis.
From participant video segments, we used RetinaFace \citep{serengil2024lightface}, a facial detection model, for detecting, tracking, and segmenting participants' faces.
For the language analysis, the participants' speeches were transcribed with the transcription models \citep{RefineASR} specifically developed for older adults.

\begin{addedfigure}[t]
  \centering

  \begin{adjustbox}{max totalsize={\textwidth}{0.9\textheight},center}

    \newlength{\imgw}   \setlength{\imgw}{0.485\linewidth} 
    \newlength{\imgsep} \setlength{\imgsep}{0.5em}         
    \newlength{\rowskip}\setlength{\rowskip}{0.45em}       

    \begin{minipage}{\textwidth}
      \begin{subfigure}[t]{\imgw}
        \includegraphics[width=\linewidth]{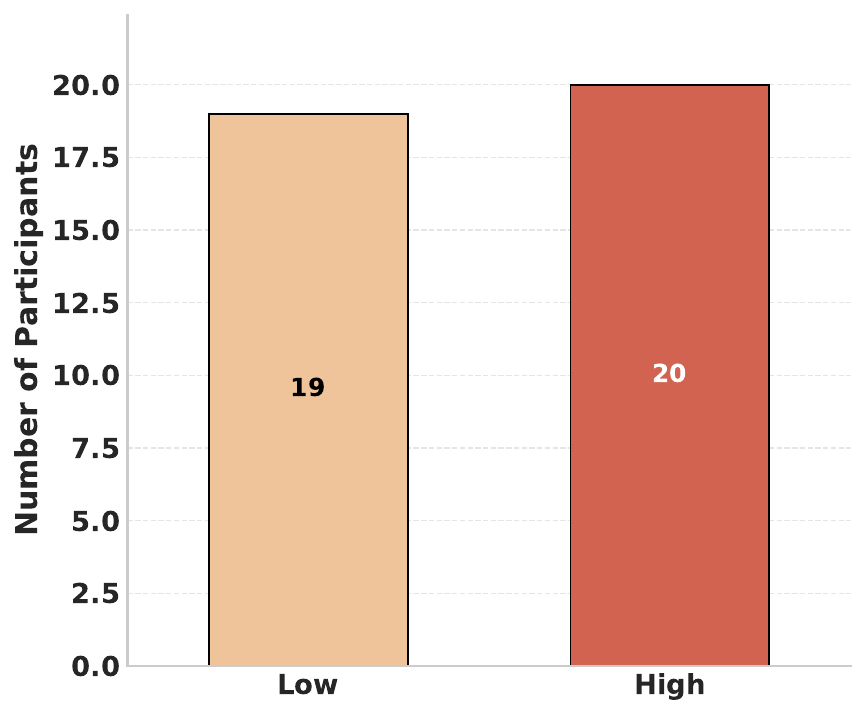}
        \caption{LSNS-6}\label{fig:g1}
      \end{subfigure}\hspace{\imgsep}%
      \begin{subfigure}[t]{\imgw}
        \includegraphics[width=\linewidth]{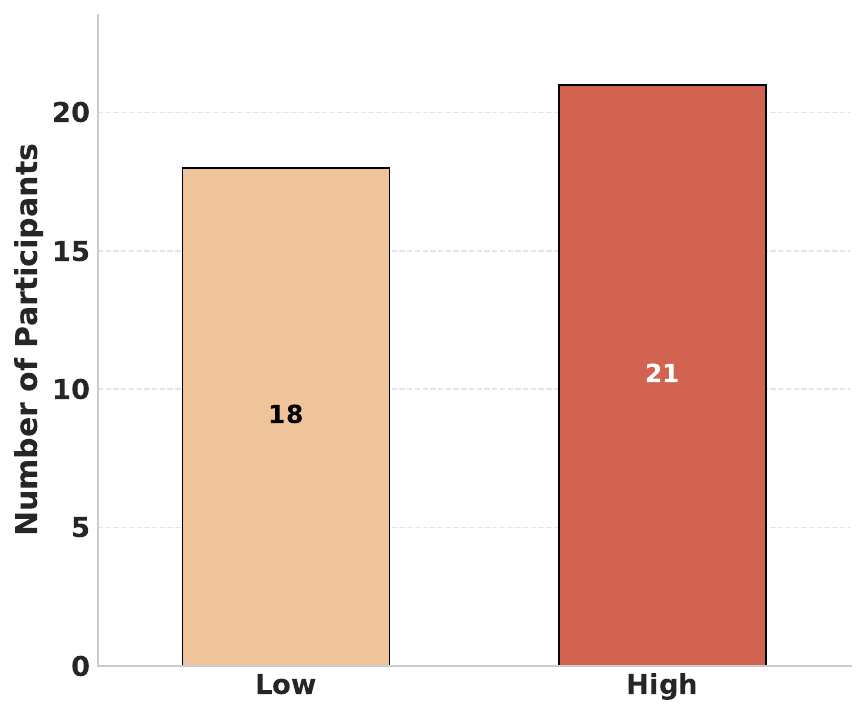}
        \caption{MoCA}\label{fig:g2}
      \end{subfigure}

      \vspace{\rowskip}

      \begin{subfigure}[t]{\imgw}
        \includegraphics[width=\linewidth]{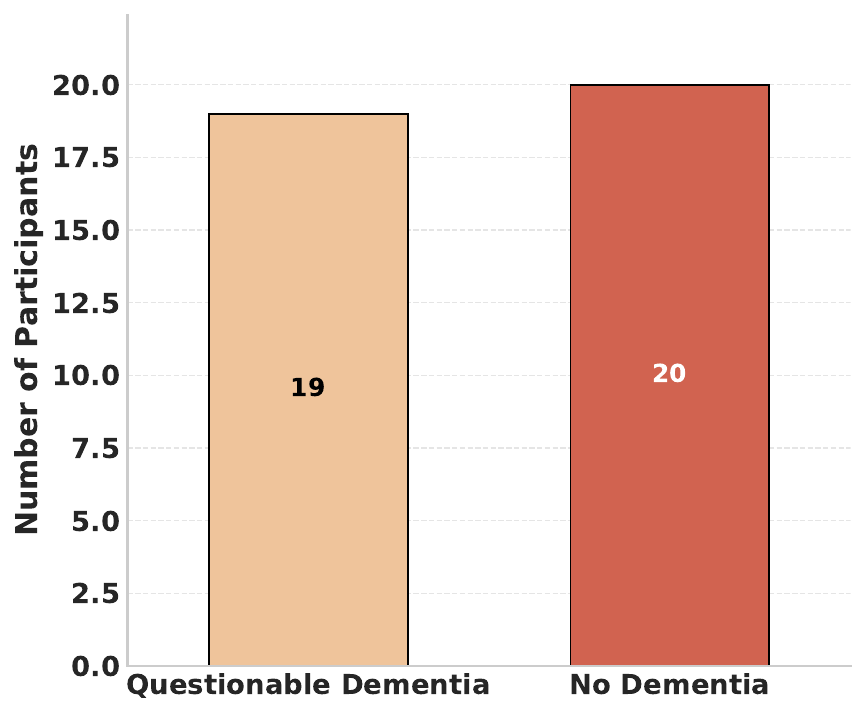}
        \caption{CDR}\label{fig:g3}
      \end{subfigure}\hspace{\imgsep}%
      \begin{subfigure}[t]{\imgw}
        \includegraphics[width=\linewidth]{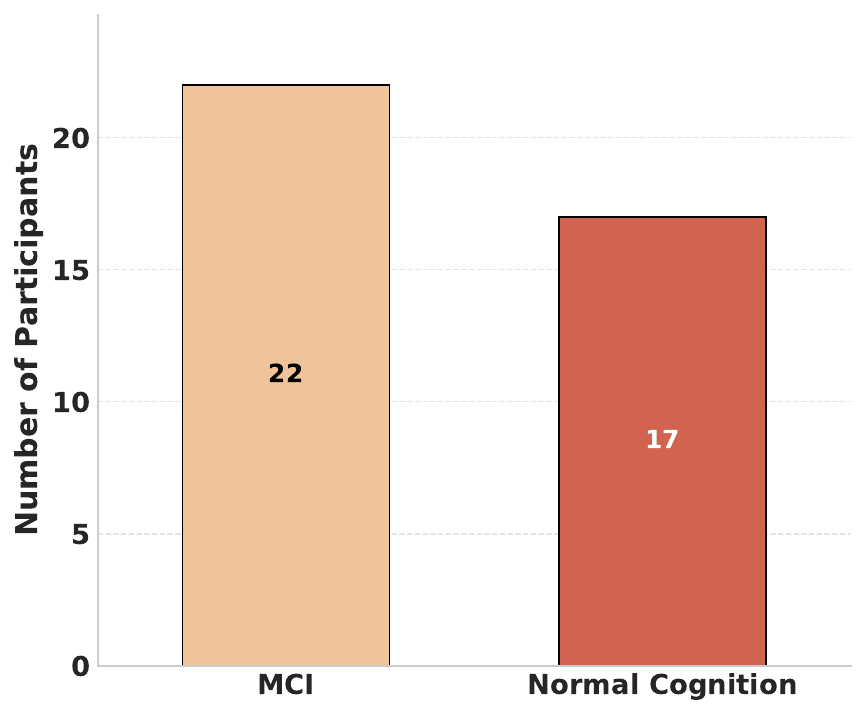}
        \caption{Clinical Diagnosis of MCI}\label{fig:g4}
      \end{subfigure}

      \vspace{\rowskip}

      \begin{subfigure}[t]{\imgw}
        \includegraphics[width=\linewidth]{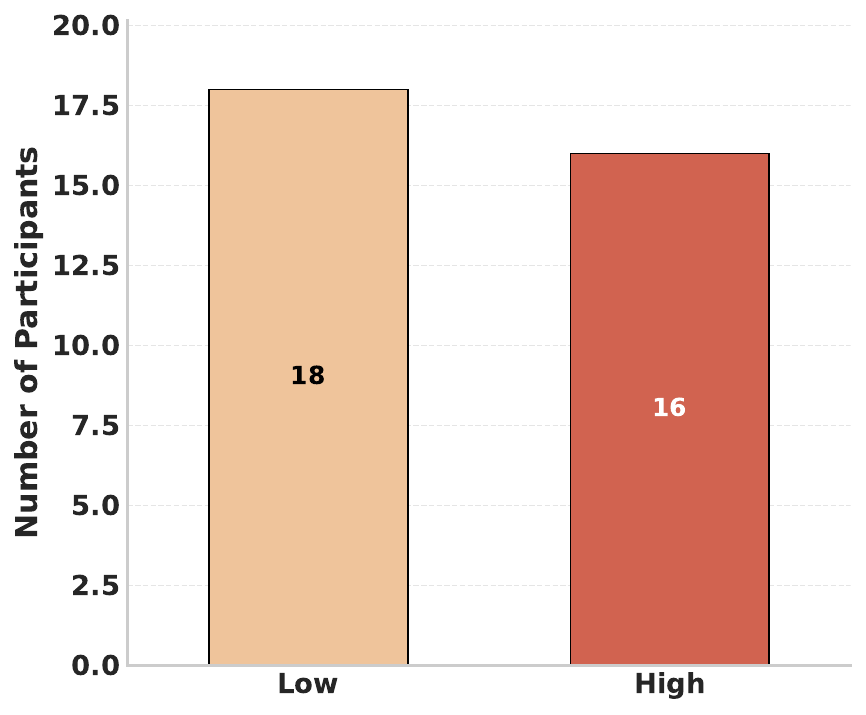}
        \caption{Negative Affect}\label{fig:g5}
      \end{subfigure}\hspace{\imgsep}%
      \begin{subfigure}[t]{\imgw}
        \includegraphics[width=\linewidth]{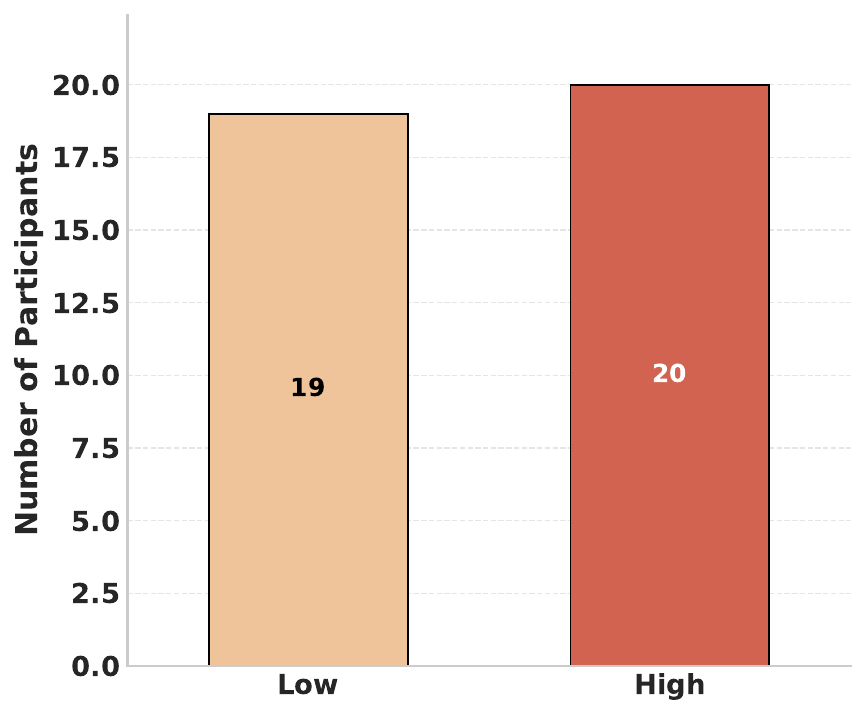}
        \caption{Neuroticism}\label{fig:g6}
      \end{subfigure}

      \vspace{\rowskip}

      \begin{subfigure}[t]{\imgw}
        \includegraphics[width=\linewidth]{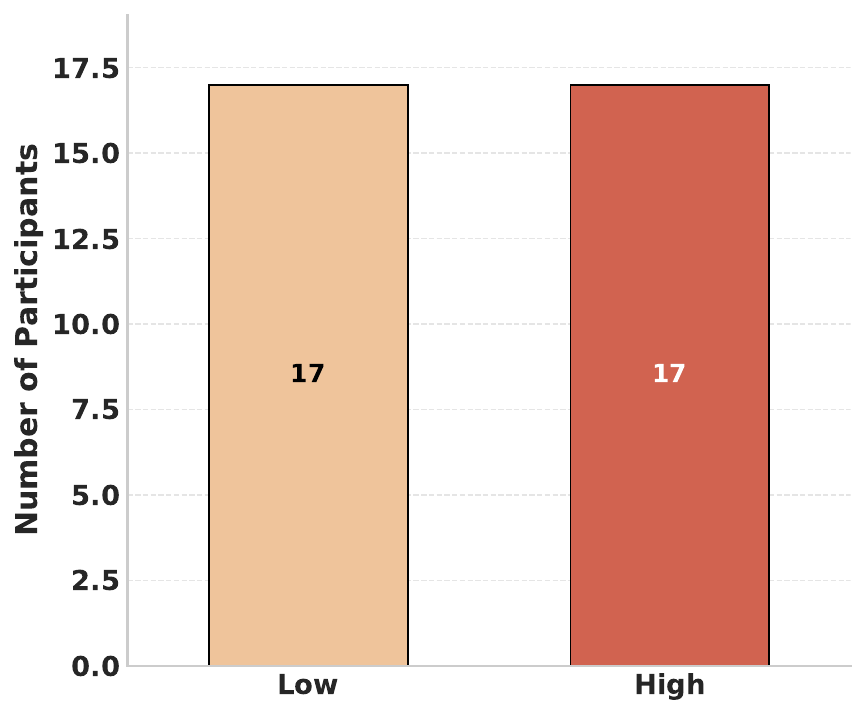}
        \caption{Psychological Well-being}\label{fig:g7}
      \end{subfigure}\hspace{\imgsep}%
      \begin{subfigure}[t]{\imgw}
        \includegraphics[width=\linewidth]{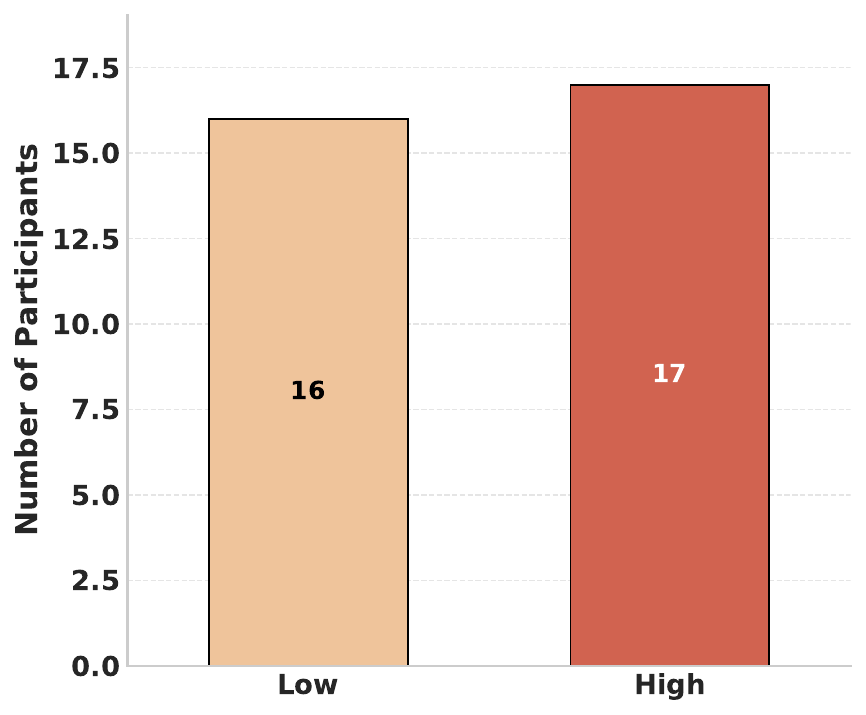}
        \caption{Social Satisfaction}\label{fig:g8}
      \end{subfigure}
    \end{minipage}

  \end{adjustbox}

  \caption{Distribution of participants' labels across the different outcomes and clinical assessments after dichotomization.}
  \label{fig:grid}
\end{addedfigure}
\clearpage

\paragraph{Facial Biomarkers}
We extracted generic facial expression features using DINOv2 \citep{oquab2023dinov2}, a foundation model for facial representation with 1024-dimensional visual embeddings.
We also extracted facial emotion, landmark\new{s}, and action units (AUs) with facial analysis pipelines used for previous mental health studies \citep{jiang2024multimodal,ekman,shao2018deep}.
Facial emotion included \removed{7}\new{seven} categories\removed{included being}\new{:} neutral, happy, sad, surprised, fearful, disgusted, and angry \citep{ekman}.
Overall, our facial features were extracted at \new{a sampling rate of 1 Hz}\removed{a 1 Hz sampling rate}.

\paragraph{Cardiovascular Biomarkers}
rPPG signals capture subtle variations in skin color that result from blood volume changes in peripheral blood vessels. These signals are extracted from video recordings of a person’s face. We extracted rPPG signals using the pyVHR package \citep{boccignone2022pyvhr}\removed{, a rPPG extraction model}.
To estimate heart rate from rPPG signals, we analyzed the power spectral density of rPPG signals~\new{at}\removed{every} ~six-second intervals, advancing \removed{1}\new{one} second at a time.
The final HRV features were derived by \removed{taking}\new{calculating} the 5th, 25th, 50th, 75th, and 95th quantile\new{s} of the estimated beats per minute, representing \new{the} statistical properties of HRV during the interview, using the pyVHR package.

\paragraph{Acoustic Biomarkers}
We first downsampled the audio to 16 kHz for the acoustic feature\new{s} extraction. Then, we extracted generic acoustic features from vocal tones using the WavLM \citep{chen2022wavlm} model, a foundation model for human speech analysis, \removed{every}\new{at a rate of} 20 ms.
We also extracted hand-crafted statistical acoustic features every 100 ms using \new{the} PyAudioAnalysis package \citep{giannakopoulos2015pyaudioanalysis}, such as spectral energy \removed{or}\new{and} entropy, which \removed{was}\new{were shown to be} effective for depression and anxiety analysis \citep{jiang2024multimodal}.

\paragraph{Linguistic Biomarkers}
We encoded the entire interview transcript for participants using LLaMA-65B \citep{touvron2023llama} to capture the high-level context of text representations in an 819\removed{6}\new{2}-dimensional vector.
We also extracted \removed{7}\new{seven} emotions (neutral, happiness, sadness, surprise, fear, disgust, and anger) and positive and negative sentiments \removed{in}\new{at the} utterance level using RoBERTa models \citep{liu2019roberta,hartmann2022emotion,hartmann2023more}. Both models are large language foundation models (LLMs). 

\paragraph{Participant-level Feature Aggregation}
\new{The extracted feature time series were processed differently for the shallow ML and deep learning models. \autoref{tab:marker-modality} shows a summary of the features extracted for each biomarker group.}

\vspace{-0.2in}
\subparagraph{\new{Shallow ML Models}}
\new{The temporal features were aggregated over the entire video sequence by extracting statistical features, such as mean and standard deviation, as shown in \autoref{fig:pipeline}, as well as summarized by a four-state Hidden Markov Model (HMM) with Gaussian observation models, using the \textit{SSM} package \citep{Linderman_SSM_Bayesian_Learning_2020}. The four states were chosen based on Jiang et al.'s study~\citep{jiang2024multimodal}, where they were effective in modeling depression and anxiety in psychiatric populations. HMMs were also successfully used in modeling emotional states with behavioral wearable data \citep{sukei_hmm_emotion_2021}.}
\removed{Once all modality features are extracted, they are aggregated over time to represent the entire interview sequence for the participant outcomes. Specifically, we extracted statistical features, such as average or standard deviation, over the entire video sequence.
We also trained a two-state HMM with Gaussian observation models to capture the temporal dynamics of each biomarker time-series using the \textit{SSM} package \citep{Linderman_SSM_Bayesian_Learning_2020}.} 
\removed{The dimensionality of the observations was determined by the length of the feature set corresponding to the patient with the longest sequence.} 
\new{For HMM, we applied the following standard processes to capture the temporal dynamics of the multimodal (facial, vocal, linguistic, and cardiovascular) feature time series: 
(1) each subject's sequence was padded to the maximum temporal length in our training dataset, $T_{max}$, calculated from the participant with the longest video length; (2) a four-state HMM was fit to all training subjects' data, which was then used to infer each subject's most likely hidden states ($T_{max}\times4$); (3) the temporal dynamics of the four hidden states were then summarized by calculating the duration and frequency of each state. We later used the HMM-based features to further train a shallow ML classifier (high vs. low outcome) for each feature and task.}

\vspace{-0.2in}
\subparagraph{\new{Deep Learning Models}}
\new{We directly processed the raw feature time series using sequential analysis models, like Long Short-Term Memory (LSTM) networks and Transformers \citep{zou_lstm_trans_2023, aina_hybrid_lstm_2024}. The multimodal feature time series extracted from each participant's video data had different lengths, as the interview duration differed for each participant, ranging from 25 to 47 minutes.
To model the overall temporal dynamics equally for all participants' video data, we temporally uniformly sampled the multimodal feature time series to a fixed length for each modality feature across all participants.
For uniform sampling, we used the minimum sequence length for each modality feature across our participants' data as follows: rPPG at 1510; Emotion$+$AUs at 587; DINOv2 at 587; RoBERTa Sentiment at 93; and for WavLM and Acoustic features, the length was set to 1500.}

\newcolumntype{M}[1]{>{\centering\arraybackslash}m{#1}} 

\begin{addedtable}[t]
  \centering
  \caption{Biomarker modality and the corresponding feature sets extracted. `X+HMM' means the HMM-based feature extracted from the `X' feature time series.}
  \label{tab:marker-modality}
  \footnotesize
  \setlength{\tabcolsep}{12pt} 
  \renewcommand{\arraystretch}{1.1}
  \begin{tabular}{M{0.19\linewidth} M{0.30\linewidth}} 
    \hline
    \textbf{Biomarker Modality} & \textbf{Features} \\
    \hline
    Demographics &
      \begin{tabular}[c]{@{}c@{}}Age+Sex+Years of education \\ Age \\ \end{tabular} \\
    \hline
    {Facial} &
      \begin{tabular}[c]{@{}c@{}}DINOv2 \\ Emotion{+}AUs \\ Emotion{+}AUs{+}HMM \\ \end{tabular} \\
    \hline
    {Cardiovascular} &
      \begin{tabular}[c]{@{}c@{}}rPPG \\ rPPG{+}HMM \\ \end{tabular} \\
    \hline
    {Audio} &
      \begin{tabular}[c]{@{}c@{}}WavLM \\ Acoustic \\ Acoustic{+}HMM \\ \end{tabular} \\
    \hline
    {Language} &
      \begin{tabular}[c]{@{}c@{}}LLaMA-65B \\ RoBERTa Sentiment \\ RoBERTa Sentiment{+}HMM \\ \end{tabular} \\
    \hline
  \end{tabular}
\end{addedtable}

\paragraph{Multimodal Fusion and Classification}

We applied a late~\removed{-} fusion approach, which aggregate\removed{s}\new{d} the classification outcomes from each modality \new{feature based on \autoref{tab:marker-modality}}, which was more effective than the early~\removed{-} fusion approaches in previous studies, especially with a small sample size (N=39) \citep{jiang2024multimodal}.
\new{\autoref{tab:model-modalities} shows the combinations of models and corresponding features used in our experiments to study the effectiveness of variants of ML models, including both shallow ML and deep learning models.}

\vspace{-0.2in}
\subparagraph{\new{Shallow ML Models}}
\new{We studied the effect of shallow ML models popularly used in previous mental health and human behavior analysis studies, including logistic regression classifier with L2 regularization (LR), gradient-boosting decision tree (GBDT) classifier, Support Vector Machines (SVM), and Random Forest (RF)~\citep{mohamed_svm_rf_2023,saad_development_svm_rf_2024,jiang2024multimodal}.
For SVM and RF, we used all the features from each biomarker modality. But, for LR and GBDT, we used feature sets from each modality that were low- (<100) and high ($\geq$100)-dimensional features, respectively, following Jiang et al. \citep{jiang2024multimodal}, which was effective in quantifying depression and anxiety in the psychiatric population.
}

\vspace{-0.2in}
\subparagraph{\new{Deep Learning Models}}
\new{
Following previous work in mental health screening \citep{zou_lstm_trans_2023, aina_hybrid_lstm_2024}, we used LSTM and Transformers directly on the raw feature time series data, rather than the aggregated statistical and HMM-based features.
For the non-temporal feature sets, including LLaMA text embeddings and demographics, we used a multilayer perceptron (MLP).
}

\vspace{-0.2in}
\subparagraph{\new{Late Fusion from Per-feature Classification}}

\new{Final decisions were obtained by fusing per-feature classification outputs using a late fusion approach. 
We employed three different aggregation rules for fusion: (i) \emph{majority voting} over hard class predictions produced by the per-feature models, (ii) \emph{average-probability} fusion that averaged class probabilities produced by the per-feature models and selected the class with the highest probability, and (iii) \emph{selective voting}, which first selected features whose AUC exceeded 0.5 for validation sets in a nested cross-validation to exclude classifiers performing poorly due to irrelevant modality features for a classification task. The selected classifiers were then used for majority voting on the left-out test sets.
}


\begin{addedtable}[t]
  \centering
  \small
  \caption{Classifiers and corresponding features used in our experiments, including the HMM-processed variants (entries ending with ``$+$ HMM''). }
  \begin{tabular}{p{0.30\textwidth} p{0.62\textwidth}}
    \hline
    \textbf{Model} & \textbf{Features} \\
    \hline
    LR &
    Emotion \& AUs; Demographics; RoBERTa sentiment; Acoustic; Emotion \& AUs $+$ HMM; RoBERTa sentiment $+$ HMM; Acoustic $+$ HMM \\[2pt]
    \hline
    GBDT &
    DINOv2; rPPG; LLaMA-65B; WavLM; rPPG $+$ HMM \\
    \hline
    SVM &
    All features \\[2pt]
    \hline
    RF &
    All features \\
    \hline
    LSTM &
    Emotion \& AUs; DINOv2; Acoustic; WavLM; RoBERTa sentiment; rPPG;  \\[2pt]
    \hline
    Transformer &
    Emotion \& AUs; DINOv2; Acoustic; WavLM; RoBERTa sentiment; rPPG;\\[2pt]
    \hline
    MLP &
    Demographics; LLaMA-65B \\[2pt]
    \hline
  \end{tabular}
  
  \label{tab:model-modalities}
\end{addedtable}

\removed{We applied a late-fusion approach for classification, which was more effective than the early-fusion approaches in previous studies \citep{jiang2024multimodal}.
For each modality, we first used a logistic regression classifier with L2 regularization or a gradient-boosting classifier to classify the dichotomized ratings (high and low) for cognitive impairment and other outcomes. 
Then, we aggregated the classification scores from all modalities using majority voting or average scores approaches.
The majority voting outputs the final prediction by voting classification results from all modalities.
The average score first averages the probability of the predicted class for all modalities and makes the final decision with the class with the higher probability.}

\subsection{Experiment}





\subsubsection{Quantifying Cognitive Impairment and Psychological Well-being}
For evaluating our multimodal analysis system, we used participant-wise 5-fold nested cross-validation with \removed{20}\new{100} runs.
For each fold, we split 64\%, 16\%, and 20\% of our participants into training, validation, and testing splits, respectively. 
\new{This evaluation aimed to assess the generalizability of our trained model for unseen test subjects not included in the training sets, which is important in clinical applications.}
For the evaluation metric\new{s}, we used \removed{an} area under the receiver operating characteristic curve (AUROC or AUC), \removed{and} accuracy, \new{and macro-F1 score}, following previous work in human behavior analysis \citep{jiang2024multimodal}. \removed{We also evaluated the standard deviation of the performance from all cross-validated models to measure the statistical significance of the model performances.}
\new{For statistical significance analysis, we also report their 95\% confidence intervals (CI), aggregated from 100 runs of 5-fold nested cross-validation.}

We evaluated unimodal and multimodal fusion models in various combinations to understand the relevance of each modality and the interplay between the modalities for assessing cognitive function and psychological well-being in older adults.
For multimodal fusion, we explored 1) acoustic and language fusion, 2) face and cardiovascular fusion, and 3) all modalities combined. 
\removed{Differently from majority voting, which considers all modalities, the selected voting only includes the classification results for the modality that achieved AUC > 0.5 for the validation set.
This was to exclude noise from classifiers performing poorly due to irrelevant features for multimodal fusion.}
We also evaluated the classification performance using demographic variables (years of education, gender, \new{and} age), \removed{and race,} \new{shown in \autoref{tab:marker-modality}}, to understand their predictability in cognitive impairment and other outcomes.
\removed{Years of education and age are used as continuous real-valued variables, and gender and race are represented as categorical one-hot encoders for our participants.} \new{All demographic variables were one-hot encoded (treating age as categorical via one-hot encoding of unique values) and then standardized using a z-score transformation to ensure comparability across variables.}
We separately evaluated the age-only classifier in addition to the demographic classifier, according to previous work using age for prescreening MCI \citep{celsis2000age}.
We also included demographic variables for all multimodal fusion analyses, assuming this information is usually available from participants' input in real-world deployment scenarios.

\subsubsection{Feature Importance Analysis}
\new{
Quantifying the relevance of each feature for model predictions is important to build clinicians' trust in clinical decision support \citep{ahmad_interpretable_2018}. 
We applied feature importance analysis for \textit{modality feature-level} for the best-performing multimodal fusion model for each classification task to quantify cognitive impairment and psychological well-being. Specifically, we adapted the MM-SHAP technique \citep{parcalabescu-frank-2023-mm-shap}, which was developed for multimodal classification models.}
\new{MM-SHAP first computed per-sample per-modality (feature set) scores from the per-sample per-feature SHAP (SHapley Additive Explanations) values, $\phi_j(x_i)$, which were then normalized across the modality features. We divided the per-sample per-modality (feature set) scores by the modality feature dimension to normalize the modality feature SHAP value to account for  the different dimension sizes of each modality feature:
\begin{equation}
\label{eq:modality-share-sample}
s_m(x_i) \;=\;
\frac{ \dfrac{1}{|\mathcal{F}_m|}\sum_{j\in\mathcal{F}_m} \big| \phi_j(x_i) \big| }
     { \sum\limits_{m'=1}^{M} \dfrac{1}{|\mathcal{F}_{m'}|}\sum\limits_{j\in\mathcal{F}_{m'}} \big| \phi_j(x_i) \big| },
\qquad \sum_{m=1}^{M} s_m(x_i)=1.
\end{equation}}
\new{\noindent where, we index modality feature sets by $m\in\{1,\dots,M=12\}$, shown in \autoref{tab:marker-modality}, with feature index sets $\{\mathcal{F}_m\}$, and each modality feature $m$ has dimensionality $=|\mathcal{F}_m|$.}
\new{ The final per-modality feature global contribution was calculated by averaging $s_m(x_i)$ over all samples:
\begin{equation}
\label{eq:modality-share-mean}
\bar{s}_m \;=\; \frac{1}{N} \sum_{i=1}^{N} s_m(x_i).
\end{equation}
This yielded per-modality feature contributions that sum to one, giving equal weight to each sample and mitigating dominance by high-dimensional features.
For statistical significance of feature importance, we report the average MM-SHAP values from 100 runs of participant-wise nested 5-fold cross-validation.
}

\subsubsection{Fairness and Bias Analysis}

\new{Fairness in ML concerns whether models make equitable predictions across participant sub-groups, based on sensitive attributes, such as age, sex, disease conditions, etc \citep{yuan_algorithmic_fairness_2023, jiang2024evaluating}. 
We evaluated the bias of our multimodal classification pipeline for quantifying psychological well-being and cognitive impairment and studied the effect of the bias mitigation technique on the trade-off between model bias and performance.
}

\paragraph{\new{Bias Evaluation}}

\new{We evaluated the bias of our multimodal classification model with respect to four sensitive attributes: sex, age, years of education (YOE), and clinical diagnosis of MCI (MCI vs.\ normal cognition), shown in \autoref{table:demo}. 
Age was categorized into two groups based on the median value, with $\leq$ 78.9 and > 78.9. 
YOE was divided into three groups: $\leq$15 years (“Below college”), 16 years (“College graduate”), and $\geq$17 years (“Graduate+”). 
For each sensitive attribute, we calculated two metrics to quantify the bias of our model \citep{yuan_algorithmic_fairness_2023, jiang2024evaluating} using the Fairlearn library \citep{weerts_2023_fairlearn}:  Equalized Odds Ratio (EOR)~\citep{hardt_2016_eo}, which captures the balance of True Positive Rate (TPR) and False Positive Rate (FPR) across all groups, and  Demographic Parity Ratio (DPR), which is the least-favored group’s selection rate divided by the most-favored group’s selection rate.
For the statistical significance of model biases, we report the average DPR and EOR, along with the 95\% confidence interval, aggregated from 100 runs of nested 5-fold participant-wise cross-validation for the best-performing models for each classification task.
}

\paragraph{\new{The Four-fifths Rule}}
\new{
There is no existing threshold of fairness specific to AI for cognitive impairment and psychological well-being in the aging population.
To determine the fairness of the model, we applied the four-fifths rule (threshold = 0.8) from the U.S. Equal Employment Opportunity Commission \citep{eeoc_adverse_impact} to our bias metrics, following the standard practice in Fair ML research. 
The model is considered perfectly fair when DPR = 1 and EOR = 1.
}


\paragraph{\new{Bias Mitigation}}

\new{
For the classification tasks with significant bias (DPR<0.8 or EOR<0.8), we studied the effect of Equalized Odds (EO) post-processing \citep{hardt_2016_eo} using Fairlearn’s \texttt{ThresholdOptimizer} \citep{weerts_2023_fairlearn} for bias mitigation. 
During the 100 runs of participant-wise 5-fold nested cross-validation, this method learned group-specific thresholds on the validation set to align TPR and FPR across groups while maximizing the balanced accuracy. We then applied these adjusted decision threshold rules on the left-out test folds for evaluation. 
After applying bias mitigation, we also report average DPR and EOR, as well as the F1 score of the classification performance, along with the 95\% CI, aggregated from 100 runs of participant-wise 5-fold nested cross-validation for statistical significance.
We compared the F1 scores before and after applying mitigation, considering bias mitigation for each sensitive attribute, as the EO post-processing \citep{hardt_2016_eo} learned group-specific thresholds per attribute. 
F1 scores before and after applying mitigation were calculated by averaging the F1 scores for each sub-group concerning the sensitive attributes of interest.
}

\subsubsection{Hyperparameters and Software}
\new{
Here, we detail the software environment, random seeds, model hyperparameters, and training protocol used across all experiments.}
\new{
All models shared the \emph{exact} same training, validation, and testing subject splits in each fold for fair comparison during 100 runs of participant-wise 5-fold nested cross-validation.
}

\new{\paragraph{HMMs (per feature)}
We used Gaussian 4-state HMMs, implemented using the \texttt{SSM} package \citep{Linderman_SSM_Bayesian_Learning_2020}, with K-means initialization and 20 EM iterations, following our participant-wise nested cross-validation setting.
}

\new{\paragraph{Shallow ML Models}
We used default parameters specified in \texttt{scikit-learn} \citep{scikit-learn}, otherwise stated below.
For random seed, all classifiers used \texttt{random\_state= 42+run\_number}.
\begin{itemize}
  \item \textbf{LR:} \texttt{penalty=L2}, \texttt{max\_iter=1000}.
  \item \textbf{GBDT:} \texttt{n\_estimators=100}, \texttt{learning\_rate=1.0}, \texttt{max\_depth=3}.
  \item \textbf{SVM:} RBF kernel; \texttt{probability=True}.
  \item \textbf{RF:} \texttt{n\_estimators=100}.
\end{itemize}}

\new{\paragraph{Deep Learning Models}
Deep learning models were implemented by \texttt{PyTorch} \citep{pytorch}.
Across all folds, we trained the models for 10 epochs with cross-entropy loss for binary classifications, which was generally sufficient to fit the models before overfitting significantly. 
For a fair comparison, we tried to maintain a similar number of parameters across models and used standard hyperparameter optimizations, including model size, learning rate, weight decay, and batch size.
Our final models with the best performances were as follows:
\begin{itemize}
  \item \textbf{LSTM:} hidden size = 128; one layer.
  \item \textbf{Transformer (ViT-style encoder) \citep{vit}:} embedding dim = 128; MLP hidden dim = 32; 4 heads; 2 encoder layers; dropout = 0.1; CLS token for classification.
  \item \textbf{MLP (non-temporal features):} one hidden layer (64 units), ReLU, dropout = 0.3.
\end{itemize}
}

\section{Results}\label{sec:results}


\begin{addedtable}[t]
\centering
\caption{Best pipelines per task with AUROC/F1 ($\pm$ 95\% CI) and runner-up AUROC deltas $\Delta$.}
\setlength{\tabcolsep}{3.5pt}
\scriptsize
\setlength{\extrarowheight}{5pt}   
\renewcommand{\arraystretch}{1.25} 
\begin{tabular}{@{}%
p{0.14\textwidth}%
p{0.25\textwidth}%
>{\centering\arraybackslash}p{0.14\textwidth}%
>{\centering\arraybackslash}p{0.14\textwidth}%
p{0.22\textwidth}@{}}
\toprule
\textbf{Task} & \textbf{Pipeline} & \multicolumn{2}{c}{\textbf{Metrics}} & \makecell[c]{\textbf{Runner-Up}\\\textbf{/ AUROC $\Delta$}} \\
\cmidrule(lr){3-4}
& & \textbf{AUROC} & \textbf{F1} & \\
\midrule
CDR &
\makecell[l]{RF\\(Audio+Language+Demo)} &
$0.77 \pm 0.008$ & $0.72 \pm 0.008$ &
\makecell[l]{RF (All)\\$\Delta=-0.01$} \\
\hline

MoCA &
\makecell[l]{LR\\(Demographics)} &
$0.65 \pm 0.01$ & $0.60 \pm 0.012$ &
\makecell[l]{LR (Acoustic)\\$\Delta=-0.03$} \\
\hline

\makecell[l]{Clinical Diagnosis\\ of MCI\\(NACC UDS)}
 &
\makecell[l]{SVM\\(Audio+Language+Demo)} &
$0.69 \pm 0.008$ & $0.64 \pm 0.01$ &
\makecell[l]{RF (WavLM)\\$\Delta=-0$} \\
\hline

LSNS &
\makecell[l]{LR\\(RoBERTa Sentiment)} &
$0.74 \pm 0.006$ & $0.71 \pm 0.008$ &
\makecell[l]{RF (RoBERTa Sentiment)\\$\Delta=-0.08$} \\ 
\hline

Neuroticism &
\makecell[l]{LR\\(RoBERTa Sentiment+HMM)} &
$0.66 \pm 0.012$ & $0.60 \pm 0.012$ &
\makecell[l]{RF (Emotion \& AUs)\\$\Delta=-0.01$} \\
\hline

Negative Affect &
\makecell[l]{LR\\(Acoustic+HMM)} &
$0.74 \pm 0.006$ & $0.70 \pm 0.006$ &
\makecell[l]{LR (Sentiment HMM)\\$\Delta=-0.03$} \\
\hline

\makecell[l]{Social\\Satisfaction} &
\makecell[l]{LR/GBDT\\(Facial+Cardio+Demo)} &
$0.75 \pm 0.014$ & $0.60 \pm 0.016$ &
\makecell[l]{RF (Emotion \& AUs)\\$\Delta=-0.11$} \\ 
\hline

Psychological Well-being &
\makecell[l]{SVM\\(Facial+Cardio+Demo)} &
$0.72 \pm 0.014$ & $0.67 \pm 0.014$ &
\makecell[l]{RF (Facial+Cardio+Demo)\\$\Delta=-0.08$} \\ 
\bottomrule
\end{tabular}
\label{tab:best-pipelines}
\end{addedtable}
\subsection{Quantifying Cognitive Impairment and Psychological Well-being}

\new{
\autoref{tab:best-pipelines} shows the best-performing classification approaches for all the classification tasks, including both unimodal and multimodal fusion, as well as the second-best approach (or runner-up) with absolute differences in AUC from the best model (AUROC $\Delta$).
\autoref{fig:aurocs_heatmap} reports AUROC values for the three shallow models (LR/GBDT, SVM, and RF) across all tasks (rows) and feature modalities (columns). 
Each cell shows the AUROC with the corresponding best unimodal feature or multimodal fusion strategy. 
A darker color indicates higher classification performance.
In our experiment, all the deep learning models (LSTM and Transformers) showed lower classification performance compared to the best shallow ML models. 
The detailed experiment results of the deep learning models are shown in \ref{app:deep} in Table \ref{tab:best-dl} and Figure \ref{fig:dl-heatmap}.
}

\subsection{Feature Importance Analysis}

\new{
\autoref{fig:shap-3x1-cog-a}, \autoref{fig:shap-2x1-psych}, and \autoref{fig:shap-3x1-cog-b} show the modality feature importance from MM-SHAP \citep{parcalabescu-frank-2023-mm-shap} for the best-performing multimodal model for each classification task.
For CDR (\autoref{fig:shap-cdr-mult}) and clinical diagnosis of MCI (\autoref{fig:shap-normcog-mult}), we only show the importance of audio, language, and demographic modalities, as classification performance decreased when adding facial expression and cardiovascular features.
Likewise, for neuroticism (\autoref{fig:shap-neuro-mult}), social satisfaction (\autoref{fig:shap-socsat-mult}), and psychological well-being (\autoref{fig:shap-psych-mult}), we only show the importance of facial expression and cardiovascular features, as classification performance decreased when adding audio and language features.
%
%
%
}

\subsection{Fairness and Bias Analysis}

\new{
\autoref{fig:dp-cog-eor-dpr}, \autoref{fig:dp-psych-eor-dpr}, and \autoref{fig:dp-psych-eor-f1} show the bias and fairness analysis (EOR and DPR) before (\textcolor{blue}{BLUE}) and after (\textcolor{orange}{ORANGE}) applying bias mitigation for the best-performing models from \autoref{tab:best-pipelines} concerning each sensitive attribute.
\autoref{fig:dp-cog-eor-dpr} shows the results for quantifying cognitive impairments (CDR, MoCA, and clinical diagnosis of MCI).
\autoref{fig:dp-psych-eor-dpr} shows the results for quantifying LSNS and neuroticism, and \autoref{fig:dp-psych-eor-f1} shows the results for quantifying negative affect, social satisfaction, and psychological well-being from NIHTB-EB. 
\autoref{fig:dp-cog-f1-f1} and \autoref{fig:dp-psych-f1-f1} show the F1 score comparisons for before (\textcolor{blue}{BLUE}) and after applying bias mitigation (\textcolor{orange}{ORANGE}) for each attribute for all the classification tasks for the best-performing models from \autoref{tab:best-pipelines}.
\autoref{fig:dp-cog-f1-f1} shows the results for CDR, MoCA, clinical diagnosis of MCI, and LSNS, and \autoref{fig:dp-psych-f1-f1} shows the results for neuroticism, negative affect, social satisfaction, and psychological well-being.
%
}

\removed{Our experiment results for quantifying cognitive functions are shown in \mbox{\autoref{table:uni_cog}} for each modality and \mbox{\autoref{table:multi_cog}} for multimodal fusion analysis.}
\removed{
The results for quantifying social network, neuroticism, and psychological well-being are shown in \mbox{\autoref{table:uni_psych}} for each modality and \mbox{\autoref{table:multi_psych}} for multimodal fusion.}

\removed{Quantifying MCI diagnosis (5th column in \mbox{\autoref{table:uni_cog}} \& \mbox{\autoref{table:multi_cog}}) was also most effective when using language-based sentiment and emotion features (0.66 AUC and 0.69 accuracy), followed by an acoustic and language fusion model (0.66 AUC and 0.63 accuracy).}

\section{Discussion}\label{sec:discussion}





\new{
Here, we first provide an overview of the experiment results across all the classification tasks, followed by a detailed discussion of each task. Then, we discuss the limitations and future research directions.
}

\subsection{Classification Performance}

\subsubsection{Shallow ML Models}

\new{Overall, from \autoref{tab:best-pipelines} and \autoref{fig:aurocs_heatmap}, LR/GBDT models dominated five tasks: LSNS (0.74 AUC), neuroticism (0.66 AUC), negative affect (0.74 AUC), social satisfaction (0.75 AUC), and MoCA (0.65 AUC). 
RF performed best in CDR (0.77 AUC), while SVM outperformed all models in psychological well-being (0.72 AUC), and tied with RF for the clinical diagnosis of MCI (0.69 AUC).
SVM underperformed relative to LR/GBDT and RF. 
Across tasks, the best models tended to rely on the HMM-based features over the statistical features. 
}

\subsubsection{Deep Learning Models}

\new{We also demonstrated that deep learning models (even with regularization and hyperparameter tuning) overfit due to the limited dataset size (N=39), having difficulty capturing behavioral patterns related to cognitive impairment and psychological well-being that are generalizable across different subjects. 
}
\new{
From Figure \ref{fig:dl-heatmap} in \ref{app:deep}, compared to the best-performing shallow ML models, when using LSTM, absolute AUC values decreased by 21\% for classifying CDR, 15\% for classifying MoCA, and 6\% for predicting the clinical diagnosis of MCI, 20\% for classifying LSNS, 13\% for classifying neuroticism, 10\% for classifying negative affect, 21\% for classifying social satisfaction, and 15\% for classifying psychological well-being, with a median drop of 15\%. 
When using Transformer, absolute AUC values decreased by 21\% for classifying CDR, 11\% for classifying MoCA, 4\% for predicting the clinical diagnosis of MCI, 17\% for classifying LSNS, 9\% for classifying neuroticism, 13\% for classifying negative affect, 23\% for classifying social satisfaction, and 18\% for classifying psychological well-being, with a median drop of 15\%. 
While our successful classification results with shallow ML models in \autoref{tab:best-pipelines} demonstrate the feasibility of our analysis, the deep learning models' results suggest that future work should focus on collecting larger datasets to effectively train deep learning-based methods, to further improve on our feasibility study.
}
\begin{addedfigure}[t]
  \centering
  \newlength{\caproom}
  \setlength{\caproom}{\abovecaptionskip+\belowcaptionskip+5.2\baselineskip}

  \begin{adjustbox}{max width=\linewidth,
                    max height=\dimexpr\textheight-\caproom\relax,
                    keepaspectratio}
    \includegraphics{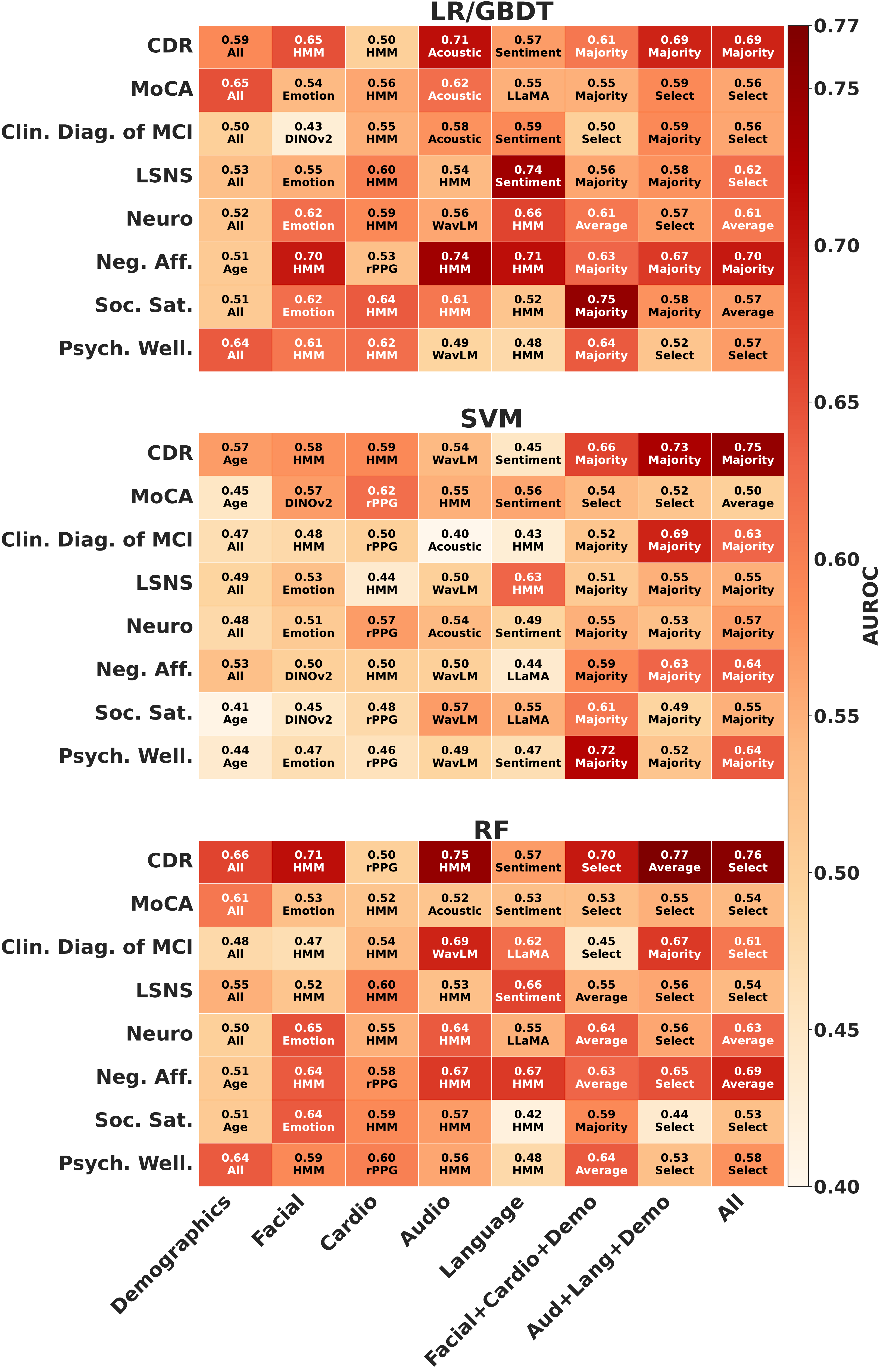}
  \end{adjustbox}

  \caption{AUROC for the shallow ML models across tasks (rows) and feature modalities (columns). Each cell shows the AUROC (best) and the best-performing feature or fusion method (bottom). 
  For example, \textit{Aud+Lang+Demo} denotes fusion of audio, language, and demographics, and \textit{All} denotes fusion of all available modalities. 
  The color scale is shared across panels and the darker color indicates higher performance.}
  \label{fig:aurocs_heatmap}
\end{addedfigure}
\clearpage

\begin{addedfigure}[t]
  \centering
  \captionsetup{font=footnotesize, labelfont=bf, skip=3pt}

  \begin{subfigure}[t]{0.98\linewidth}
    \centering
    \includegraphics[height=0.26\textheight, keepaspectratio]{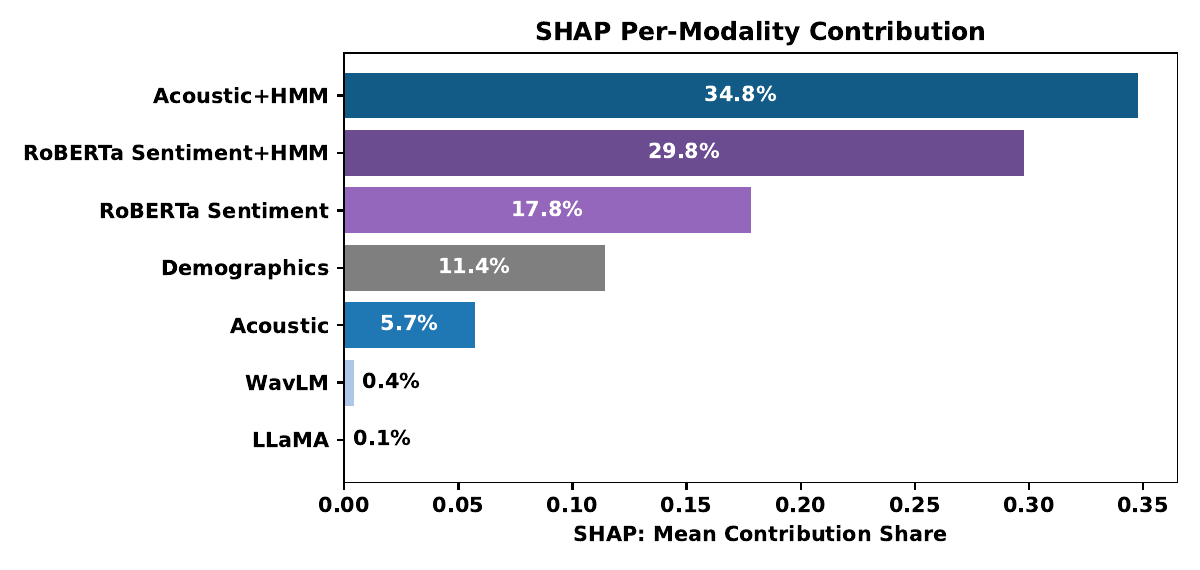}
    \caption{CDR — RF (Aud+Lang+Demo)}
    \label{fig:shap-cdr-mult}
  \end{subfigure}

  \vspace{3pt}

  \begin{subfigure}[t]{0.98\linewidth}
    \centering
    \includegraphics[height=0.32\textheight, keepaspectratio]{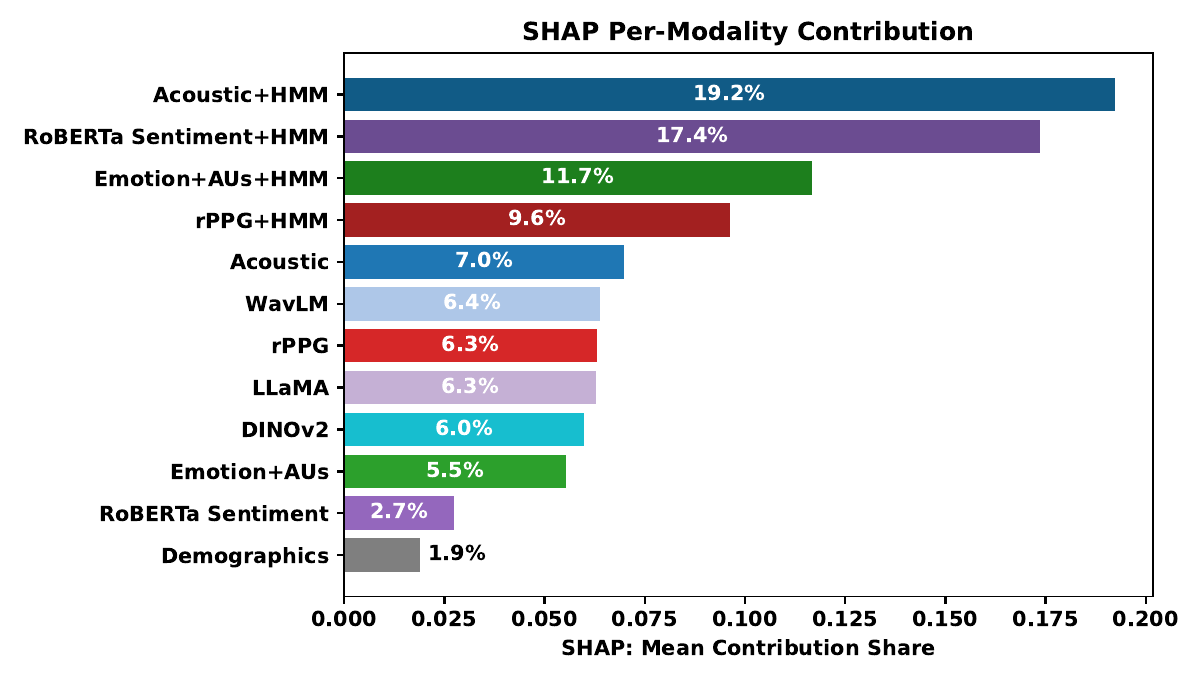}
    \caption{MoCA — LR/GBDT (All)}
    \label{fig:shap-moca-mult}
  \end{subfigure}

  \vspace{3pt}

  \begin{subfigure}[t]{0.98\linewidth}
    \centering
    \includegraphics[height=0.26\textheight, keepaspectratio]{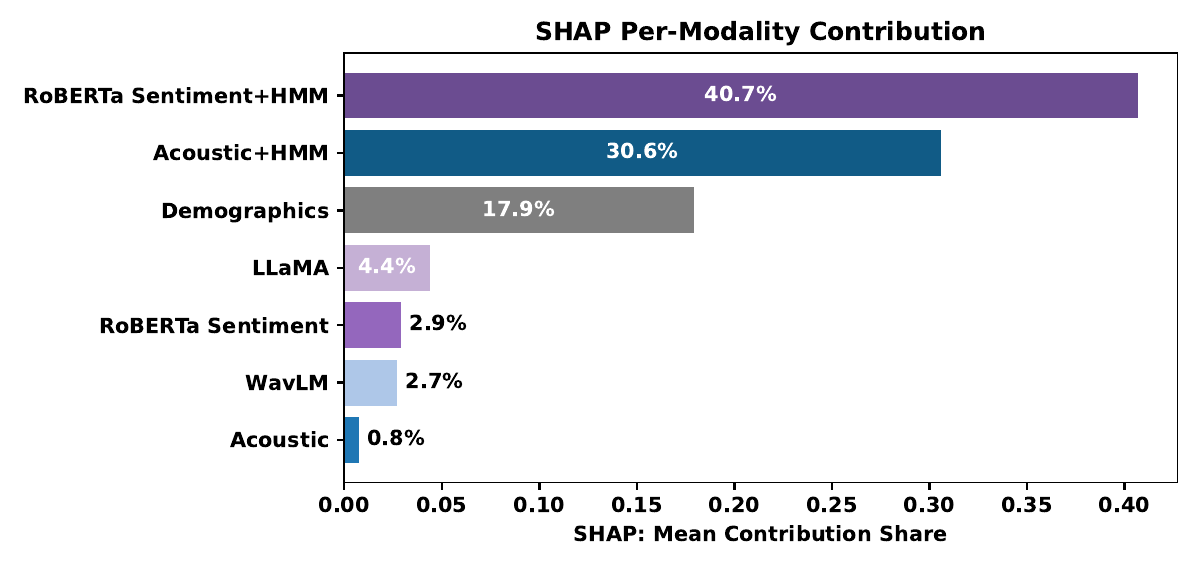}
    \caption{Clinical Diagnosis of MCI — SVM (Aud+Lang+Demo)}
    \label{fig:shap-normcog-mult}
  \end{subfigure}

  \caption{MM-SHAP bar plots for best-performing multimodal models: CDR, MoCA, and clinical diagnosis of MCI.}
  \label{fig:shap-3x1-cog-a}
\end{addedfigure}
\clearpage

\begin{addedfigure}[t]
  \captionsetup{font=normalsize}
  \captionsetup[subfigure]{font=normalsize,justification=centering,skip=2pt}
    
  \begin{subfigure}[t]{0.98\linewidth}
    \centering
    \includegraphics[height=0.34\textheight, keepaspectratio]{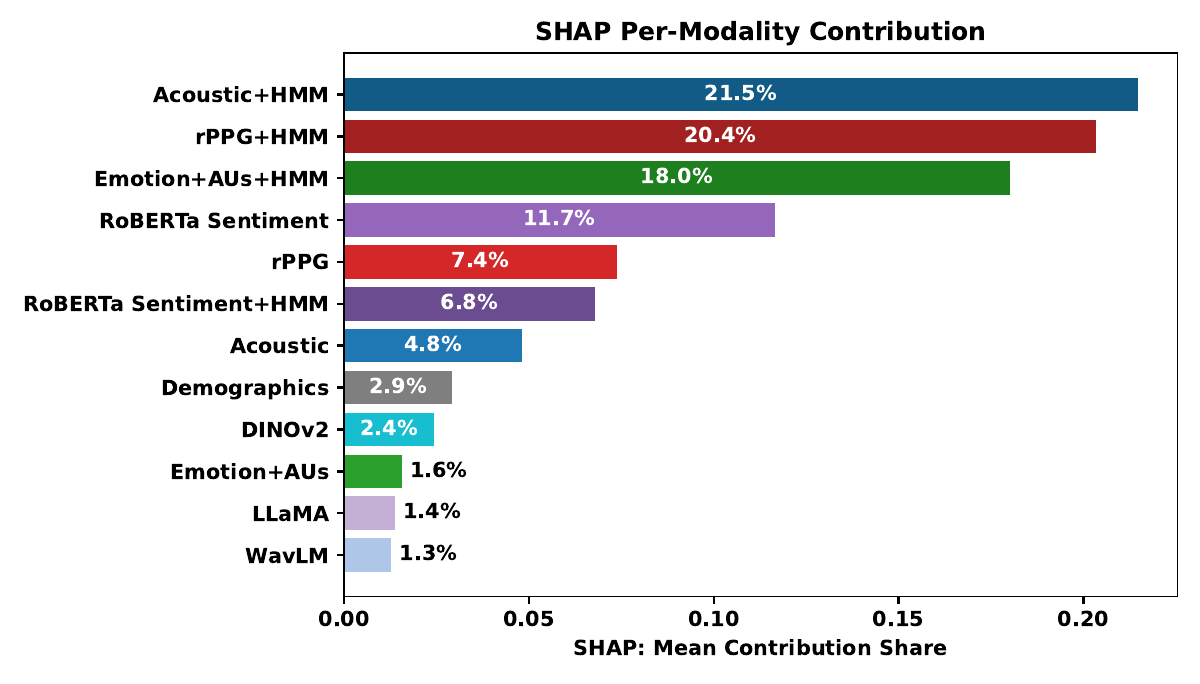}
    \caption{LSNS — LR/GBDT (All)}
    \label{fig:shap-lsns-mult}
  \end{subfigure}

  \vspace{3pt}

  \begin{subfigure}[t]{0.98\linewidth}
    \centering
    \includegraphics[height=0.26\textheight, keepaspectratio]{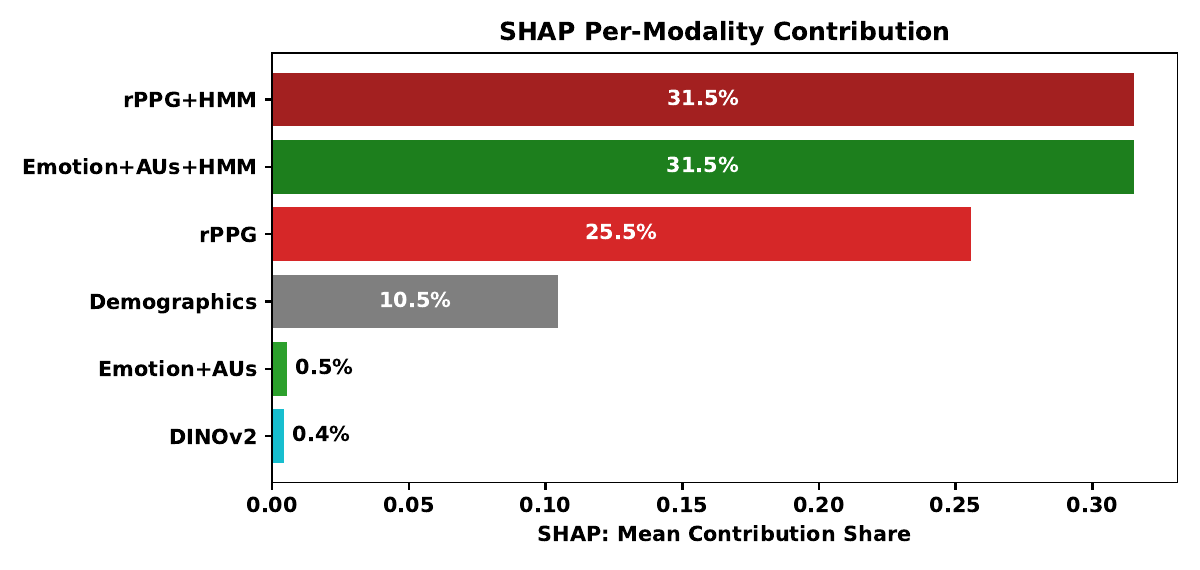}
    \caption{Neuroticism — RF (Facial+Card+Demo)}
    \label{fig:shap-neuro-mult}
  \end{subfigure}

  \caption{MM-SHAP bar plots: LSNS and Neuroticism.}
  \label{fig:shap-2x1-psych}
\end{addedfigure}

\subsection{Feature Importance Analysis}

Our findings show that facial, acoustic, and linguistic features from foundation models (DINOv2, WavLM, and LLaMA-65B) significantly underperformed, with an average absolute decrease of \removed{22\%} \new{11\%} AUC compared to the best models using handcrafted and HMM-based features from facial expression, acoustic, and linguistic-based emotion signals. 
\new{This was further confirmed via MM-SHAP analysis, where \autoref{fig:shap-3x1-cog-a}, \autoref{fig:shap-2x1-psych}, and \autoref{fig:shap-3x1-cog-b} showed that DINOv2, WavLM, and LLaMA-65B features ranked near the bottom in terms of contributions to model predictions. 
In contrast, HMM-derived features consistently ranked among the top contributors, highlighting HMM's ability to capture temporal dynamics effectively when having a small sample size.}

\removed{Those}\new{The used} foundation models \removed{are}\new{were} trained to capture generic facial appearance, acoustic waveforms, and language embeddings by training the models with a large-scale database available from the internet that contains individuals with various demographics and contexts.
We suspect such features, when directly used, are not designed to capture specific behavior patterns in cognitive impairments or psychological well-being in older adults, especially in the MCI population.
In future work, we will study the effect of transfer learning for fine-tuning those foundation models with our dataset to capture behavior patterns in older subjects with MCI \citep{Zhuang2019Comprehensive}. 

\subsection{Fairness and Bias Analysis}

\new{
Across tasks, all the best-performing models had both EOR and DPR $< 0.6$ across attributes before applying bias mitigation, indicating severe bias, except for quantifying the clinical diagnosis of MCI (\autoref{fig:grid-f-normcog-dpr}) and psychological well-being (\autoref{fig:grid-psych-dpr}), which showed $DPR_{diag} = 0.82 \pm 0.03$ and $DPR_{psych}=0.75 \pm 0.09$.
}
\new{
Years of education (YOE) suffered from the most disparities across models, with an average EOR of 0.15 and an average DPR of 0.38.
After applying bias mitigation, the fairness of quantifying cognitive impairment improved for all attributes and tasks, but with a significant absolute drop in utility ($\Delta \text{F1 score} > 0.15$).
Applying bias mitigation improved the absolute EOR by 0.41-0.65 for quantifying psychological well-being, but with a significant drop in F1 score by as much as 0.26-0.36 points.
}

\new{After applying bias mitigation, quantifying the clinical diagnosis of MCI and MoCA had the largest absolute EOR gains (+0.76).
Regardless, several tasks remained heavily biased after applying bias mitigation (e.g., CDR, LSNS, neuroticism, negative affect) with EOR < 0.6 and DPR < 0.7.
}\new{
These persistent biases across sensitive attributes for all tasks suggest that the disparities are systematic across tasks.}

\new{
We consider that biases can arise from two sources: (1) imbalanced demographic distributions, such as sex and age, shown in \autoref{table:demo}, and (2) biases inherited from pre-trained models for extracting multimodal features in our classification pipeline (DINOv2, LLaMA-65B, and WavLM), potentially encoding historic biases in facial and speech datasets available from the internet \citep{straw2020artificial,georgopoulos2021mitigating}. }\new{  Based on this feasibility study, future work needs to recruit participants with diverse demographic and cultural backgrounds, as well as explore state-of-the-art bias mitigation methods, such as adversarial debiasing \citep{yan_2020_debiasing_multimodal, seth_2023_debiasing_vlm}, for developing a robust, generalizable, and fair model.}


\begin{addedfigure}[t]
  \centering
  \captionsetup{font=footnotesize, labelfont=bf, skip=3pt}

  \begin{subfigure}[t]{0.98\linewidth}
    \centering
    \includegraphics[height=0.32\textheight, keepaspectratio]{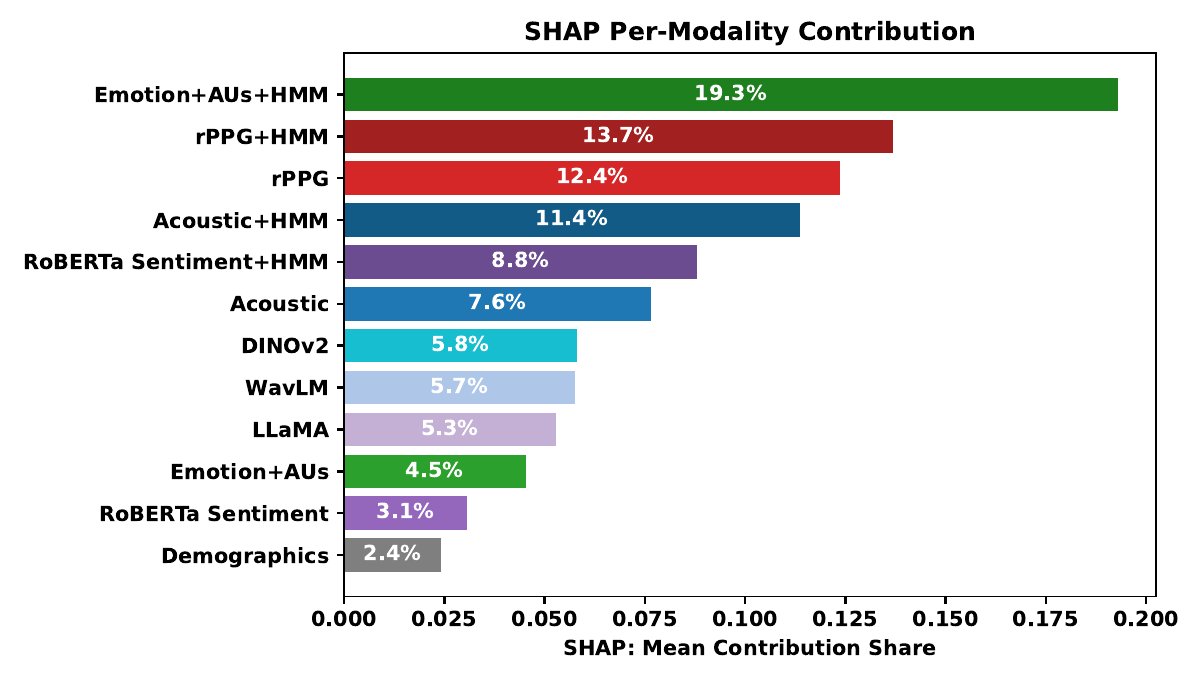}
    \caption{Negative Affect — LR/GBDT (All)}
    \label{fig:shap-negaff-mult}
  \end{subfigure}

  \begin{subfigure}[t]{0.98\linewidth}
    \centering
    \includegraphics[height=0.26\textheight, keepaspectratio]{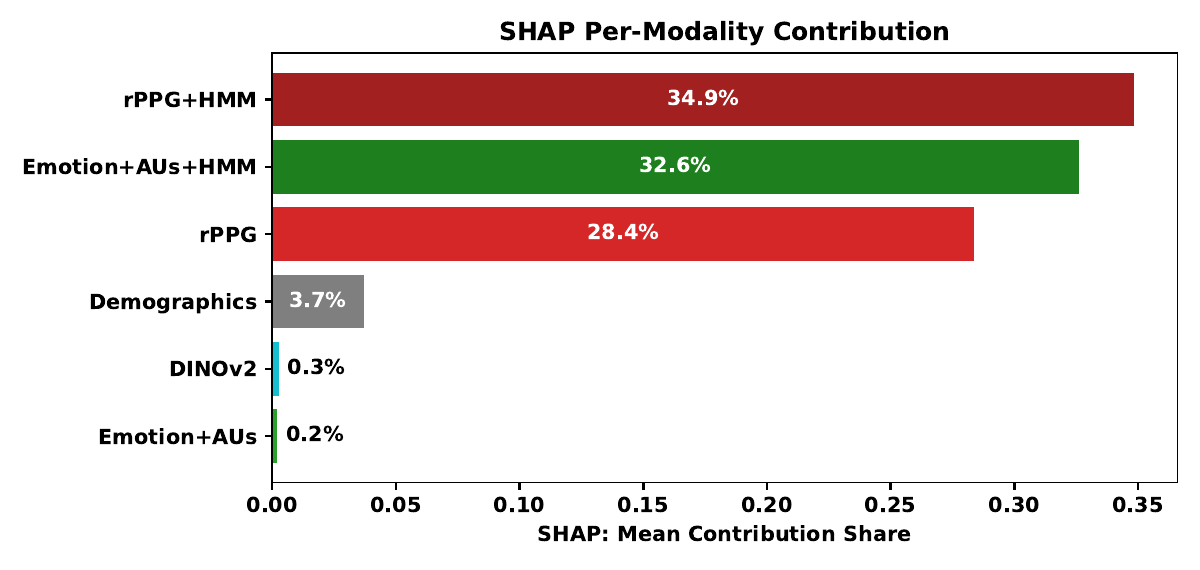}
    \caption{Social Satisfaction — LR/GBDT (Facial+Card+Demo)}
    \label{fig:shap-socsat-mult}
  \end{subfigure}
  
  \begin{subfigure}[t]{0.98\linewidth}
    \centering
    \includegraphics[height=0.26\textheight, keepaspectratio]{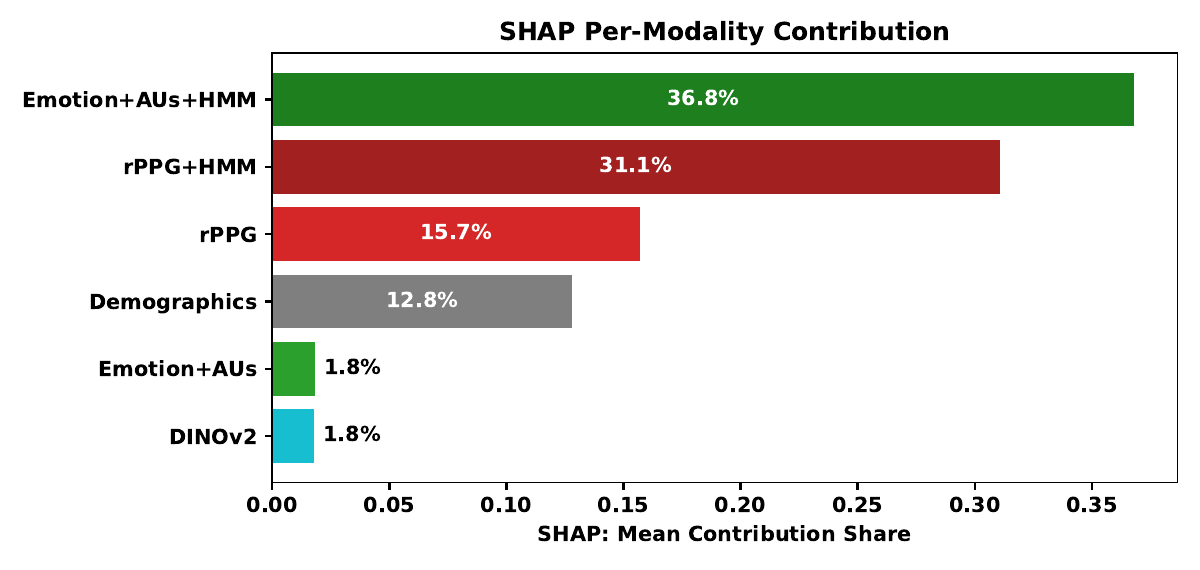}
    \caption{Psychological Well-being — SVM (Facial+Card+Demo)}
    \label{fig:shap-psych-mult}
  \end{subfigure}

  \vspace{4pt}
  \caption{MM-SHAP bar plots: negative affect, social satisfaction, and psychological well-being}
  \label{fig:shap-3x1-cog-b}
\end{addedfigure}


\begin{addedfigure}[p]
  \centering
  \begin{minipage}[c][0.9\textheight][c]{\linewidth}
    \captionsetup{font=normalsize}
    \captionsetup[subfigure]{font=normalsize,justification=centering,skip=2pt}

    \begin{subfigure}[t]{0.49\linewidth}
      \centering
      \includegraphics[width=\linewidth,keepaspectratio]{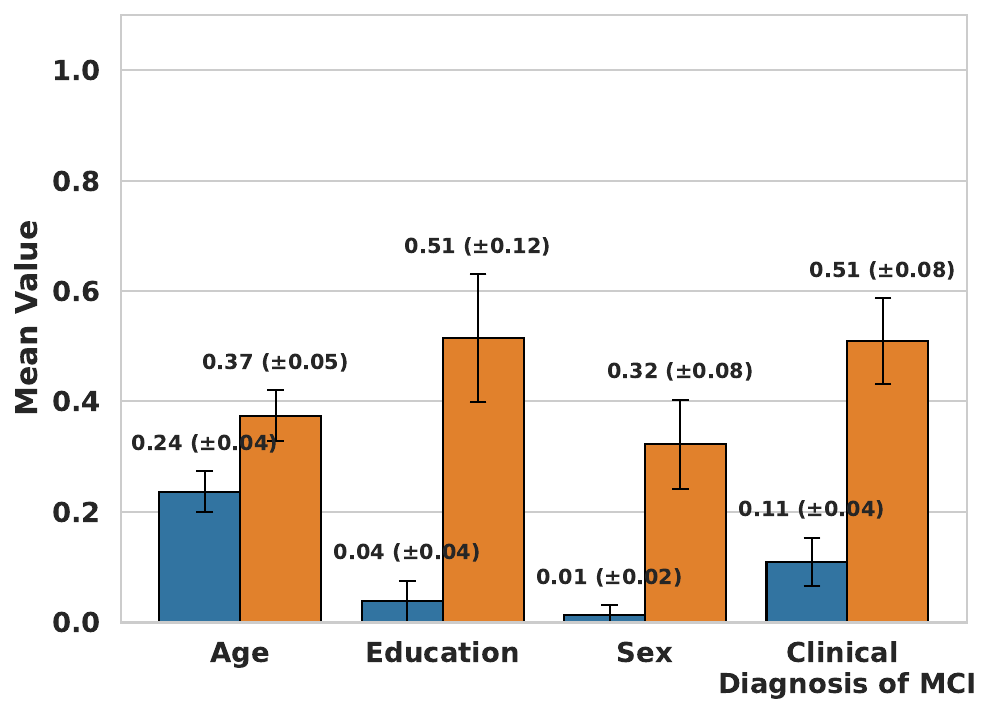}
      \caption{EOR of CDR}\label{fig:grid-a-cdr-eor}
    \end{subfigure}\hfill
    \begin{subfigure}[t]{0.49\linewidth}
      \centering
      \includegraphics[width=\linewidth,keepaspectratio]{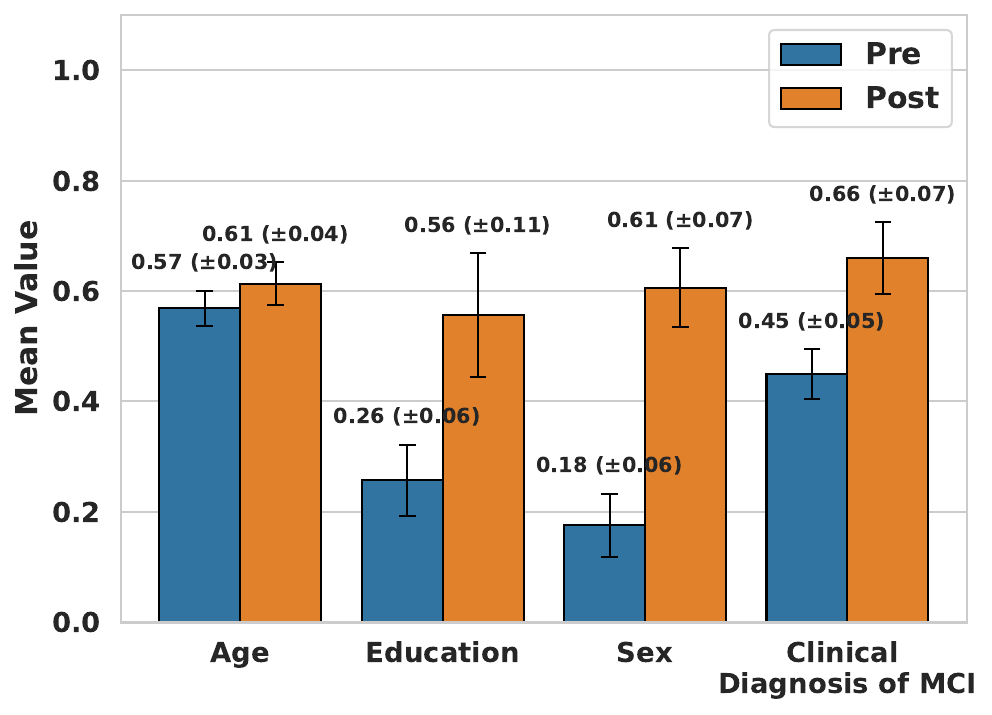}
      \caption{DPR of CDR}\label{fig:grid-b-cdr-dpr}
    \end{subfigure}

    \medskip

    \begin{subfigure}[t]{0.49\linewidth}
      \centering
      \includegraphics[width=\linewidth,keepaspectratio]{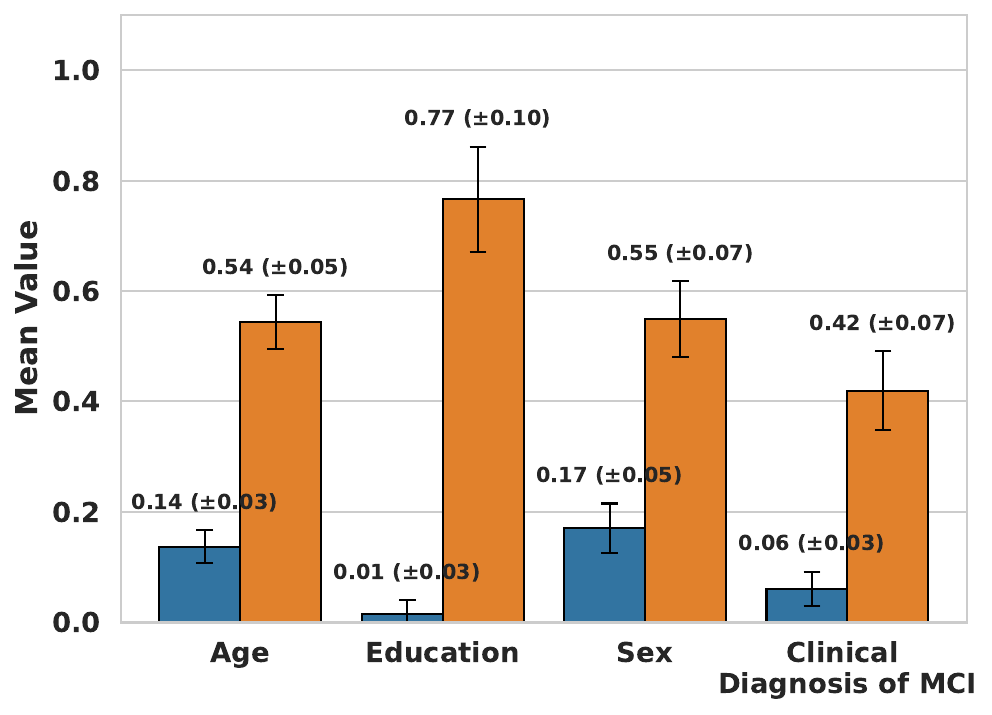}
      \caption{EOR of MoCA}\label{fig:grid-c-moca-eor}
    \end{subfigure}\hfill
    \begin{subfigure}[t]{0.49\linewidth}
      \centering
      \includegraphics[width=\linewidth,keepaspectratio]{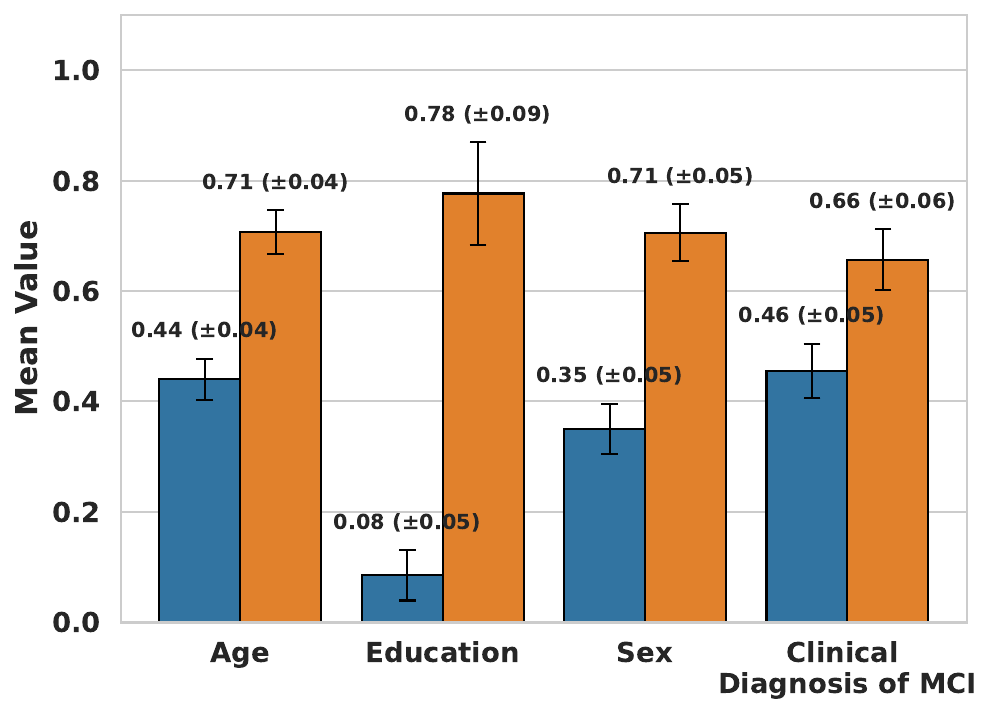}
      \caption{DPR of MoCA}\label{fig:grid-d-moca-dpr}
    \end{subfigure}

    \medskip

    \begin{subfigure}[t]{0.49\linewidth}
      \centering
      \includegraphics[width=\linewidth,keepaspectratio]{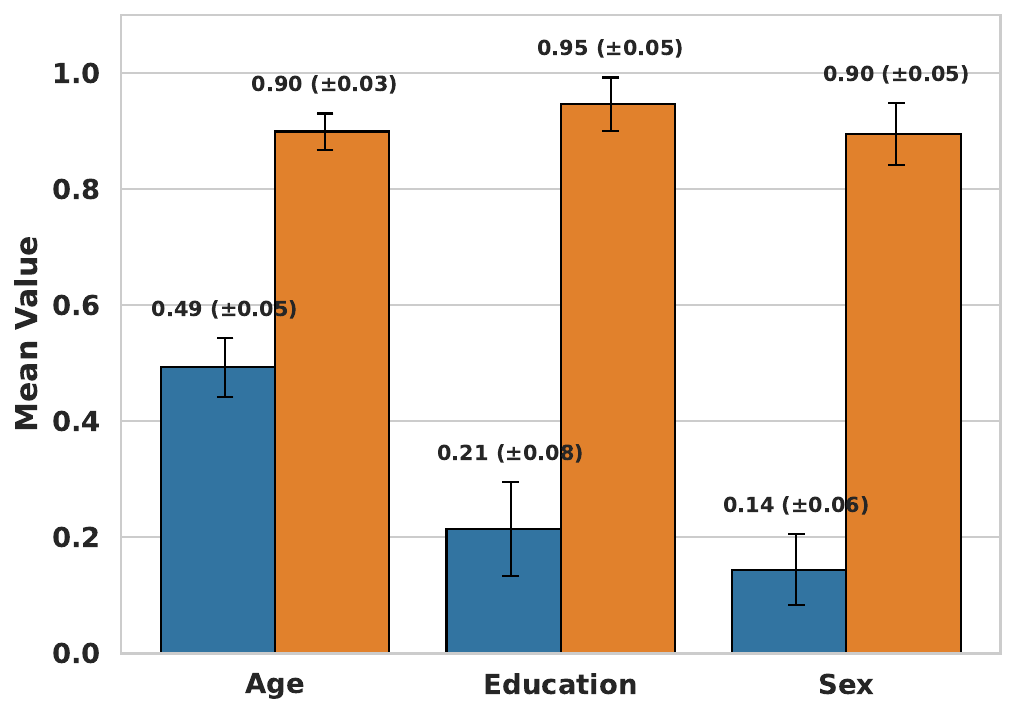}
      \caption{EOR of Clinical Diagnosis of MCI}\label{fig:grid-e-normcog-eor}
    \end{subfigure}\hfill
    \begin{subfigure}[t]{0.49\linewidth}
      \centering
      \includegraphics[width=\linewidth,keepaspectratio]{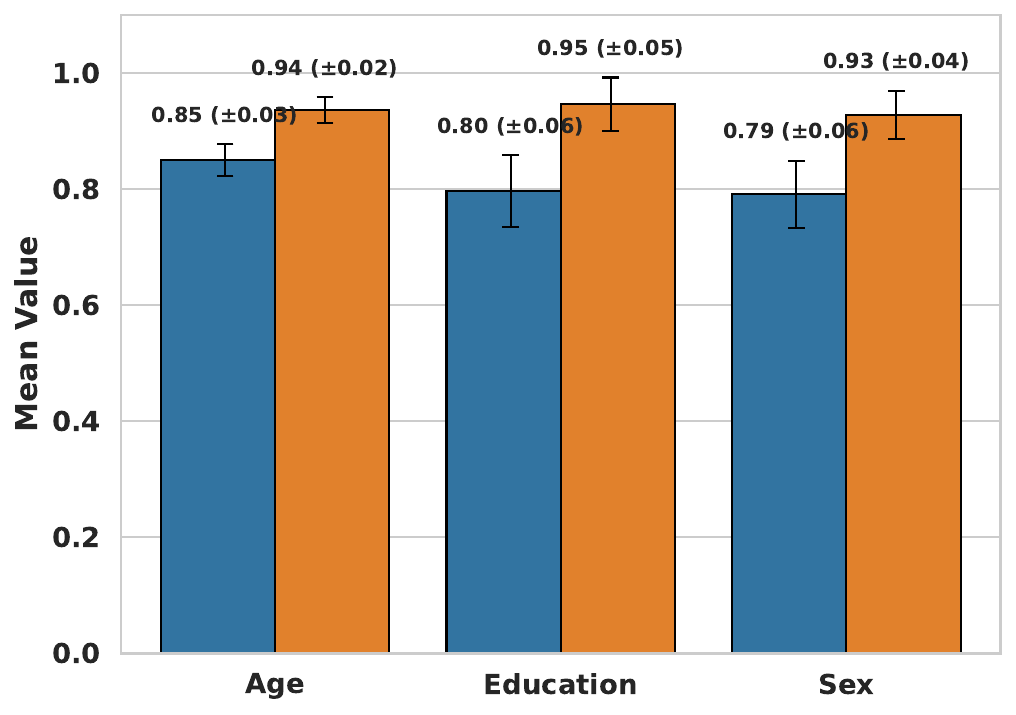}
      \caption{DPR of Clinical Diagnosis of MCI}\label{fig:grid-f-normcog-dpr}
    \end{subfigure}

    \caption{Fairness and bias analysis for quantifying cognitive impairments before and after applying bias mitigation.}
    \label{fig:dp-cog-eor-dpr}
  \end{minipage}
\end{addedfigure}
\clearpage

\begin{addedfigure}[t]
  \centering
  \captionsetup{font=normalsize}
  \captionsetup[subfigure]{font=normalsize,justification=centering,skip=2pt}

  \begin{adjustbox}{max totalsize={\textwidth}{0.92\textheight},center}
    \begin{minipage}{\textwidth}

      \begin{subfigure}[t]{0.48\textwidth}
        \centering
        \includegraphics[width=\linewidth,keepaspectratio]{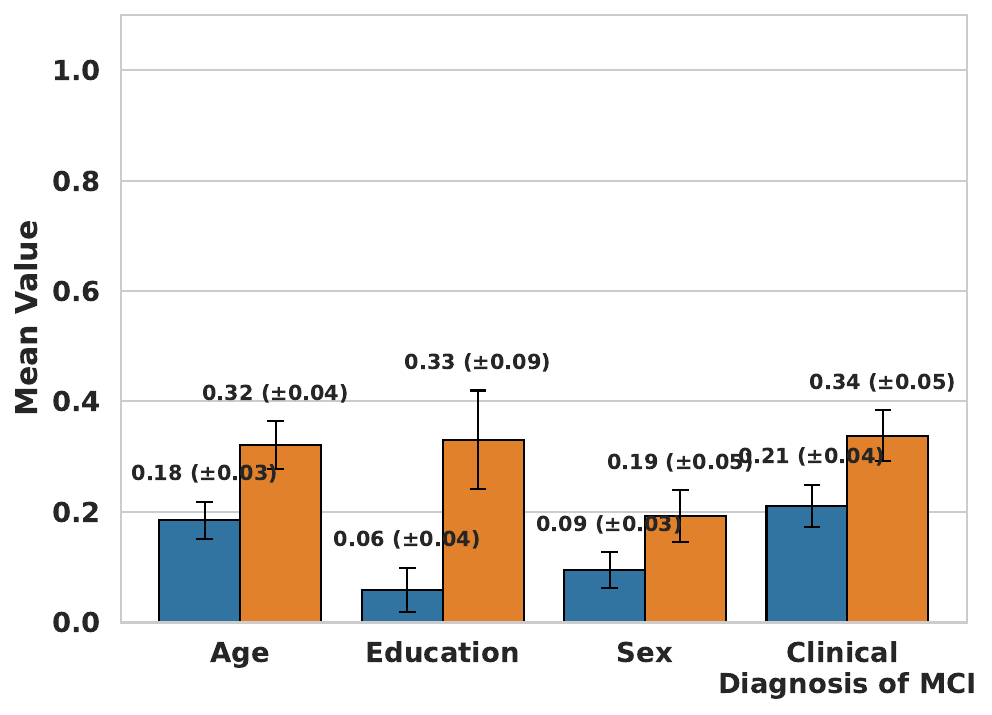}
        \caption{EOR of LSNS}
        \label{fig:grid-lsns-eor}
      \end{subfigure}\hfill
      \begin{subfigure}[t]{0.48\textwidth}
        \centering
        \includegraphics[width=\linewidth,keepaspectratio]{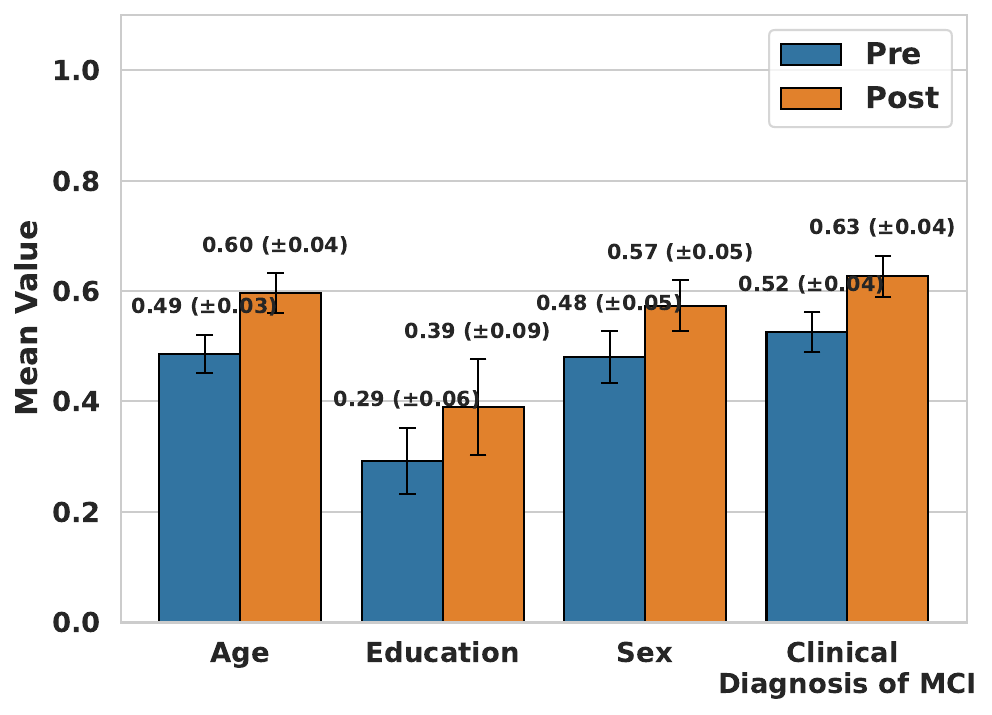}
        \caption{DPR of LSNS}
        \label{fig:grid-lsns-dpr}
      \end{subfigure}
\medskip
      \begin{subfigure}[t]{0.48\textwidth}
        \centering
        \includegraphics[width=\linewidth,keepaspectratio]{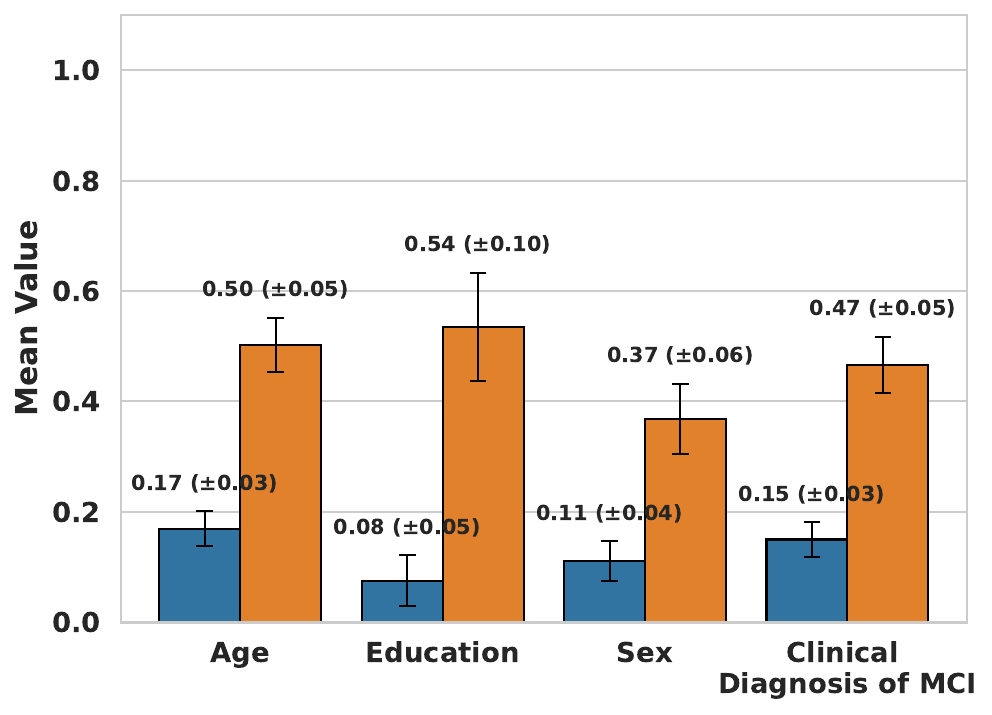}
        \caption{EOR of Neuroticism}
        \label{fig:grid-neuro-eor}
      \end{subfigure}\hfill
      \begin{subfigure}[t]{0.48\textwidth}
        \centering
        \includegraphics[width=\linewidth,keepaspectratio]{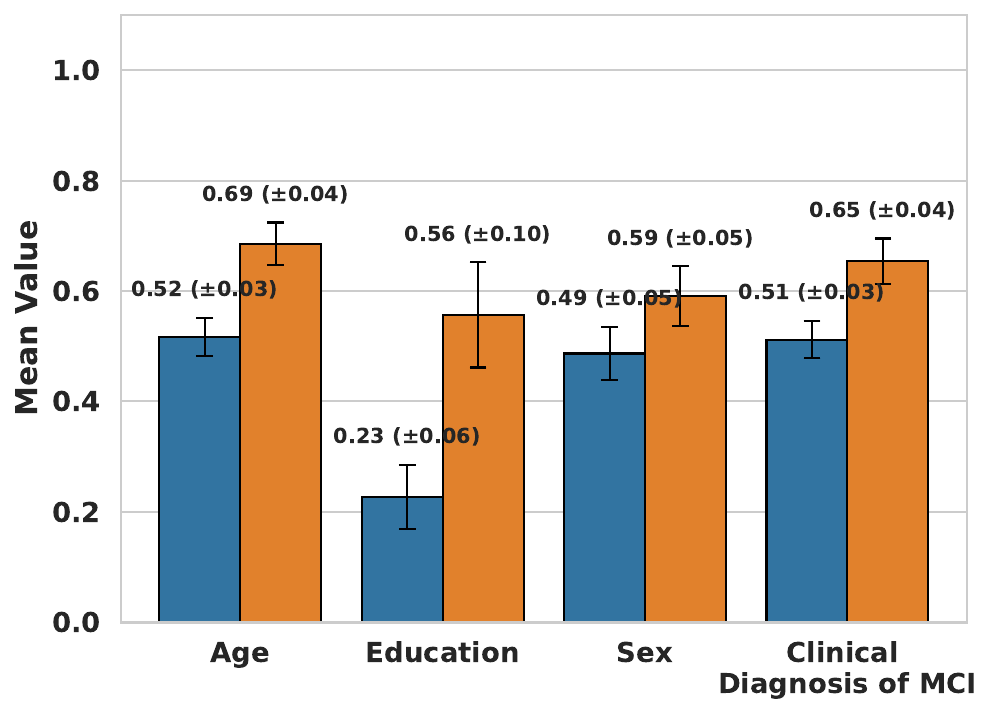}
        \caption{DPR of Neuroticism}
        \label{fig:grid-neuro-dpr}
      \end{subfigure}

    \end{minipage}
  \end{adjustbox}

  \caption{Fairness and bias analysis for quantifying social network and neuroticism assessments before and after applying bias mitigation.}
  \label{fig:dp-psych-eor-dpr}
\end{addedfigure}

\subsection{Quantifying Cognitive Assessment}
\label{sec:result_cog}


Quantifying MoCA scores and clinical diagnosis of MCI was challenging with our multimodal analysis system, showing \removed{0.64} \new{0.65} and \removed{0.66} \new{0.69} AUCs, respectively.
This is possibly due to the narrow range of MoCA \new{scores} in this group. 
The majority (68\%) of our subjects had MoCA scores between 21 and 28, with a median score of 24, having limited variability of scores to detect distinguishable features associated with MoCA $\leq24$ vs.  $>24$.
Conversely, quantifying CDR (0 vs. 0.5) proved more effective, achieving an AUC of \removed{0.78} \new{0.77}. The CDR assesses functional outcomes in daily life (e.g., forgetfulness of events, \removed{functionality in shopping} \new{shopping functionality}, and participation in volunteer or social groups), reflecting cognitive performance \removed{but} that's not necessarily linked to cognitive testing scores. 
These functional aspects \removed{are}\new{were} likely better captured by acoustic and linguistic features from remote interviews compared to measures like MoCA.

It is also worth noting that, while age has been a significant factor for predicting MoCA scores \citep{Nasreddine2005,Ciesielska2016} in the general population, it \removed{does}\new{did} not show \removed{any} \new{a strong} predictive capacity in our analysis \new{(AUC = 0.53)}. This discrepancy may be due to the participants' narrow age range (Overall, 80.69 $\pm$ 4.6) \new{as} shown in \autoref{table:demo}. 

\begin{addedfigure}[t]
  \centering
  \captionsetup{font=normalsize}
  \captionsetup[subfigure]{font=normalsize,justification=centering,skip=2pt}

  \begin{adjustbox}{max totalsize={\textwidth}{\textheight},center}
    \begin{minipage}{\textwidth}

      \begin{subfigure}[t]{0.48\textwidth}
        \centering
        \includegraphics[width=\linewidth,keepaspectratio]{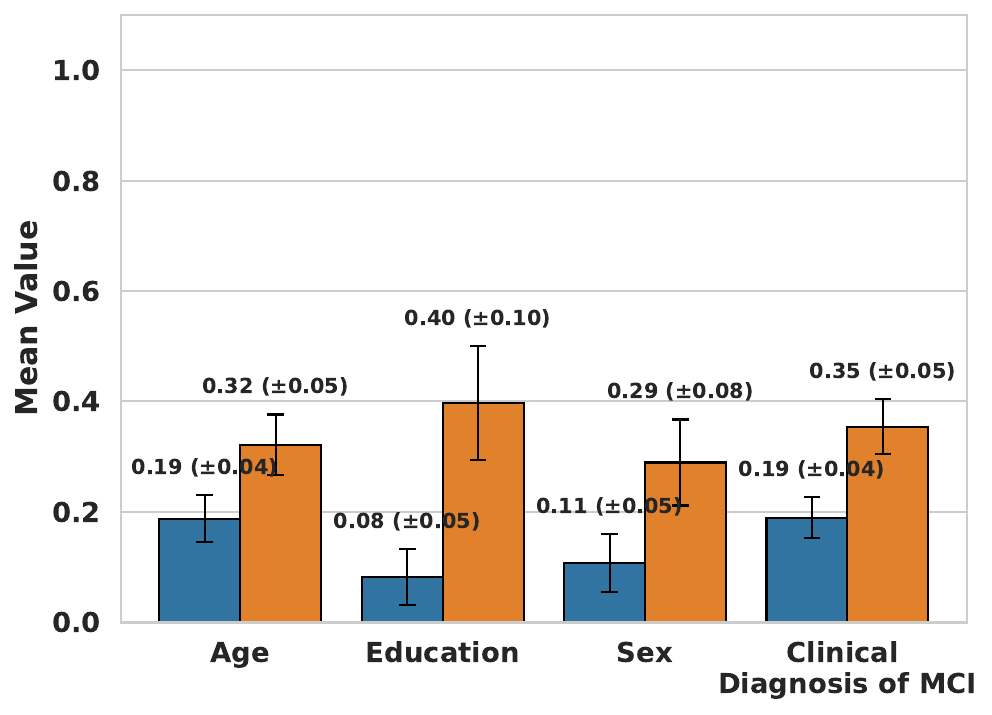}
        \caption{EOR of Negative Affect}
        \label{fig:grid-negaff-eor}
      \end{subfigure}\hfill
      \begin{subfigure}[t]{0.48\textwidth}
        \centering
        \includegraphics[width=\linewidth,keepaspectratio]{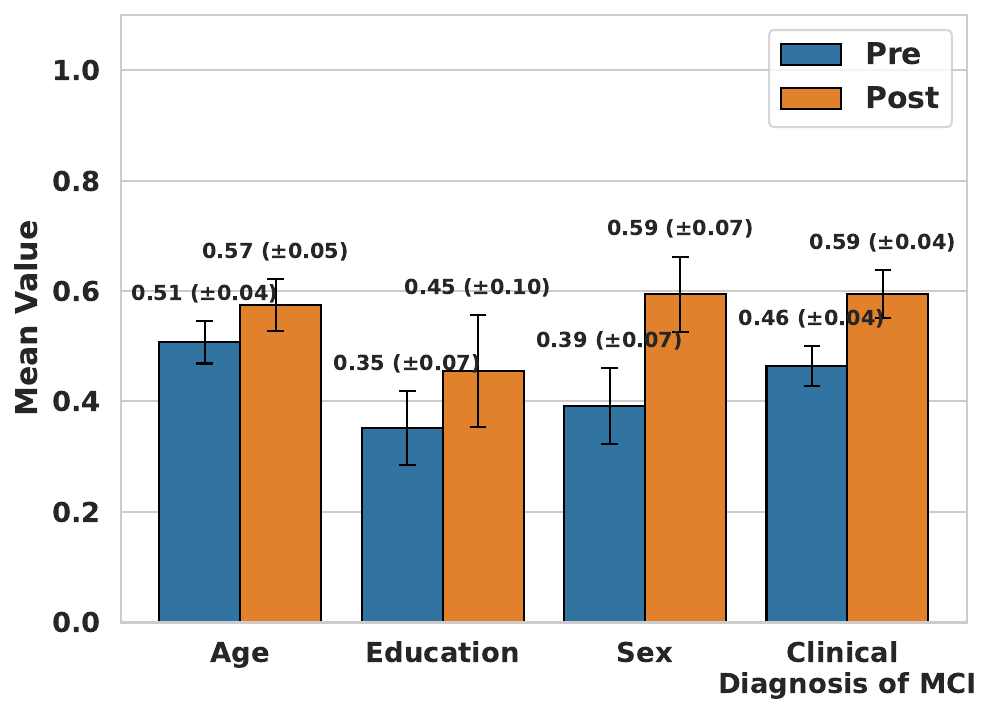}
        \caption{DPR of Negative Affect}
        \label{fig:grid-negaff-dpr}
      \end{subfigure}
       
        \medskip
      \begin{subfigure}[t]{0.48\textwidth}
        \centering
        \includegraphics[width=\linewidth,keepaspectratio]{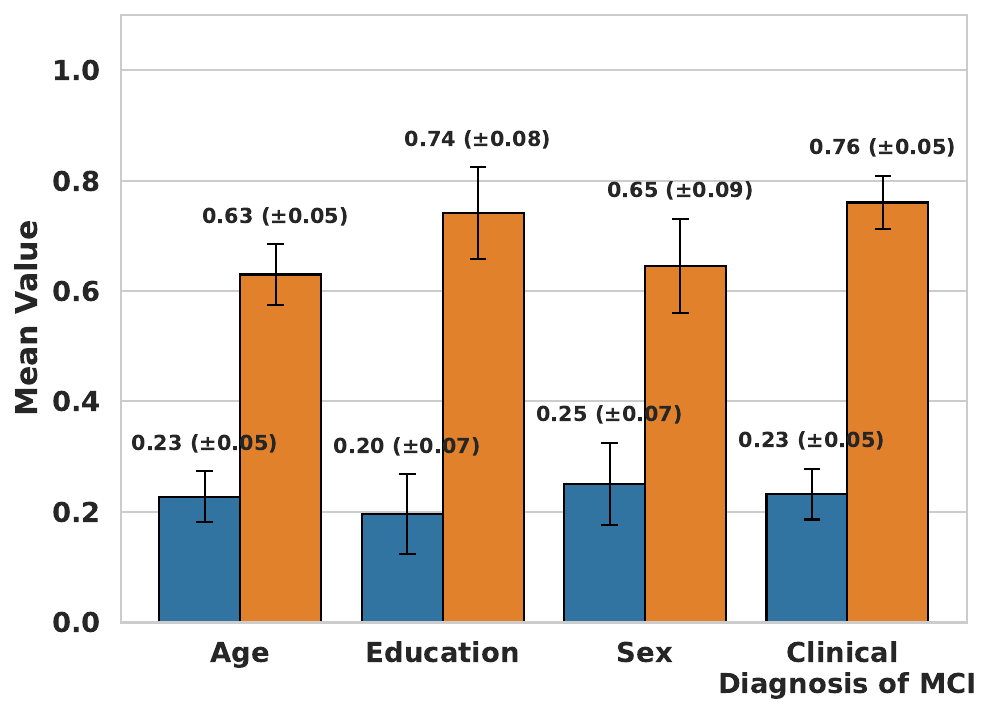}
        \caption{EOR of Social Satisfaction}
        \label{fig:grid-socsat-eor}
      \end{subfigure}\hfill
      \begin{subfigure}[t]{0.48\textwidth}
        \centering
        \includegraphics[width=\linewidth,keepaspectratio]{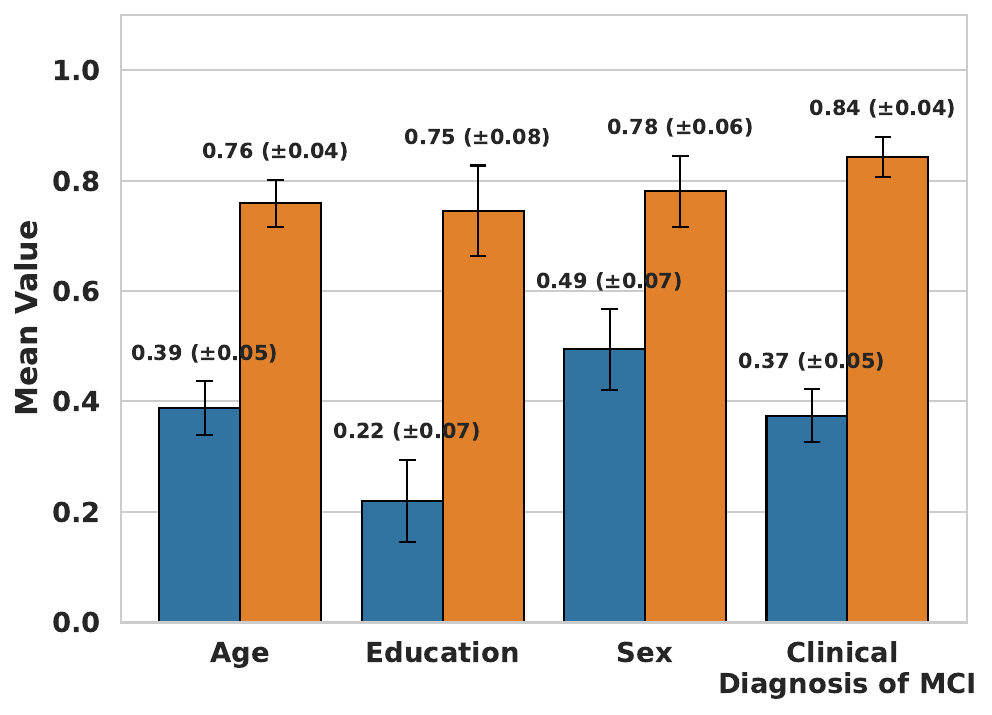}
        \caption{DPR of Social Satisfaction}
        \label{fig:grid-socsat-dpr}
      \end{subfigure}
        \medskip

      \begin{subfigure}[t]{0.48\textwidth}
        \centering
        \includegraphics[width=\linewidth,keepaspectratio]{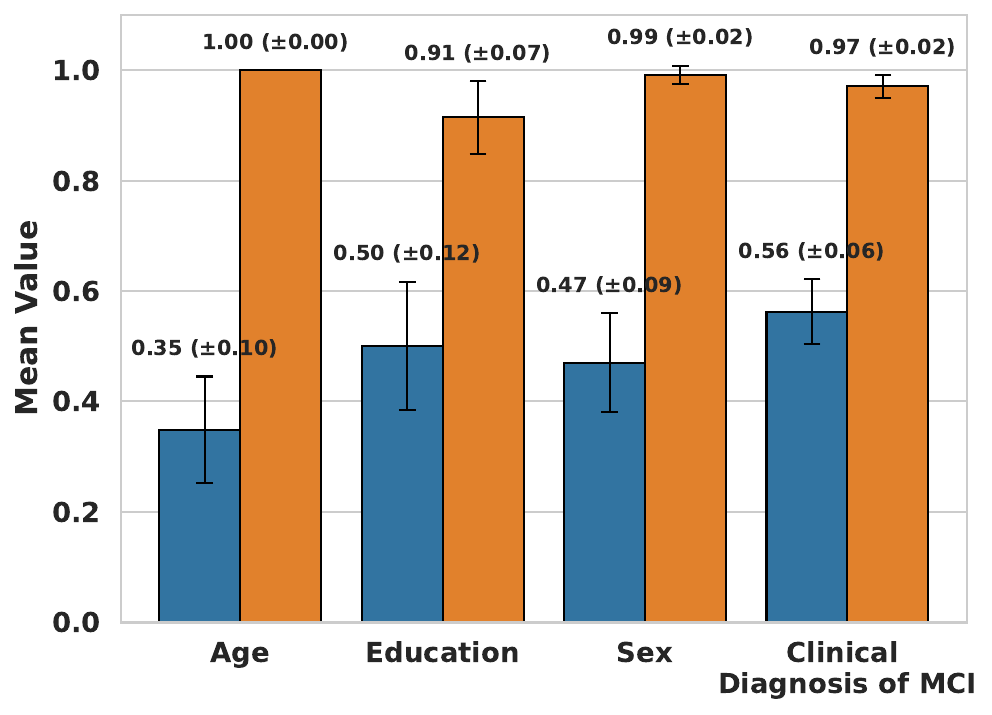}
        \caption{EOR of Psych. Well-being}
        \label{fig:grid-psych-eor}
      \end{subfigure}\hfill
      \begin{subfigure}[t]{0.48\textwidth}
        \centering
        \includegraphics[width=\linewidth,keepaspectratio]{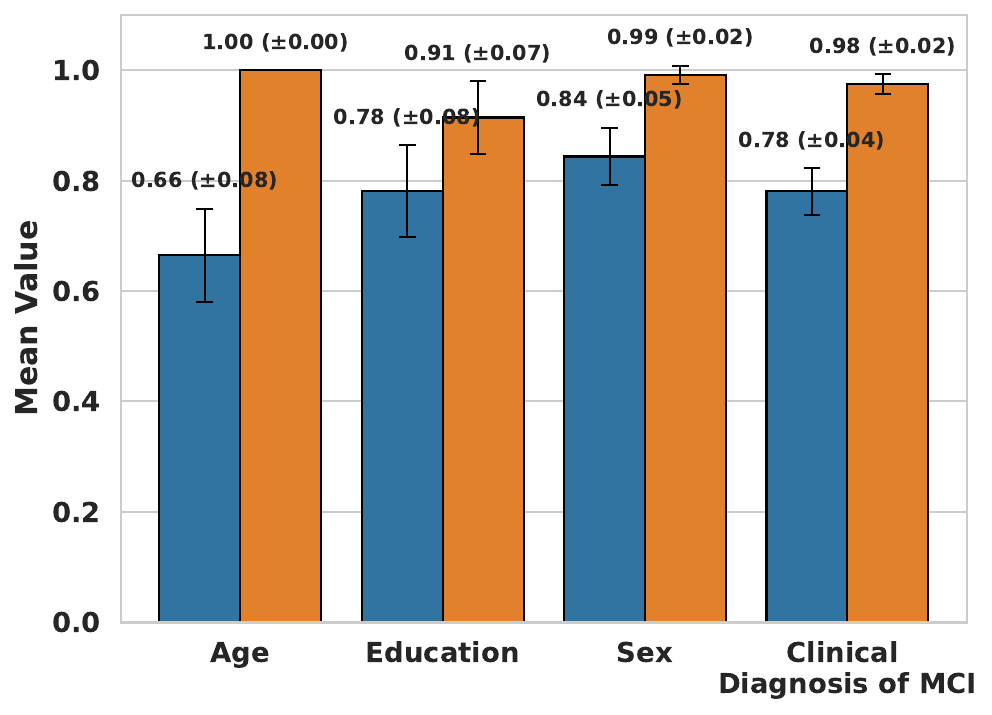}
        \caption{DPR of Psych. Well-being}
        \label{fig:grid-psych-dpr}
      \end{subfigure}

    \end{minipage}
  \end{adjustbox}

  \caption{Fairness and bias analysis for quantifying negative affect, social satisfaction, and psychological well-being before and after applying bias mitigation.}
  \label{fig:dp-psych-eor-f1}
\end{addedfigure}
\clearpage


\begin{addedfigure}[t]
  \centering
  \captionsetup{font=normalsize}
  \captionsetup[subfigure]{font=normalsize,justification=centering,skip=2pt}

  \begin{subfigure}[t]{0.48\textwidth}
    \centering

    \includegraphics[width=\linewidth,height=0.26\textheight, keepaspectratio]{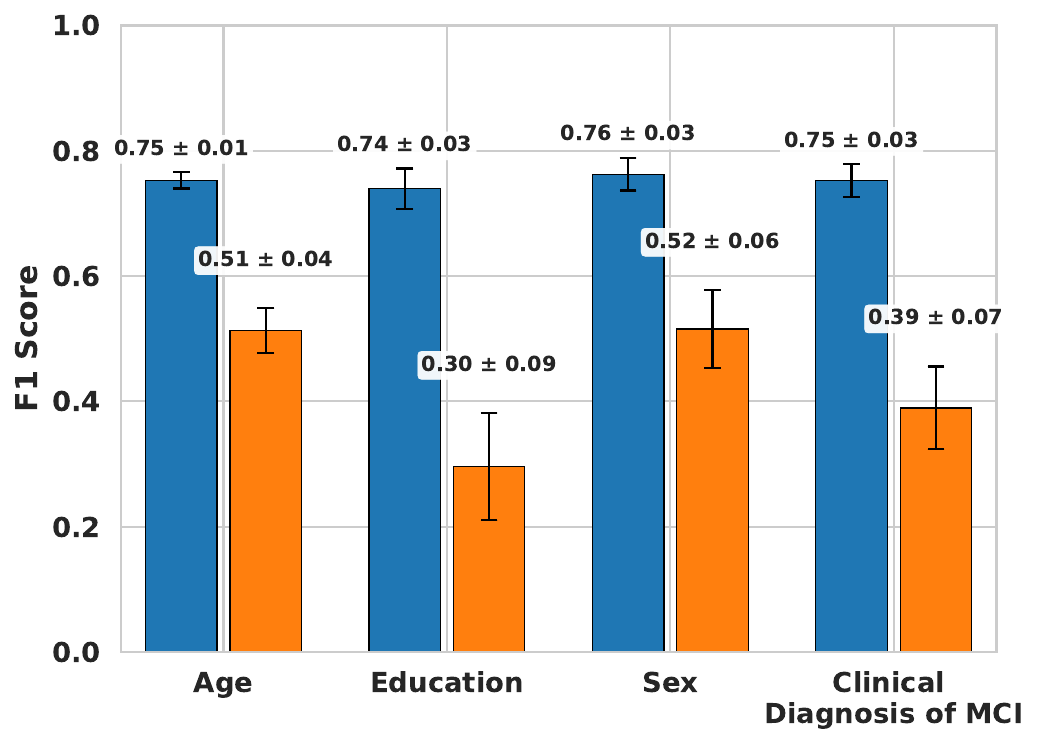}
    \caption{CDR}
    \label{fig:grid-a-cdr-f1}
  \end{subfigure}\hfill
  \begin{subfigure}[t]{0.48\textwidth}
    \includegraphics[width=\linewidth,height=0.26\textheight, keepaspectratio]{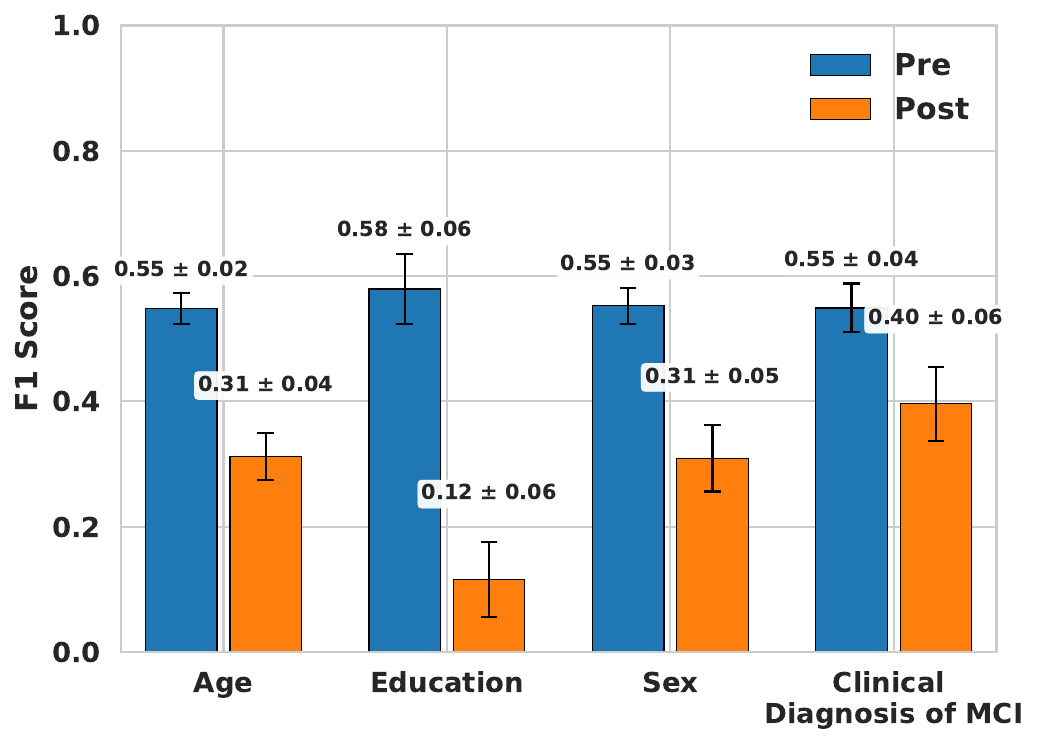}
    \caption{MoCA}
    \label{fig:grid-b-moca-f1}
  \end{subfigure}

  \medskip

  \begin{subfigure}[t]{0.48\textwidth}
    \centering
    \includegraphics[width=\linewidth,height=0.26\textheight, keepaspectratio]{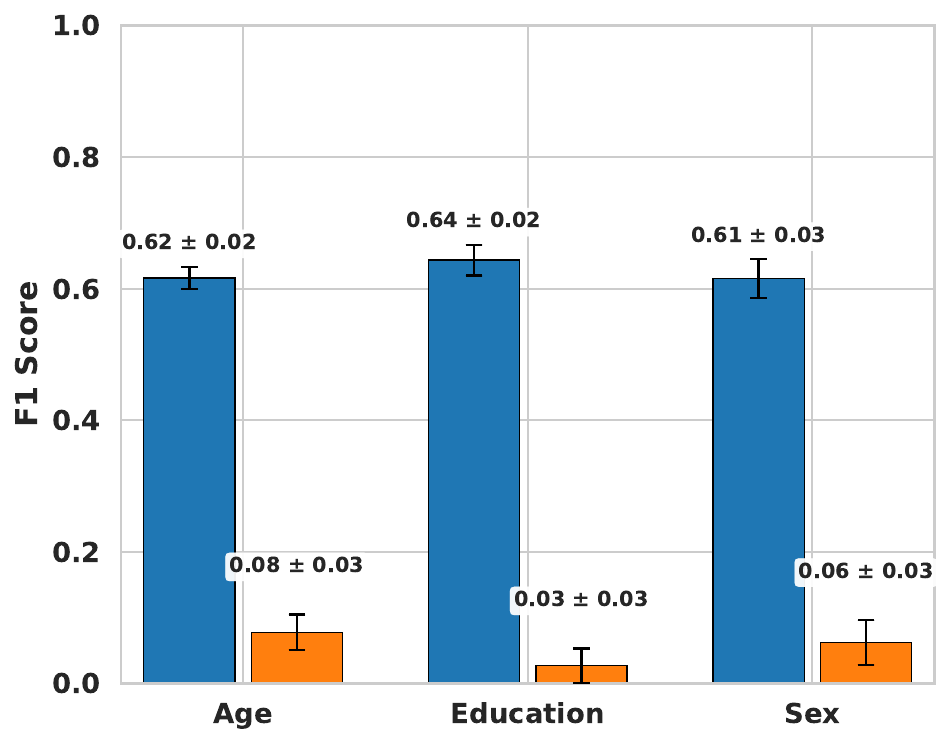}
    \caption{Clinical Diagnosis of MCI}
    \label{fig:grid-c-normcog-f1}
  \end{subfigure}\hfill
  \begin{subfigure}[t]{0.48\textwidth}
    \centering
    \includegraphics[width=\linewidth,height=0.26\textheight, keepaspectratio]{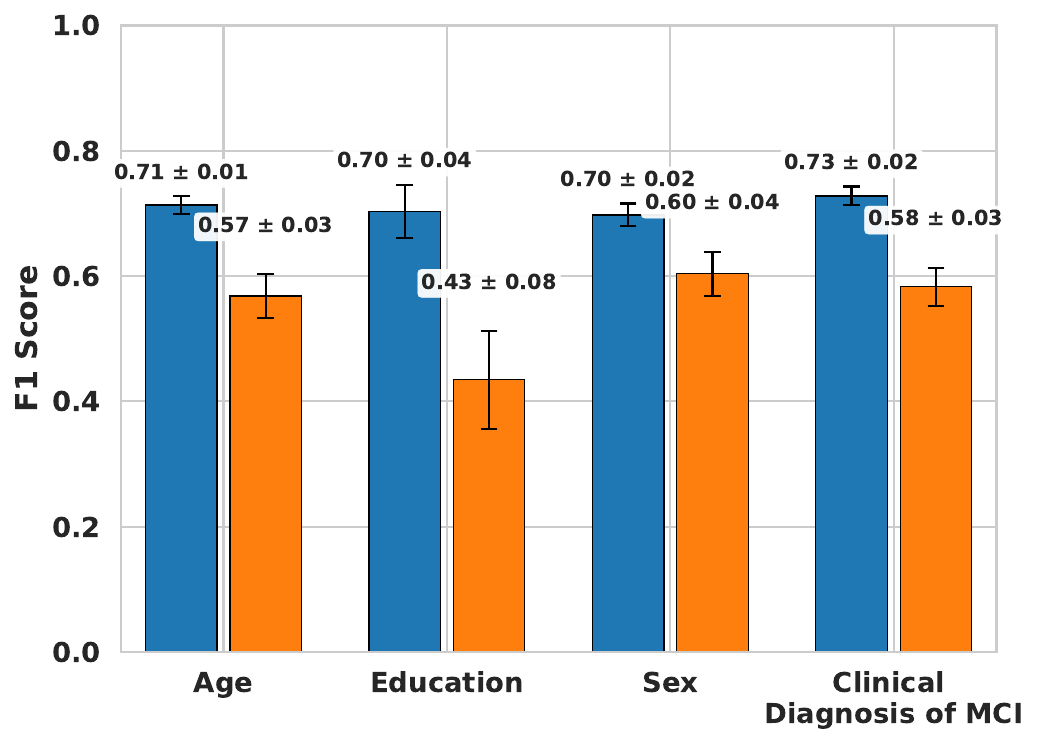}
    \caption{LSNS}
    \label{fig:grid-d-lsns-f1}
  \end{subfigure}

  \medskip

  \caption{F1 scores before and after applying bias mitigation for CDR, MoCA, clinical diagnosis of MCI, and LSNS.
  }
  \label{fig:dp-cog-f1-f1}
\end{addedfigure}

\begin{addedfigure}[t]
  \centering
  \captionsetup{font=normalsize}
  \captionsetup[subfigure]{font=normalsize,justification=centering,skip=2pt}

  \begin{adjustbox}{max totalsize={\textwidth}{0.92\textheight},center}
    \begin{minipage}{\textwidth}

      \begin{subfigure}[t]{0.48\textwidth}
        \centering
        \includegraphics[width=\linewidth,keepaspectratio]{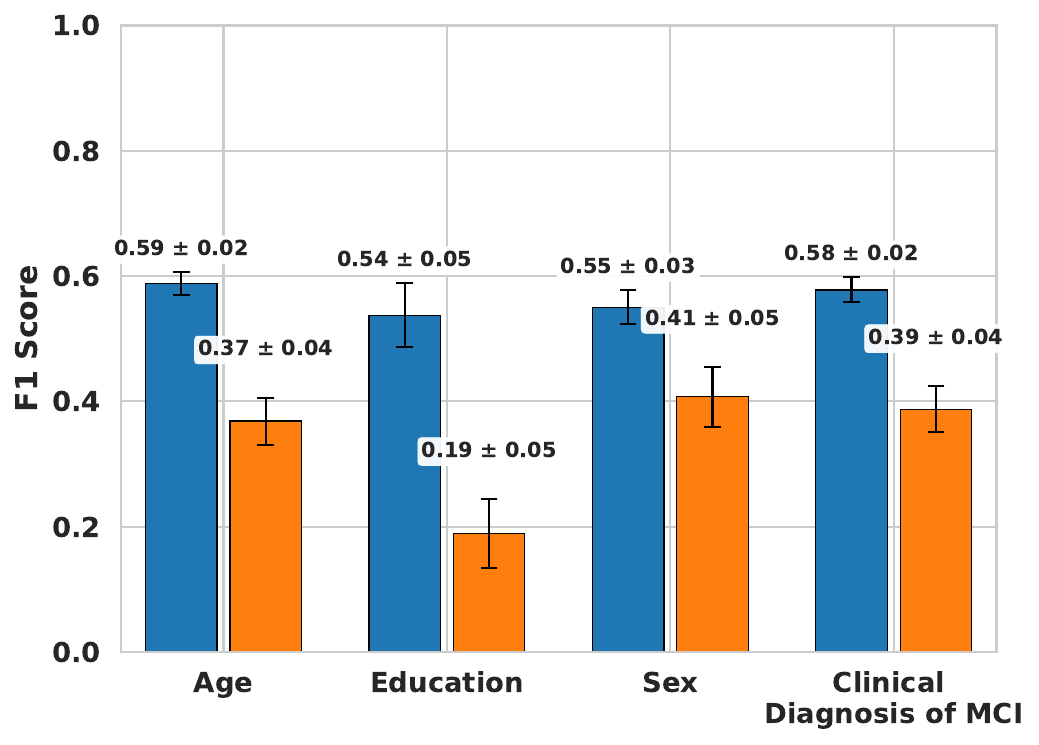}
        \caption{Neuroticism}
        \label{fig:grid-neuro-f1}
      \end{subfigure}\hfill
      \begin{subfigure}[t]{0.48\textwidth}
        \centering
        \includegraphics[width=\linewidth, keepaspectratio]{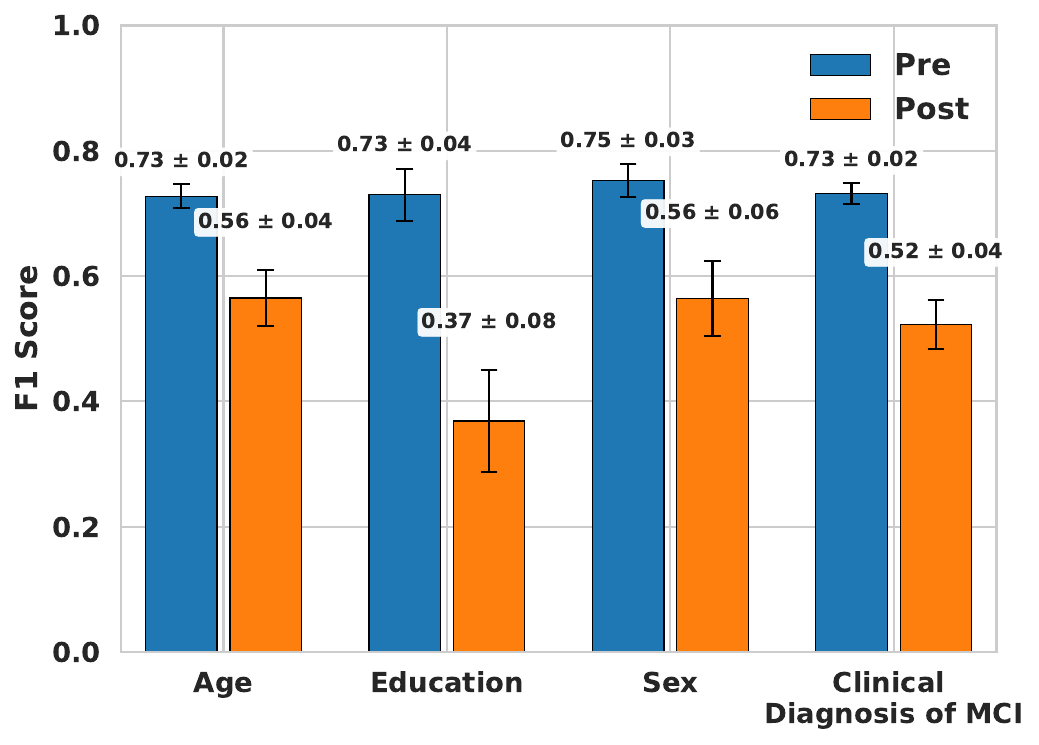}
        \caption{Negative Affect}
        \label{fig:grid-e-negaff-f1}
      \end{subfigure}

      \begin{subfigure}[t]{0.48\textwidth}
        \centering
        \includegraphics[width=\linewidth,keepaspectratio]{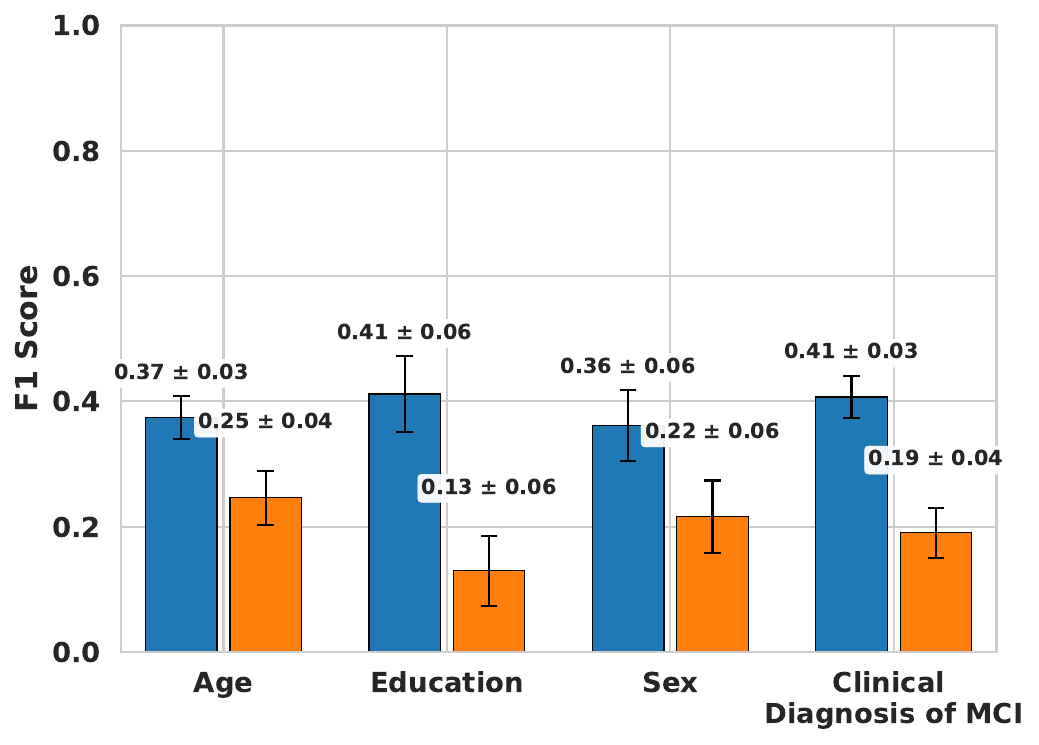}
        \caption{Social Satisfaction}
        \label{fig:grid-socsat-f1}
      \end{subfigure}\hfill
      \begin{subfigure}[t]{0.48\textwidth}
        \centering
        \includegraphics[width=\linewidth, keepaspectratio]{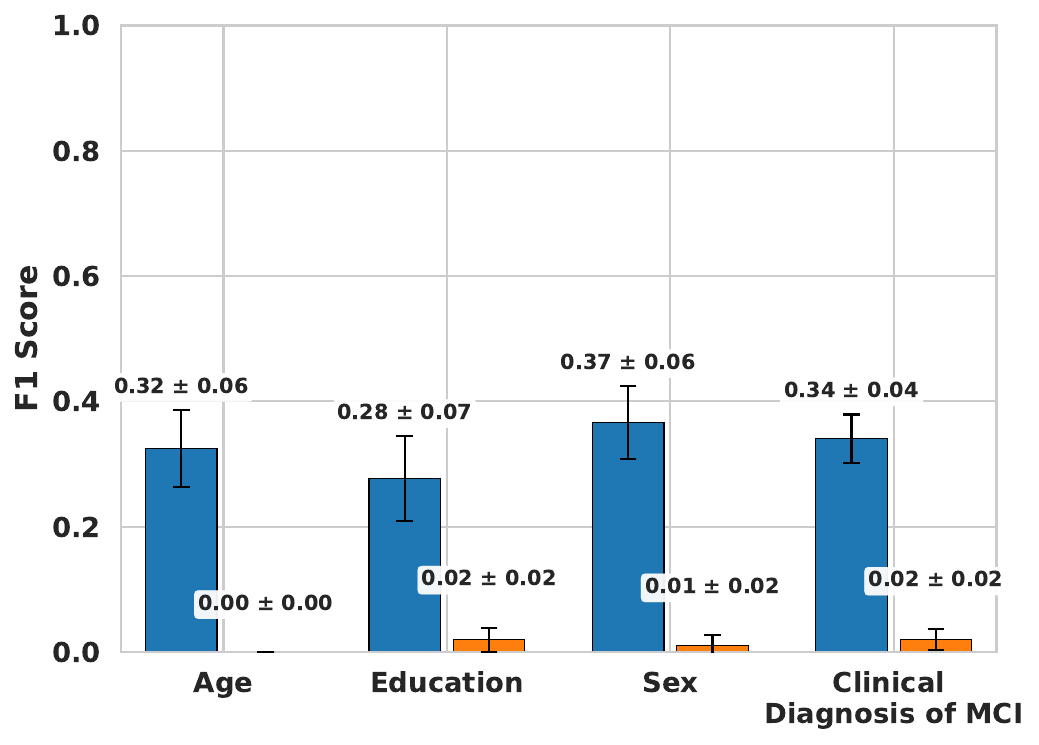}
        \caption{Psychological Well-being}
        \label{fig:grid-f-psych-f1}
      \end{subfigure}

    \end{minipage}
  \end{adjustbox}

  \caption{F1 scores before and after applying bias mitigation for neuroticism, and NIHTB-EB (negative affect, social satisfaction, and psychological well-being.}
  \label{fig:dp-psych-f1-f1}
\end{addedfigure}

Our analysis indicate\removed{s}\new{d} that \removed{the} language- and audio-based approaches were effective for quantifying cognitive impairment, compared to facial and HRV features. \new{From MM-SHAP analysis of multimodal models for CDR, MoCA, and clinical diagnosis of MCI (\autoref{fig:shap-3x1-cog-a}), HMM-processed acoustic features and RoBERTa text sentiment consistently emerged as the leading contributors.
}
This finding aligns with previous studies demonstrating the potential of speech and text analysis for quantifying cognitive impairment~\citep{themistocleous2020voice,laguarta2021longitudinal,asgari2017predicting}.
The proposed pipeline can potentially serve marginalized communities without internet connectivity for video transmission, such as rural areas or low- and middle-income countries \citep{graves2021disparities,dodoo2021telemedicine,kyei2024emergence}, for prescreening older adults with high risk of cognitive impairment, using only acoustic and linguistic features. 
Regarding the facial analysis, previous work reported that facial expressions are significantly different and heterogeneous among individuals with MCI \citep{jiang2022automated,morellini2022emotion}. We consider \new{that} this heterogeneity made the model underperform through facial features.
\new{
Still, future work needs to further evaluate our model with large-scale participant recruitment with various demographic and disease conditions, as our fairness and bias analysis showed significant biases before applying bias mitigation, with EOR < 0.5 across attributes.
}

\subsubsection{CDR}
\paragraph{Classification Performance}

For CDR \removed{(6th column in \mbox{\autoref{table:uni_cog}} \& \mbox{\autoref{table:multi_cog})}} \new{(1st row in \autoref{tab:best-pipelines} \& \autoref{fig:aurocs_heatmap})}, the acoustic and language fusion model performed the best (\removed{0.78 AUC and 0.74 accuracy}\new{0.77 AUC}). 
\new{Overall, multimodal pipelines outperformed all unimodal models,  with best performance using either Audio+Language+Demographics (A+L+D) or All-modalities configurations across RF, LR/GBDT, and SVM. 
We can see that both A+L+D and All-modalities followed the audio-only model by 0.02 difference in absolute AUC points in LR/GBDT. While in SVM and RF, they outperformed all other configurations. The best-performing unimodal model was audio in RF (0.75 AUC) and LR/GBDT (0.71 AUC). }

\paragraph{Feature Importance Analysis} 

\new{For the best-performing multimodal model (RF-A+L+D), HMM-processed acoustic and RoBERTa text sentiment features were the strongest contributors (each > 0.25), followed by the non-HMM RoBERTa sentiment feature \autoref{fig:shap-cdr-mult}. Foundation model embeddings (WavLM and LLaMA) contributed the least (each $<0.025$).}

\new{\paragraph{Fairness and Bias Analysis} The best-performing model showed significant bias before applying bias mitigation, with EOR clustered around \(0.125 \pm 0.115\), and DPR clustered around $0.4 \pm 0.2$ across all attributes (\autoref{fig:grid-a-cdr-eor}, \autoref{fig:grid-b-cdr-dpr}). 
Sex exhibited the greatest disparity (EOR = 0.01, DPR = 0.18), followed by YOE (EOR = 0.04, DPR = 0.26). Age was the least biased (EOR = 0.24, DPR = 0.57), comparable to the clinical diagnosis of MCI. After applying bias mitigation, EOR and DPR improved across all attributes to $0.415 \pm 0.095$ and $0.61\pm 0.05$, respectively, with each of YOE and the clinical diagnosis of MCI reaching an EOR of 0.51. 
The largest absolute DPR gain was exhibited by sex (+0.43), and the smallest by age (+0.04). YOE experienced the largest absolute EOR gain (\(+0.47\)), but age showed the smallest EOR gain (\(+0.13\)). 
Yet, as shown in \autoref{fig:grid-a-cdr-f1}, these shifts in EOR followed with large absolute F1 score drops of \(\Delta \text{F1 score}_{\text{YOE}} = -0.44\) and \(\Delta \text{F1 score}_{\text{age}} = -0.24\).}

\subsubsection{MoCA}


\new{\paragraph{Classification Performance} 
Our results show that differentiating those with high vs. low MoCA scores (2nd row in \autoref{fig:aurocs_heatmap} \& \autoref{tab:best-pipelines} was best when using demographics for LR/GBDT and RF models (0.65 AUC). SVM performed best with rPPG (0.62 AUC). 
The second-best modalities were audio for LR/GBDT (0.62 AUC), facial features for SVM (0.57 AUC), and Audio+Language+Demographics fusion (0.55 AUC) in RF.}
Facial features and HRV features derived from rPPG were not useful indicators of quantifying MoCA \new{for RF and LR/GBDT, but were more informative for SVM}.
All multimodal fusion approaches in \removed{\mbox{\autoref{table:multi_cog}}} \new{\autoref{fig:aurocs_heatmap}} showed statistically similar performance compared to the best unimodal approach using \removed{the linguistic feature, LLaMA-65B} \new{demographics across all models}.

\removed{Our results show that differentiating those with high vs. low MoCA scores (4th column in \mbox{\autoref{table:uni_cog}} \& \mbox{\autoref{table:multi_cog}}) is best when using linguistic modality, LLaMA-65B alone (0.64 AUC and 0.63 accuracy).
This was followed by acoustic features (0.63 AUC and 0.58 accuracy) and demographic variables (0.62 AUC and 0.59 accuracy).} 

\paragraph{Feature Importance Analysis} 


\new{\autoref{fig:shap-moca-mult} shows the top contributing modality features for the LR/GBDT All-modalities fusion model. Notably, all the HMM-processed variants occupied the top four positions, with Acoustic and RoBERTa text sentiment ranking first and second, followed by Emotion+AUs and rPPG. In contrast, foundation models' embeddings (DINOv2, LLaMA, and WavLM) contributed little (each < 0.07), and demographics had the smallest overall impact on the model output (<0.02).
}

\new{\paragraph{Fairness and Bias Analysis} \autoref{fig:grid-c-moca-eor} and \autoref{fig:grid-d-moca-dpr} report EOR and DPR, before and after applying bias mitigation, for classifying MoCA. 
Before applying bias mitigation, EOR was less than 0.2 and DPR was less than 0.5 for all attributes. 
YOE exhibited the greatest bias before applying bias mitigation (EOR = 0.01 and DPR = $0.08$). Sex showed the least disparity in EOR ($0.17$), whereas clinical diagnosis of MCI had the least disparity in DPR (0.46). 
After applying bias mitigation, all attributes' EOR and DPR improved, with YOE increasing the most, reaching 0.77 in EOR and 0.78 in DPR. As shown in \autoref{fig:grid-b-moca-f1}, this gain coincided with the largest absolute F1 score decrease ($\Delta$F1 score $=-0.46$). All attributes experienced lower F1 scores after bias mitigation was applied. The smallest drop ($\Delta$F1 score $=-0.15$) occurred for clinical diagnosis of MCI, which also had the lowest EOR ($0.42$), and was most biased after applying bias mitigation.}

\subsubsection{Clinical Diagnosis of MCI}

\new{\paragraph{Classification Performance.} 
Quantifying clinical diagnosis of MCI (3rd row in \autoref{tab:best-pipelines} \& \autoref{fig:aurocs_heatmap}) was effective (0.69 AUC) when using either Audio+Language+Demographics fusion with SVM or WavLM (Audio) with RF. 
Among unimodal models, audio and language had the highest performance in RF (0.69 AUC and 0.62 AUC, respectively) and LR/GBDT (0.58 AUC and 0.59 AUC, respectively). Fusing audio and language yielded the best multimodal AUC scores (SVM: 0.69, RF: 0.67, LR/GBDT: 0.59). The fusion model outperformed the unimodals in both LR/GBDT and SVM, while for RF, it trailed unimodal WavLM by an absolute difference of 0.02 AUC points.}

\paragraph{Feature Importance Analysis} 

\new{The best-performing fusion model (SVM - Audio + Language + Demographics) showed that the HMM-processed acoustic and RoBERTa text-sentiment features ranked as the most influential contributors (audio and language), similar to the pattern observed in the MoCA fusion model. Foundation model embeddings (LLaMA and WavLM) ranked low (each contributing < 0.05), and the non-HMM processed acoustic features had the smallest overall contribution \autoref{fig:shap-normcog-mult}.}

\new{\paragraph{Fairness and Bias Analysis} \autoref{fig:grid-e-normcog-eor} and \autoref{fig:grid-f-normcog-dpr} show that the best-performing model (Audio+Language+Demographics) exhibited significant bias across all attributes according to its EOR values (<0.5) before applying bias mitigation. 
DPR showed opposing results with $0.82\pm0.03$ before applying bias mitigation.
This highlights the importance of using multiple metrics to analyze the bias of models from varying perspectives.
EOR is considered to be a stricter measure of bias compared to DPR, as it is based on TPR and FPR \citep{jiang2024evaluating}.
Sex showed the greatest disparity (EOR = 0.14, DPR = 0.79), whereas age was the least biased (EOR = 0.49, DPR = 0.85). 
After applying bias mitigation, EOR and DPR improved significantly for all attributes, reaching EOR and DPR $\ge$0.9 . 
The most significant absolute gains were observed for sex ($\Delta EOR = +0.76$, $\Delta DPR = +0.14$) and YOE ($\Delta EOR = +0.74$, $\Delta DPR = +0.15$). As shown in \autoref{fig:grid-c-normcog-f1}, these gains coincided with notable utility losses. 
YOE experienced the largest absolute F1 score decrease ($\Delta \text{F1 score} = -0.61$, final F1 score=0.03). Overall, after applying bias mitigation, F1 scores clustered around 0.055
$\pm$0.025, which demonstrated the significant trade-off between bias and performance.
}

\subsection{Quantifying Social Network and Psychological Well-being Assessment}
\label{sec:result_psych}

\removed{Overall, quantifying social isolation (LSNS) was most effective using language features.
This is consistent with prior research indicating the utility of text sentiment analysis in understanding the social health of individuals.}
\new{When quantifying social engagement (LSNS), the fusion model’s SHAP analysis in \autoref{fig:shap-lsns-mult} showed the strongest contributions from the HMM-processed acoustic and rPPG features, with text sentiment also among the top four. 
This is consistent with our classification results for the unimodal text-sentiment model from \autoref{tab:best-pipelines}, achieving a 0.74 AUC, supporting that text sentiment analysis is most informative for social health} \citep{liu2022relationship,zhou2022distinguishing,asgari2017predicting}. 
\removed{Other scales for psychological well-being required facial videos to quantify them, and cardiovascular measures were most effective at quantifying negative affect, similar to those reported in mental health studies for various populations using wearable-based cardiovascular health monitoring} 
\new{Neuroticism had the HMM-processed variants of rPPG and Emotions+AUs as the top SHAP contributors in \autoref{fig:shap-neuro-mult}, with an AUC of 0.64. 
Yet, the best overall model was the unimodal text-HMM (AUC 0.66), whose performance was statistically similar to when using the emotion features (an absolute difference of 0.01 in AUC). 
This finding is consistent with neuroticism capturing tendencies toward emotional distress, which can be reflected in cardiovascular signals~\citep{hrv_neuroticism_2024} and facial expressions~\citep{facs_neuroticism_2015}}. 
\new{This} \removed{Our study} demonstrates that contactless cardiovascular measures have the potential to quantify behavioral symptoms often accompanied by MCI, which calls for further exploration.

\new{Across the NIHTB-EB tasks (negative affect, social satisfaction, and psychological well-being), multimodal models generally outperformed unimodal models. 
Consistent with the classification results, multimodal SHAP analysis from \autoref{fig:shap-3x1-cog-b} indicates heavier reliance on the HMM-processed emotion (facial) and rPPG features, suggesting that quantification of NIHTB-EB benefits from both facial video and cardiovascular measures 
\citep{diCampliSanVito2023StressDetection,s20030718,Chen2019Artificial}.
\removed{Consistent with this, prior work links cardiovascular features to markers of social support and well-being} 
For negative affect, the best unimodal model (HMM-acoustic) outperformed the multimodal model with an absolute AUC difference of 0.04, shown in \autoref{fig:aurocs_heatmap}.
}~\removed{Negative affects, such as mood disturbances, anxiety, and depression, can be an early sign of dementia, and these emotional changes often precede noticeable cognitive decline and may be linked to neurodegenerative processes in the brain~\citep{ismail2017mild}}\new{Negative affect, encompassing unpleasant mood states like sadness, fear/anxiety, and anger \citep{Babakhanyan2018}, and related emotional changes often precede noticeable cognitive decline. 
They may be linked to neurodegenerative processes in the brain ~\citep{ismail2017mild}.
Cardiovascular measures were reported to be effective in quantifying such emotions in mental health studies using wearables \citep{diCampliSanVito2023StressDetection,s20030718,Chen2019Artificial}.} 

\new{Before applying bias mitigation, all the best-performing models had EOR $\le$ 0.56 across attributes, indicating severe bias in quantifying social engagement and psychological well-being. 
LSNS (text sentiment) and negative affect (Acoustic+HMM) were especially skewed, with EOR $\le$ 0.4 across attributes both before and after applying bias mitigation, suggesting that threshold post-processing was insufficient. 
The model for quantifying psychological well-being showed the least disparities overall, with its worst attribute at $EOR_{age}$ = 0.35. The EOR of all models improved after applying bias mitigation, with quantifying psychological well-being achieving a high EOR (0.955 $\pm$0.045), but at the cost of significant F1 score drops, as shown in \autoref{fig:dp-psych-f1-f1}. Notably, it had the highest EOR after applying bias mitigation (1.00 for age), but with its F1 score dropping to 0, entirely losing the utility of the model.}

\subsubsection{LSNS}
\paragraph{Classification Performance}
Our results showed that language-based emotion and sentiment features were most effective for quantifying social engagement, LSNS (4th \removed{column in \mbox{\autoref{table:uni_psych}} \& \mbox{\autoref{table:multi_psych}}} \new{row in \autoref{tab:best-pipelines} \& \autoref{fig:aurocs_heatmap}}), \new{across all models and modality configurations, having RoBERTa sentiment as the best-performing (LR:0.74 AUC, RF: 0.66 AUC) or its HMM-variant (SVM: 0.63 AUC).} \removed{with 0.75 AUC and 0.73 accuracy.}
\removed{Facial emotion, landmark, and action unit features showed the second-best performance with 0.6 AUC and 0.59 accuracy.
When fusing modalities, only the all-modality fusion with the majority vote matched the facial model with 0.6 AUC and 0.56 accuracy.}
\new{Cardiovascular features showed the second-best performance for RF, while All-modality fusion was the second-best for SVM (0.63 AUC) and LR/GBDT (0.62 AUC).}

\paragraph{Feature Importance Analysis} 

\new{For the best-performing multimodal model (LR/GBDT - All modalities), HMM-processed variants of acoustic, rPPG, and emotion features dominated the contributions, with acoustic and rPPG showing the largest impacts (each \(> 0.25\)), as shown in \autoref{fig:shap-lsns-mult}. In contrast, foundation model embeddings (WavLM, LLaMA, and DINOv2), as well as the non-HMM versions of emotion and acoustic features, and demographics, contributed the least (each < 0.025).}

\new{\paragraph{Fairness and Bias Analysis} The best-performing text model (LR-RoBERTa text sentiment) was biased across all attributes before applying bias mitigation, with EOR $\le 0.21$ and DPR $\le 0.52$ across all attributes (\autoref{fig:grid-lsns-eor} and \autoref{fig:grid-lsns-dpr}). YOE showed the greatest bias (EOR$=0.06$, DPR$=0.29$), whereas the clinical diagnosis of MCI was the least biased (EOR $= 0.21$, DPR=0.52). Sex also displayed significant bias (EOR $= 0.09$, DPR=0.48). 
After applying bias mitigation, EOR and DPR improved for all attributes (EOR $\ge 0.19$, DPR $\ge 0.39$). YOE had the largest absolute EOR gain ($\Delta \text{EOR} = +0.27$, $\Delta \text{DPR} = +0.1$), while sex had the smallest EOR and DPR gain ($\Delta \text{EOR} = +0.10$, $\Delta \text{DPR} = +0.09$). As shown in \autoref{fig:grid-d-lsns-f1}, these improvements were accompanied by utility losses: F1 scores declined across all attributes, with the largest absolute decline for YOE ($\Delta \text{F1 score} = -0.27$) and the smallest absolute decline for sex ($\Delta \text{F1 score} = -0.10$).}

\subsubsection{Neuroticism} 
\paragraph{Classification Performance}
\removed{For quantifying neuroticism (5th column in \mbox{\autoref{table:uni_psych}} \& \mbox{\autoref{table:multi_psych}}), multimodal fusion from all modalities performed effectively with 0.71 AUC and 0.65 accuracy.
This was followed by features from facial emotion, landmark, and action units (0.69 AUC and 0.66 accuracy), which contributed the most when fusing multiple modalities.}
\new{For quantifying neuroticism (5th row in \autoref{tab:best-pipelines} \& \autoref{fig:aurocs_heatmap}), the best single pipeline was LR with RoBERTa Sentiment+HMM (0.66 AUC), followed by RF with unimodal (facial emotion/landmarks/AUs) features with a 0.01 difference (0.65 AUC). 
Among multimodal fusion-based approaches, the Facial+Cardiovascular+Demographic (F+C+D) setting showed the highest performance (RF: 0.64 AUC, LR/GBDT: 0.61 AUC), matching the All-modalities fusion for both.
For SVM, F+C+D ranked second (0.55 AUC) behind All-modalities fusion (0.57 AUC). Overall, F+C+D and All-modality fusion methods consistently ranked as the best two multimodal options.}

\new{\paragraph{Feature Importance Analysis} 
The multimodal RF (F+C+D) model had the HMM-processed rPPG and facial emotion, landmarks, and AUs as the top contributing features (each with a value greater than 0.3), shown in \autoref{fig:shap-neuro-mult}. 
DINOv2 features and the non-HMM processed facial, emotion, and AUs had the lowest contributions (each < 0.025).}

\new{\paragraph{Fairness and Bias Analysis} The best model (LR-HMM text sentiment) exhibited significant biases across all attributes (EOR $\le$ 0.17 and DPR $\le0.52$) as shown in \autoref{fig:grid-neuro-eor} and \autoref{fig:grid-neuro-dpr} before applying bias mitigation. YOE had the largest bias (EOR = 0.08 and DPR = 0.23), while age had the smallest (EOR=0.17 and DPR=0.52). 
After applying bias mitigation, all attributes exhibited decreased biases, with an EOR clustered around $0.455\pm0.085$ and a DPR around $ 0.6\pm0.065$. YOE and sex had the highest and smallest absolute gains in EOR (+0.46, +0.26) and DPR (+0.33, +0.1). This coincided with the largest and smallest absolute drops in F1 scores (-0.35, -0.14 respectively). All other attributes also showed drops in F1 scores as shown in \autoref{fig:grid-neuro-f1}. }

\subsubsection{Negative Affect}

\paragraph{Classification Performance} 

\removed{Facial and cardiovascular fusion performed effectively when quantifying negative affect (6th column in \mbox{\autoref{table:uni_psych}} \& \mbox{\autoref{table:multi_psych}}) with 0.79 AUC and 0.75 accuracy.}
\removed{Cardiovascular features alone showed 0.76 AUC and 0.67 accuracy, contributing most when fused with other facial features.}
\new{
The Acoustic HMM-based unimodal approach showed the best performance overall (0.74 AUC).
Then, models trained on all modalities performed strongly (LR/GBDT: 0.70 AUC, SVM: 0.64 AUC, RF: 0.69 AUC) when quantifying negative affect (6th row in \autoref{tab:best-pipelines} \& \autoref{fig:aurocs_heatmap}).
}

\paragraph{Feature Importance Analysis} 
\new{The best-performing multimodal model (LR/GBDT - all modalities) exhibited heavier dependence on the HMM-processed emotion (facial emotion, landmarks, and AUs) and rPPG features as shown in \autoref{fig:shap-negaff-mult}. With Emotion+HMM features contributing the most (>0.19), followed by rPPG+HMM and the non-HMM rPPG features. All the HMM-processed features ranked in the top five (each having a contribution > 0.08), while demographics showed the least contribution (<0.025). 
Foundation models' (WavLM, DINOv2, and LLaMA) embeddings also showed low contributions (<0.06).}

\new{\paragraph{Fairness and Bias Analysis} The best-performing model (LR-Acoustic+HMM) exhibited strong bias across all attributes before applying mitigation (EOR $\le 0.19$ and DPR $\le0.51$), as shown in \autoref{fig:grid-negaff-eor} and \autoref{fig:grid-negaff-dpr}. YOE showed the greatest bias (EOR $= 0.08$, DPR=$0.35$), followed by sex (EOR $= 0.11$, DPR=0.39). After applying bias mitigation, EOR and DPR improved for all attributes, with EOR clustering around $0.345 \pm 0.055$ and DPR clustering around $0.52\pm0.07$. YOE had the largest absolute EOR and DPR gains ($\Delta EOR=+0.32, \Delta DPR=+0.1$), whereas age had the smallest EOR and DPR gains ($+0.13, +0.06$). As shown in \autoref{fig:grid-e-negaff-f1}, these fairness gains came with utility losses. F1 scores decreased across attributes, with the largest decrease for YOE ($\Delta \text{F1 score} = -0.36$) and the smallest for age ($\Delta \text{F1 score} = -0.17$).}

\vspace{50pt}

\subsubsection{Social Satisfaction}

\paragraph{Classification Performance}
\removed{Facial emotion, landmark, and action unit features were most useful when quantifying social satisfaction (7th column in \mbox{\autoref{table:uni_psych}} \& \mbox{\autoref{table:multi_psych}}) during remote interviews (0.68 AUC and 0.63 accuracy).}
\new{When quantifying social satisfaction (7th row in \autoref{tab:best-pipelines} \& \autoref{fig:aurocs_heatmap}), facial (emotion, landmarks, and AUs) and cardiovascular features consistently ranked as the best two unimodal modalities in RF and LR/GBDT in terms of performance. Fusing them yielded the best multimodal performance in all three models (LR/GBDT: 0.75 AUC, RF: 0.59 AUC, SVM: 0.61 AUC).}

\paragraph{Feature Importance Analysis}

\new{For the best-performing multimodal model (LR/GBDT-Facial + Cardiovascular + Demographics), HMM-processed features dominated the feature importances (\autoref{fig:shap-socsat-mult}). rPPG+HMM ranked first with a contribution > 0.30, followed by the HMM-processed facial emotion, landmarks, and AUs. The non-HMM rPPG modality ranked third overall (contribution > 0.25). 
All the remaining features contributed minimally (each contributing < 0.04), showing the importance of capturing the temporal dynamics of facial and cardiovascular measures for quantifying social satisfaction. 
Foundation model embeddings (DINOv2) and the non-HMM variants of facial emotion, landmarks, and AUs had near-zero influence.}

\new{\paragraph{Fairness and Bias Analysis} 
The best-performing model exhibited bias across all attributes before applying mitigation, with EOR clustered around 0.225$\pm$0.025 and DPR clustered around $0.355\pm 0.135$ as shown in \autoref{fig:grid-socsat-eor} and \autoref{fig:grid-socsat-dpr}.
YOE exhibited the largest bias overall (EOR =0.2, DPR=0.22), while sex showed the smallest (EOR=0.25, DPR=0.49). 
Following bias mitigation, EOR and DPR of all attributes increased significantly. The clinical diagnosis of MCI exhibited the least bias after applying mitigation with EOR/DPR of 0.76/0.84. YOE had the highest absolute gain in EOR and DPR ($\Delta EOR = +0.54, \Delta DPR = +0.53$), while age and sex had the smallest gains ($\Delta EOR_{age,sex} = +0.4, \Delta DPR_{age}=0.37, \Delta DPR_{sex}=0.29$). This matched the highest and lowest absolute declines in F1 scores ($\Delta \text{F1 score}_{YOE} = -0.28, \Delta \text{F1 score}_{age} = -0.12$), as shown in \autoref{fig:grid-socsat-f1}. All other attributes also showed declines in F1 scores after applying bias mitigation, with F1 scores clustered around $0.19\pm0.06$.}

\subsubsection{Psychological Well-being}

\paragraph{Classification Performance}

\removed{When quantifying overall psychological well-being (8th column in \mbox{\autoref{table:uni_psych}} \& \mbox{\autoref{table:multi_psych}}), facial emotion, landmark, and action unit features were most effective with 0.66 AUC and 0.61 accuracy.}
\removed{This was followed by cardiovascular features (0.62 AUC and 0.60 accuracy), but fusing facial and cardiovascular features showed no improvement compared to when only using cardiovascular features.}
\new{For overall results for psychological well-being (8th row in \autoref{tab:best-pipelines} \& \autoref{fig:aurocs_heatmap}), the Facial+Cardiovascular+Demographic fusion performed the best (LR/GBDT: 0.64 AUC, SVM: 0.72 AUC, RF: 0.64 AUC). 
For unimodal models, the cardiovascular feature ranked second in both RF (0.60 AUC) and LR/GBDT (0.62 AUC).}

\paragraph{Feature Importance Analysis}

\new{The best-performing multimodal model (SVM- Facial + Cardiovascular + Demographics) displayed heavy dependence on the HMM-processed variants of facial emotion, landmarks, and AUs (contribution > 0.35), and rPPG (contribution > 0.3), as shown in \autoref{fig:shap-psych-mult}. 
The non-HMM rPPG feature ranked third overall (contribution > 0.15). In contrast, the non-HMM facial emotion, landmarks, and AUs, as well as the DINOv2 embeddings, contributed minimally to the model predictions (each < 0.02).}

\new{\paragraph{Fairness and Bias Analysis} 
The best-performing model exhibited biases before applying mitigation with EOR clustering around 0.455$\pm$0.105 and DPR clustering around $0.75 \pm 0.09$, as shown in \autoref{fig:grid-psych-eor} and \autoref{fig:grid-psych-dpr}. Age had the largest bias (EOR = 0.35, DPR = 0.66), whereas clinical diagnosis of MCI was the least biased in EOR (0.56), followed by YOE (EOR = 0.50). Sex was the least biased in DPR (0.84), followed by YOE and clinical diagnosis of MCI (DPR = 0.78). 
After applying bias mitigation, EOR and DPR increased across all attributes (EOR and DPR $\approx 0.955\pm$0.045). Age and YOE showed the largest and smallest absolute gains in EOR/DPR, respectively ($\Delta EOR/DPR_{Age}=+0.65/+0.34, \Delta EOR/DPR_{YOE}=+0.41/+0.13$). Similar to the other classification tasks, utility losses accompanied these gains in fairness.
As shown in \autoref{fig:grid-f-psych-f1}, the age attribute's F1 score fell to 0, and the remaining attributes also yielded very low F1 scores ($\le0.02$).}

\subsection{Limitations \& Future work}



The proposed work studie\removed{s}\new{d} the association of facial, acoustic, linguistic, and cardiovascular patterns with cognitive impairment and social and psychological well-being in older adults with normal cognition or MCI. 
Our feasibility study demonstrated that features extracted from remotely conducted \new{video} conversations \removed{can}\new{could} detect a broad range of symptoms linked to cognitive decline, including social engagement and emotional well-being. This remote assessment approach holds promise for the early identification of individuals at risk \removed{for}\new{of} cognitive decline, ultimately creating opportunities for timely interventions to slow or prevent further deterioration.
Yet, our feasibility study ha\removed{s}\new{d} several limitations.

\paragraph{Limited Sample Size and Varying Disease Conditions}
This study include\removed{s}\new{d} subjects with normal cognition and MCI, which help\removed{s}\new{ed} to quantify and contrast the behavioral characteristics related to the early \new{detection} of MCI. However, the proposed method \removed{is}\new{was} evaluated on a small number of participants (N=39).~\new{Due to the small dataset size, using participant-wise nested cross-validation may have hindered the model's ability to learn patterns that are generalizable for the classification tasks, instead, being overly focused on participant-specific variations. 
Nevertheless, our models demonstrated the feasibility of generalization, achieving AUCs of 0.77 for distinguishing CDR, 0.75 for social satisfaction, and 0.74 for the negative affect scales and LSNS. Running 100 runs of the participant-independent nested cross-validation also improved the statistical significance of these estimates.}
\removed{Also}\new{In future studies}, comprehensively quantifying cognitive impairment requires including various subtypes of MCI, such as amnestic or non-amnestic and single-domain or multi-domain MCI~\citep{busse2006mild,rapp2010subtypes,bradfield2023mild}.
We \new{also} excluded individuals diagnosed with ADRD\removed{ -T}\new{, as t}heir behavioral patterns could differ significantly from those examined in this study~\citep{davis2018estimating,de2019stages}.
Individuals also experience various co-morbidit\removed{y}\new{ies} that influence behaviors or progression in cognitive impairments~\citep{makizako2016comorbid,stephan2011occurrence,menegon2024mild}.
This requires our future studies to expand to \new{a larger and more diverse cohort } \removed{diverse ethnic, racial, and gender groups with a more significant number of participants} with varying conditions associated with aging, MCI, and ADRD to \new{develop a robust and fair model and to} test the generalizability of our findings.

\paragraph{Limited Demographic Diversity and Bias Mitigation}
Our findings may not fully \removed{apply}\new{generalize} to larger populations with diverse gender, race, and ethnic backgrounds, \new{as shown in our fairness and bias analysis}.~\new{Our fairness and bias analysis found substantial biases across the best-performing models in quantifying cognitive states and overall well-being in older adults. In order to mitigate these biases, we mainly studied post-processing (group-specific threshold adjustments) as part of our feasibility study, which reduced disparities but resulted in significant F1 score declines. Furthermore, our fairness analysis focused on sex, age, years of education, and clinical diagnosis of MCI. We did not include race due to the racial imbalance in the dataset, which was predominantly white, as shown in \autoref{table:demo}, limiting our understanding of the impact of skin color on our facial analysis pipeline for the aging population. Future work should focus on recruiting the aging population with various cultural and demographic backgrounds and studying state-of-the-art techniques in bias mitigation to develop robust and fair AI models for monitoring cognitive impairment and psychological well-being in the aging population
\citep{yan_2020_debiasing_multimodal, seth_2023_debiasing_vlm}.}



\paragraph{Extension to Longitudinal Monitoring and Personalized Models}
\removed{Second} The model was validated on the conversations captured during the 1st week out of 48 weeks of intervention \new{to remove confounding factors due to the behavioral intervention designed in the I-CONECT (NCT02871921) study \citep{Yu2021,dodge2024internet}, as discussed in \autoref{protocol}}\removed{, i.e., which constitutes a cross-sectional study}. With cognitive decline, older adults can \new{also} manifest behavior\new{al} changes over time \citep{Schwertner2022Behavioral,Islam2019Personality}. Our next exploration will take advantage of all the data and aim to identify longitudinal changes in outcomes using the features examined here.  For this, we will explore model adaptation methods to tackle potential model degradation due to behavior\new{al} changes over time \citep{Liang2023Comprehensive}. \new{This approach will allow us to assess cognitive function and psychological well-being adaptively over time, taking into account changes induced by external factors, such as interventions or disease progression.}
Moreover, model personalization needs to be studied, as each patient's rate of cognitive decline and behavior\new{al} changes could depend on their personality\new{, disease condition,} and \new{demographic} background \citep{Ferrari2022Deep,Li2024Comparison}.

\paragraph{Impact of Data Quality}
\removed{Third,} \new{A} few video recordings had significantly low quality with low resolutions or pixelation due to weak internet connectivity.
The effect of those internet connectivity issues on model bias and performance degradation needs to be quantified, as internet connectivity can be less than ideal when the system is deployed in low-resource communities.
\new{As mentioned above}\removed{Fourth}, we utilized the baseline interview data recorded during the first week of the trial to capture behaviors closest to the time of assessments on our subjects' cognitive function and psychological well-being.
However, it is important to note that participants were still becoming familiar with the study procedures during the initial sessions. 
As a result, participants may have exhibited heightened levels of nervousness or unfamiliarity, potentially influencing their behavior and responses in ways that do not fully represent their typical cognitive or emotional state. This consideration was not factored into our analysis. Future studies will explore data from later sessions to understand and mitigate these potential data quality issues.


\paragraph{Multimodal Fusion and Transfer Learning Techniques}
\new{
In this feasibility study, we mainly applied late fusion of multimodal features, as it demonstrated superior performance compared to early fusion and mid fusion in previous mental health research, especially when only a small sample size was available~\citep{jiang2024multimodal}.
With larger recruitment and longitudinal data collection, future work needs to explore advanced multimodal time\removed{-} series methods \citep{jiang2025multi} for modeling complex temporal relationships in facial, cardiovascular, audio, and language features related to cognitive impairment and psychological well-being in aging populations.
Furthermore, our findings showed that existing state-of-the-art foundation models for extracting facial, audio, and language features had minimal contributions for quantifying cognitive impairment and psychological well-being in aging populations.
This calls for further study on the impact of transfer learning the pre-trained models with few-shot learning techniques \citep{li2024survey} to finetune those models for aging populations living with mental health problems.
We anticipate that these future research directions will unlock the potential of deep learning models for developing generalizable and robust models for aging populations.
}

\removed{
We will also explore other methods for modeling the temporal dynamics of facial, cardiovascular, audio, and language features.
In this work, we mainly used HMM, which showed varying performance across modalities.} 
\removed{Contrary to our original hypothesis that modeling temporal dynamics would increase the performance, we observed significant changes in absolute AUC from -29\% (\mbox{\autoref{table:uni_cog}}, Acoustic) to +30\% (\mbox{\autoref{table:uni_psych}}, Acoustic) in performances after adding HMM-based features.}
\removed{We chose a two-state HMM with Gaussian observation to ensure the convergence of the model when trained on our datasets.
Yet, this HMM model can be too simple and suboptimal for modeling complex temporal dynamics of facial, cardiovascular, audio, and language variation over the entire interview period relating to various rating scales used in this study.} 
\removed{To investigate this, we will explore state-of-the-art sequential models such as recurrent neural networks to extract temporal features in future work \citep{Lipton2015Critical}.}

\section{Conclusions}\label{sec:conclusion}

With the increasing prevalence of Alzheimer's disease in an aging society, it is important to quantify the cognitive impairment and psychological well-being\new{, such as neuroticism or negative affect linked to various mental health problems,} of older adults, which helps to facilitate appropriate interventions to slow down cognitive decline \citep{stateofthescienceonmild,managementofmildcognitive}. 
The disparity and the lack of supply in dementia care services significantly challenge the early detection and monitoring of MCI \citep{detectionratesofmildcognitive,dementianeurologydeserts}.
To tackle this challenge, \new{recent studies} proposed a remote interview \new{video} analysis that can potentially transform telehealth platforms into behavior\new{al} monitoring systems for MCI in resource-limited communities \citep{poor_transfrmer_multimodal_mci_2024, sun_vivit_mci_facial_24}.
\new{This study made further innovations and demonstrated for the first time, to the best of our knowledge, that it was feasible to quantify various gold-standard measures of cognitive impairment and psychological well-being in aging populations through a multimodal analysis pipeline integrating facial, acoustic, linguistic, and cardiovascular patterns observed in daily conversations.
Our models could quantify CDR (0.77 AUC), social satisfaction (0.75 AUC), negative affect (0.74 AUC), LSNS (0.74 AUC), and psychological well-being (0.72 AUC).
} 
Our \new{feature importance analysis} showed that acoustic and language patterns were most useful for quantifying cognitive impairments and social isolation, indicating the use of telephone-based monitoring for communities with limited internet connectivity.
Psychological measures \removed{, such as neuroticism or negative affect linked to various mental health problems,} were most associated with facial patterns and cardiovascular measures from videos. 
\new{We also found that all models exhibited significant bias.
While post-processing bias mitigation improved disparity, it significantly reduced F1 scores, leading to a lack of utility in the models. 
Our comprehensive analysis of feature importance and fairness in our models provides practical insights and calls for urgent future research by the machine learning and clinical research communities to develop clinically relevant, robust, and generalizable models for cognitive and psychological health in aging populations. 
}

\removed{
This work makes an important step toward understanding the psychological well-being of older adults through a telehealth platform. 
Our future work will explore the extent of these technologies for the older adults and MCI population on longitudinal monitoring scenarios and the potential biases in the model due to varying demographic backgrounds \citep{jiang2024evaluating}.}


\ifthenelse{\boolean{blind}}{
}{
\section{Acknowledgements}

Hyeokhyen Kwon, Salman Seyedi, Bolaji Omofojoye, and Gari Clifford are partially funded by the National Institute on Deafness and Other Communication Disorders (grant 1R21DC021029-01A1). 
Hyeokhyen Kwon and Gari Clifford are also partially supported by the James M. Cox Foundation and Cox Enterprises, Inc., in support of Emory’s Brain Health Center and Georgia Institute of Technology.
Gari Clifford is partially supported by the National Center for Advancing Translational Sciences of the National Institutes of Health (NIH) under Award Number UL1TR002378. Gari Clifford and Allen Levey are partially funded by NIH grant R56AG083845 from the National Institute on Aging. Hirko Dodge is funded by NIH grants and serves as the CEO of the I-CONNECT Foundation, a 501(c)(3) non-profit organization.  The I-CONECT study received funding from the NIH: R01AG051628 and R01AG056102.  The authors extend their gratitude to the participants of the I-CONECT study. 
}

\bibliographystyle{unsrt}
\new{\bibliography{sample_unified}}

\clearpage
\appendix


\section{Classification Performance of Deep Learning Models}\label{app:deep}

\new{
\autoref{tab:best-dl} shows the best-performing classification approaches (LSTM, Transformer, or MLP) for all classification tasks,
including both unimodal and multimodal fusion, as well as the 2nd best approach (or runner-up) with differences in AUC from the best model (AUROC $\Delta$). \autoref{fig:dl-heatmap} reports AUROC values for the two deep learning models (Transformer and LSTM) across all tasks (rows) and feature modalities (columns). The non-temporal features (LLaMA and demographic variables) were used to train an MLP model. Each cell shows the AUROC with the corresponding best unimodal feature or multimodal fusion strategy. A darker color indicates higher classification performance. In our experiment, all deep learning models (LSTM and Transformer) showed lower classification performance compared to the best shallow ML models.
}


\begin{addedtable}[H]
\centering
\caption{Best pipelines per task with AUROC/F1 ($\pm$ 95\% CI) and runner-up AUROC deltas ($\Delta$) for the deep learning models.}
\setlength{\tabcolsep}{3.5pt}
\scriptsize
\setlength{\extrarowheight}{10pt}   
\renewcommand{\arraystretch}{1.25} 
\begin{tabular}{@{}%
p{0.14\textwidth}%
p{0.25\textwidth}%
>{\centering\arraybackslash}p{0.14\textwidth}%
>{\centering\arraybackslash}p{0.14\textwidth}%
p{0.22\textwidth}@{}}
\toprule
\textbf{Task} & \textbf{Pipeline} & \multicolumn{2}{c}{\textbf{Metrics}} & \makecell[c]{\textbf{Runner-Up}\\\textbf{/ AUROC $\Delta$}} \\
\cmidrule(lr){3-4}
& & \textbf{AUROC} & \textbf{F1} & \\
\midrule
CDR &
\makecell[l]{Transformer\\(Audio+Language+Demo)} &
$0.55 \pm 0.016$ & $0.54 \pm 0.024$ &
\makecell[l]{LSTM (Demographics)\\$\Delta=-0$} \\
\hline

MoCA &
\makecell[l]{Transformer\\(Acoustic)} &
$0.54 \pm 0.006$ & $0.41\pm0.01$ &
\makecell[l]{Transformer \\(Audio+Language+Demo)\\$\Delta=-0.02$} \\
\hline

\makecell[l]{Clinical \\Diagnosis of\\ MCI\\(NACC UDS)}
 &
\makecell[l]{Transformer\\(All)} &
$0.65 \pm 0.014 $ & $0.56 \pm 0.018$ &
\makecell[l]{Transformer \\(Audio+Language+Demo)\\$\Delta=-0.01$} \\
\hline

LSNS &
\makecell[l]{Transformer\\(DINOv2)} &
$0.57 \pm 0.012$ & $0.53\pm0.014$ &
\makecell[l]{Transformer (Acoustic)\\$\Delta=-0.01$} \\ 
\hline

Neuro &
\makecell[l]{Transformer\\(Facial+Card+Demo)} &
$0.57 \pm 0.008$ & $0.55 \pm 0.014$ &
\makecell[l]{Transformer (All)\\$\Delta=-0.02$} \\
\hline

Negative Affect &
\makecell[l]{MLP\\(LLaMA)} &
$0.66 \pm 0.014$ & $ 0.60\pm0.018 $ &
\makecell[l]{LSTM \\(Audio+Language+Demo)\\$\Delta=-0.02$} \\
\hline

\makecell[l]{Social\\Satisfaction} &
\makecell[l]{LSTM\\(Facial+Card+Demo)} &
$0.54 \pm 0.014$ & $0.45 \pm 0.018$ &
\makecell[l]{LSTM \\(Demographics)\\$\Delta=-0$} \\ 
\hline

Psychological Well-being &
\makecell[l]{MLP\\(Demographics)} &
$0.6 \pm 0.027$ & $0.54 \pm 0.022$ &
\makecell[l]{LSTM \\(All)\\$\Delta=-0.03$} \\
\bottomrule
\end{tabular}
\label{tab:best-dl}
\end{addedtable}

\begin{addedfigure}[t]
  \centering
  \setlength{\caproom}{\abovecaptionskip+\belowcaptionskip+5.2\baselineskip}

  \begin{adjustbox}{max width=\linewidth,
                    max height=\dimexpr\textheight-\caproom\relax,
                    keepaspectratio}
    \includegraphics{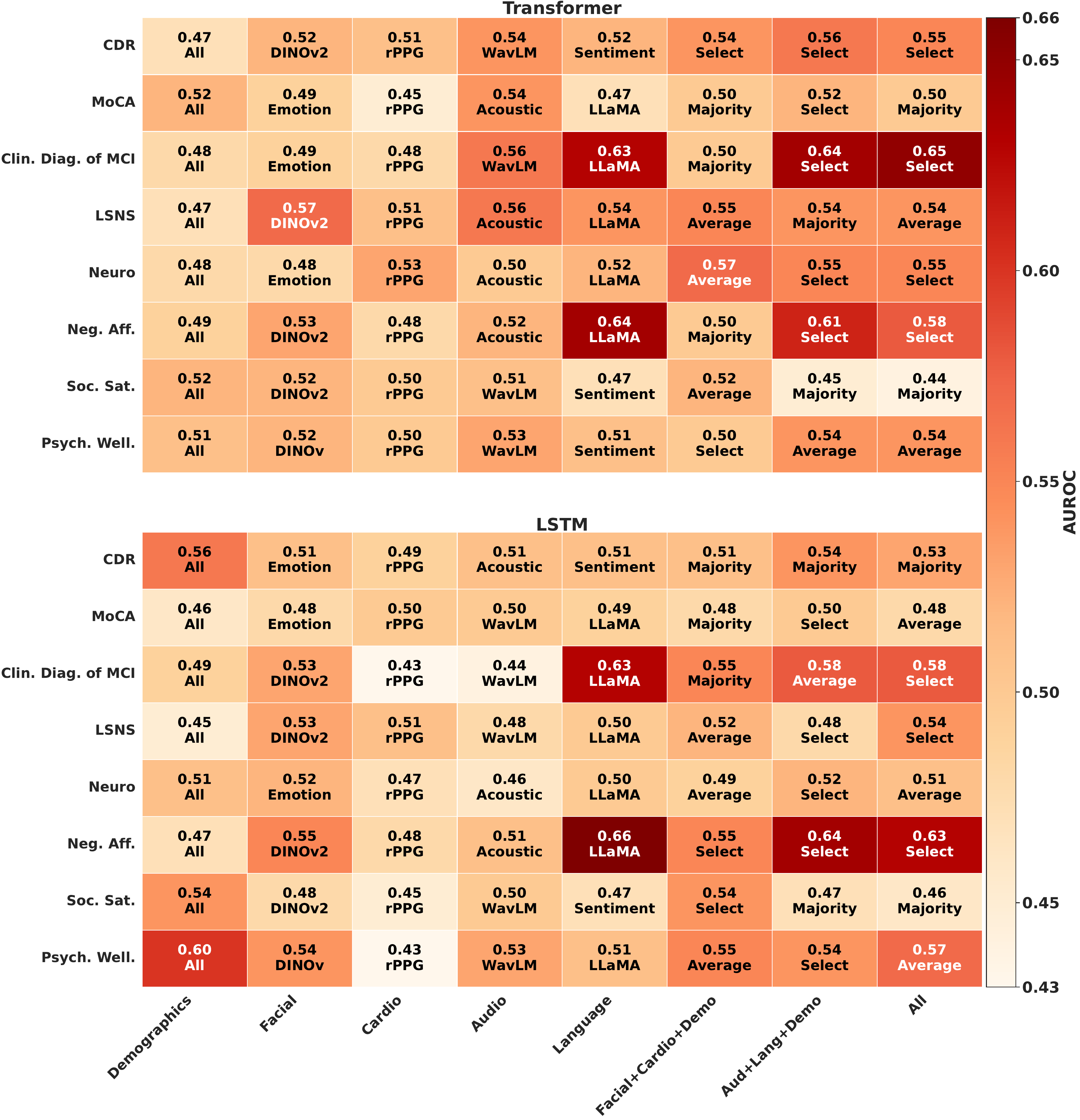}
  \end{adjustbox}

  \caption{AUROC for the deep learning models across tasks (rows) and feature modalities (columns). Each cell shows the AUROC (best) and the best-performing feature or fusion method (bottom). 
  For example, \textit{Aud+Lang+Demo} denotes fusion of audio, language, and demographics, and \textit{All} denotes fusion of all available modalities. 
  The color scale is shared across panels and the darker color indicates higher performance.}
  \label{fig:dl-heatmap}
\end{addedfigure}
\clearpage





\begin{removedtable}
\centering
\caption{Binary classification performance for unimodal analysis pipeline for dichotomized gold-standard cognitive impairments.
\textbf{Bold} number refers to the best performance for classifying each assessment outcome.
"$\ast$" number refers to the model performance better than the random chance (\textit{AUROC} $>$ 0.5). 
}
\begin{adjustbox}{max width=\linewidth}
\begin{tabular}{|c|c|c|c|c|c|}
\hline
\textbf{Modality} &
\textbf{Feature} & \textbf{Metric} & \textbf{MoCA<=24} & \textbf{Clinical Diagnosis of MCI}& \textbf{CDR=0.5}\\
\hline
\multirow{4}{*}[0em]{Demographics}
& \multirow{2}{*}[0em]{All} & AUC & 0.62$\ast$ $\pm$ 0.05 & 0.59$\ast$ $\pm$ 0.07 & 0.64$\ast$ $\pm$ 0.04 \\
& & Accuracy & 0.59 $\pm$ 0.05 & 0.58 $\pm$ 0.05 & 0.70 $\pm$ 0.04 \\
\cline{2-6}
& \multirow{2}{*}[0em]{Age} & AUC & 0.49 $\pm$ 0.04 & 0.55$\ast$ $\pm$ 0.03 & 0.58$\ast$ $\pm$ 0.02 \\
& & Accuracy & 0.57 $\pm$ 0.06 & 0.59 $\pm$ 0.00 & 0.67 $\pm$ 0.00 \\
\hline
\multirow{6}{*}[0em]{\multrow{Face}} 
& \multirow{2}{*}[0em]{DINOv2}  & AUC & 0.46 $\pm$ 0.11 & 0.41 $\pm$ 0.06 & 0.47 $\pm$ 0.11 \\
& & Accuracy & 0.47 $\pm$ 0.08 & 0.44 $\pm$ 0.07 & 0.49 $\pm$ 0.10 \\
\cline{2-6}
& \multirow{2}{*}[0em]{Emotion + AUs} & AUC & 0.40 $\pm$ 0.07 & 0.40 $\pm$ 0.07 & 0.50 $\pm$ 0.08 \\
& & Accuracy & 0.42 $\pm$ 0.07 & 0.42 $\pm$ 0.07 & 0.55 $\pm$ 0.07 \\
& \multirow{2}{*}[0em]{\multrow{Emotion + AUs\\+ HMM}} & AUC & 0.37 $\pm$ 0.05 & 0.50 $\pm$ 0.06 & 0.49 $\pm$ 0.06 \\
&  & Accuracy & 0.44 $\pm$ 0.07 & 0.58 $\pm$ 0.07 & 0.58 $\pm$ 0.06 \\
\hline
\multirow{4}{*}[0em]{\multrow{Cardiovascular}} 
& \multirow{2}{*}[0em]{rPPG} & AUC & 0.45 $\pm$ 0.06 & 0.48 $\pm$ 0.07 & 0.57$\ast$ $\pm$ 0.05 \\
& & Accuracy & 0.45 $\pm$ 0.05 & 0.50 $\pm$ 0.06 & 0.53 $\pm$ 0.05 \\
& \multirow{2}{*}[0em]{rPPG + HMM} & AUC & 0.44 $\pm$ 0.05 & 0.63$\ast$ $\pm$ 0.05 & 0.55$\ast$ $\pm$ 0.07 \\
& & Accuracy & 0.49 $\pm$ 0.04 & 0.62 $\pm$ 0.05 & 0.55 $\pm$ 0.06 \\
\hline  
\multirow{6}{*}[0em]{Audio} 
& \multirow{2}{*}[0em]{WavLM} & AUC & 0.41 $\pm$ 0.08 & 0.57$\ast$ $\pm$ 0.08 & 0.66$\ast$ $\pm$ 0.08 \\
& & Accuracy & 0.42 $\pm$ 0.07 & 0.58 $\pm$ 0.06 & 0.64 $\pm$ 0.07 \\
\cline{2-6}
& \multirow{2}{*}[0em]{Acoustic} & AUC & 0.63$\ast$ $\pm$ 0.04 & 0.61$\ast$ $\pm$ 0.04 & \textbf{0.71$\ast$ $\pm$ 0.03} \\
& & Accuracy & 0.58 $\pm$ 0.06 & 0.57 $\pm$ 0.05 & 0.65 $\pm$ 0.03 \\
& \multirow{2}{*}[0em]{Acoustic + HMM}  & AUC & 0.34 $\pm$ 0.07 & 0.33 $\pm$ 0.06 & 0.59$\ast$ $\pm$ 0.07 \\
& & Accuracy & 0.40 $\pm$ 0.05 & 0.35 $\pm$ 0.04 & 0.58 $\pm$ 0.06 \\
\hline
\multirow{6}{*}[0em]{Language}
& \multirow{2}{*}[0em]{LLaMA-65B} & AUC & \textbf{0.64$\ast$ $\pm$ 0.06} & 0.50 $\pm$ 0.05 & 0.39 $\pm$ 0.07 \\
& & Accuracy & 0.63 $\pm$ 0.07 & 0.52 $\pm$ 0.04 & 0.41 $\pm$ 0.08 \\
\cline{2-6}
& \multirow{2}{*}[0em]{\multrow{RoBERTa\\sentiment}} & AUC & 0.46 $\pm$ 0.06 & 0.62$\ast$ $\pm$ 0.06 & 0.58$\ast$ $\pm$ 0.07 \\
& & Accuracy & 0.48 $\pm$ 0.04 & 0.58 $\pm$ 0.04 & 0.58 $\pm$ 0.07 \\
& \multirow{2}{*}[0em]{\multrow{RoBERTa\\sentiment + HMM}} & AUC & 0.39 $\pm$ 0.04 & \textbf{0.66$\ast$ $\pm$ 0.02} & 0.67$\ast$ $\pm$ 0.01 \\
& & Accuracy & 0.45 $\pm$ 0.05 & 0.69 $\pm$ 0.02 & 0.66 $\pm$ 0.01 \\
\hline
\end{tabular}\label{table:uni_cog}
\end{adjustbox}
\end{removedtable}

\begin{removedtable}

\centering
\caption{Binary classification performance for multimodal  analysis pipeline for dichotomized cognitive impairments.
\textbf{Bold} number refers to the best performance for classifying each assessment.
"$\ast$" number refers to the model performance better than the random chance (\textit{AUROC} $>$ 0.5). 
}
\begin{adjustbox}{width=\textwidth}
\begin{tabular}{|c|c|c|c|c|c|}
\hline
\textbf{Modality} &
\textbf{Fusion Method} & \textbf{Metric} & \textbf{MoCA<=24} & \textbf{Clinical Diagnosis of MCI}& \textbf{CDR=0.5}\\
\hline
\multirow{6}{*}[0em]{\multrow{Face \&\\Cardiovascular\\ \& Demographics}} 
& \multirow{2}{*}[0em]{\multrow{Majority vote}} & AUC & 0.44 $\pm$ 0.08 & 0.59$\ast$ $\pm$ 0.07 & 0.61$\ast$ $\pm$ 0.08 \\
& & Accuracy & 0.46 $\pm$ 0.07 & 0.58 $\pm$ 0.06 & 0.58 $\pm$ 0.08 \\
\cline{2-6}
& \multirow{2}{*}[0em]{\multrow{Average score}} & AUC & 0.47 $\pm$ 0.09 & 0.50 $\pm$ 0.08 & 0.61$\ast$ $\pm$ 0.08 \\
& & Accuracy & 0.49 $\pm$ 0.06 & 0.52 $\pm$ 0.07 & 0.58 $\pm$ 0.07 \\
\cline{2-6}
& \multirow{2}{*}[0em]{\multrow{Selected vote}} & AUC & 0.41 $\pm$ 0.09 & 0.56$\ast$ $\pm$ 0.09 & 0.56$\ast$ $\pm$ 0.08 \\
& & Accuracy & 0.46 $\pm$ 0.07 & 0.54 $\pm$ 0.07 & 0.54 $\pm$ 0.08 \\
\hline
\multirow{6}{*}[0em]{\multrow{Audio \& \\ Language \&\\ Demographics}} 
& \multirow{2}{*}[0em]{\multrow{Majority vote}} & AUC & 0.54$\ast$ $\pm$ 0.05 & \textbf{0.66$\ast$ $\pm$ 0.04} & \textbf{0.78$\ast$ $\pm$ 0.04} \\
&   & Accuracy & 0.54 $\pm$ 0.06 & 0.63 $\pm$ 0.05 & 0.74 $\pm$ 0.04 \\
\cline{2-6}
& \multirow{2}{*}[0em]{\multrow{Average score}} & AUC & 0.51$\ast$ $\pm$ 0.07 & 0.50 $\pm$ 0.05 & 0.70$\ast$ $\pm$ 0.07 \\
& & Accuracy & 0.51 $\pm$ 0.06 & 0.50 $\pm$ 0.05 & 0.63 $\pm$ 0.07 \\
\cline{2-6}
& \multirow{2}{*}[0em]{\multrow{Selected vote}} & AUC & 0.56$\ast$ $\pm$ 0.10 & 0.64$\ast$ $\pm$ 0.07 & 0.74$\ast$ $\pm$ 0.04 \\
& & Accuracy & 0.55 $\pm$ 0.08 & 0.61 $\pm$ 0.06 & 0.68 $\pm$ 0.05 \\
\hline
\multirow{6}{*}[0em]{\multrow{Face \&\\Cardiovascular\\ \& Audio \& \\ Language \&\\ Demographics}} 
& \multirow{2}{*}[0em]{\multrow{Majority vote}} & AUC & \textbf{0.58$\ast$ $\pm$ 0.09} & 0.60$\ast$ $\pm$ 0.07 & 0.74$\ast$ $\pm$ 0.05 \\
& & Accuracy & 0.57 $\pm$ 0.09 & 0.58 $\pm$ 0.07 & 0.67 $\pm$ 0.05 \\
\cline{2-6}
& \multirow{2}{*}[0em]{\multrow{Average score}} & AUC & 0.54$\ast$ $\pm$ 0.08 & 0.46 $\pm$ 0.08 & 0.68$\ast$ $\pm$ 0.07 \\
& & Accuracy & 0.57 $\pm$ 0.07 & 0.48 $\pm$ 0.07 & 0.64 $\pm$ 0.06 \\
\cline{2-6}
& \multirow{2}{*}[0em]{\multrow{Selected vote}} & AUC & 0.52$\ast$ $\pm$ 0.11 & 0.57$\ast$ $\pm$ 0.06 & 0.67$\ast$ $\pm$ 0.06 \\
& & Accuracy & 0.52 $\pm$ 0.09 & 0.56 $\pm$ 0.07 & 0.63 $\pm$ 0.05 \\
\hline  
\end{tabular}\label{table:multi_cog}
\end{adjustbox}
\end{removedtable}

 \begin{removedtable}
\centering
\caption{Binary classification performance for unimodal analysis pipeline for the dichotomized social network, neuroticism, and psychological well-being.
\textbf{Bold} number refers to the best performance for classifying each assessment.
"$\ast$" number refers to the model performance better than the random chance (\textit{AUROC} $>$ 0.5). 5 samples were excluded from the experiment for NIH toolbox scores (Negative affect, Social Satisfaction, and Psychological Well-being) because of missing data. 
}
\begin{adjustbox}{width=\textwidth}
\begin{tabular}{|c|c|c|c|c|c|c|c|}
\hline
\textbf{Modality} &
\textbf{Feature} & \textbf{Metric} & \textbf{LSNS<=12} & \textbf{Neuro<16} & \textbf{Neg.aff.<44.10} & \textbf{Soc.sat.<48.66} & \textbf{psych.wb<53.70}\\
\hline
\multirow{4}{*}[0em]{Demographics}
& \multirow{2}{*}[0em]{All} & AUC & 0.56$\ast$ $\pm$ 0.05 & 0.58$\ast$ $\pm$ 0.04 & 0.53$\ast$ $\pm$ 0.06 & 0.40 $\pm$ 0.08 & 0.34 $\pm$ 0.06 \\
& & Accuracy & 0.54 $\pm$ 0.05 & 0.52 $\pm$ 0.05 & 0.50 $\pm$ 0.04 & 0.43 $\pm$ 0.08 & 0.36 $\pm$ 0.05 \\
\cline{2-8}
& \multirow{2}{*}[0em]{Age}  & AUC & 0.34 $\pm$ 0.06 & 0.48 $\pm$ 0.03 & 0.65$\ast$ $\pm$ 0.03 & 0.51$\ast$ $\pm$ 0.08 & 0.36 $\pm$ 0.07 \\
& & Accuracy & 0.39 $\pm$ 0.05 & 0.55 $\pm$ 0.04 & 0.57 $\pm$ 0.02 & 0.47 $\pm$ 0.07 & 0.38 $\pm$ 0.05 \\
\hline
\multirow{6}{*}[0em]{\multrow{Face}} 
& \multirow{2}{*}[0em]{DINOv2}  & AUC & 0.47 $\pm$ 0.08  & 0.51$\ast$ $\pm$ 0.06 & 0.60$\ast$ $\pm$ 0.10 & 0.57$\ast$ $\pm$ 0.07 & 0.48 $\pm$ 0.08 \\
& & Accuracy & 0.47 $\pm$ 0.08 & 0.51 $\pm$ 0.05 & 0.58 $\pm$ 0.08 & 0.55 $\pm$ 0.06 & 0.48 $\pm$ 0.06 \\
\cline{2-8}
& \multirow{2}{*}[0em]{Emotion + AUs} & AUC & 0.60$\ast$ $\pm$ 0.06 & \textbf{0.69$\ast$ $\pm$ 0.05} & 0.67$\ast$ $\pm$ 0.07 & \textbf{0.68$\ast$ $\pm$ 0.05} & 0.60$\ast$ $\pm$ 0.08 \\
& & Accuracy & 0.59 $\pm$ 0.07 & 0.66 $\pm$ 0.05 & 0.65 $\pm$ 0.06 & 0.63 $\pm$ 0.05 & 0.57 $\pm$ 0.08 \\
& \multirow{2}{*}[0em]{\multrow{Emotion + AUs\\+ HMM}} & AUC & 0.42 $\pm$ 0.08 & 0.46 $\pm$ 0.06 & 0.75$\ast$ $\pm$ 0.02 & 0.61$\ast$ $\pm$ 0.03 & \textbf{0.66$\ast$ $\pm$ 0.02} \\\
&  & Accuracy & 0.45 $\pm$ 0.07 & 0.50 $\pm$ 0.07 & 0.69 $\pm$ 0.01 & 0.61 $\pm$ 0.02 & 0.61 $\pm$ 0.02 \\
\hline
\multirow{4}{*}[0em]{\multrow{Cardiovascular}} 
& \multirow{2}{*}[0em]{rPPG} & AUC & 0.36 $\pm$ 0.06 & 0.62$\ast$ $\pm$ 0.06 & 0.53$\ast$ $\pm$ 0.06 & 0.60$\ast$ $\pm$ 0.08 & 0.48 $\pm$ 0.06 \\
& & Accuracy & 0.43 $\pm$ 0.06 & 0.61 $\pm$ 0.06 & 0.54 $\pm$ 0.07 & 0.56 $\pm$ 0.06 & 0.50 $\pm$ 0.07 \\
& \multirow{2}{*}[0em]{rPPG + HMM} & AUC & 0.56$\ast$ $\pm$ 0.04 & 0.46 $\pm$ 0.06 & \textbf{0.76$\ast$ $\pm$ 0.05} & 0.61$\ast$ $\pm$ 0.07 & 0.62$\ast$ $\pm$ 0.06 \\
& & Accuracy & 0.59 $\pm$ 0.04 & 0.50 $\pm$ 0.07 & 0.67 $\pm$ 0.05 & 0.60 $\pm$ 0.06 & 0.60 $\pm$ 0.05 \\
\hline  
\multirow{6}{*}[0em]{Audio} 
& \multirow{2}{*}[0em]{WavLM} & AUC & 0.46 $\pm$ 0.08 & 0.58$\ast$ $\pm$ 0.08 & 0.60$\ast$ $\pm$ 0.07 & 0.44 $\pm$ 0.08 & 0.45 $\pm$ 0.05 \\
& & Accuracy & 0.47 $\pm$ 0.07 & 0.58 $\pm$ 0.07 & 0.56 $\pm$ 0.05 & 0.45 $\pm$ 0.08 & 0.48 $\pm$ 0.07 \\
\cline{2-8}
& \multirow{2}{*}[0em]{Acoustic} & AUC & 0.49 $\pm$ 0.05 & 0.35 $\pm$ 0.06 & 0.45 $\pm$ 0.06 & 0.31 $\pm$ 0.06 & 0.30 $\pm$ 0.07 \\
& & Accuracy & 0.51 $\pm$ 0.06 & 0.42 $\pm$ 0.05 & 0.48 $\pm$ 0.04 & 0.36 $\pm$ 0.06 & 0.38 $\pm$ 0.06 \\
& \multirow{2}{*}[0em]{Acoustic + HMM}  & AUC & 0.52$\ast$ $\pm$ 0.07 & 0.65$\ast$ $\pm$ 0.05 & 0.72$\ast$ $\pm$ 0.05 & 0.59$\ast$ $\pm$ 0.06 & 0.57$\ast$ $\pm$ 0.06 \\
& & Accuracy & 0.53 $\pm$ 0.07 & 0.63 $\pm$ 0.04 & 0.62 $\pm$ 0.05 & 0.52 $\pm$ 0.07 & 0.56 $\pm$ 0.05 \\
\hline
\multirow{6}{*}[0em]{Language}
& \multirow{2}{*}[0em]{LLaMA-65B}  & AUC & 0.44 $\pm$ 0.09 & 0.49 $\pm$ 0.06 & 0.55$\ast$ $\pm$ 0.08 & 0.51$\ast$ $\pm$ 0.07 & 0.35 $\pm$ 0.10 \\
& & Accuracy & 0.47 $\pm$ 0.07 & 0.49 $\pm$ 0.05 & 0.55 $\pm$ 0.07 & 0.51 $\pm$ 0.06 & 0.39 $\pm$ 0.09 \\
\cline{2-8}
& \multirow{2}{*}[0em]{\multrow{RoBERTa\\sentiment}} & AUC & \textbf{0.75$\ast$ $\pm$ 0.02} & 0.55$\ast$ $\pm$ 0.04 & 0.59$\ast$ $\pm$ 0.03 & 0.47 $\pm$ 0.06 & 0.53$\ast$ $\pm$ 0.03 \\
& & Accuracy & 0.73 $\pm$ 0.04 & 0.55 $\pm$ 0.04 & 0.53 $\pm$ 0.04 & 0.48 $\pm$ 0.06 & 0.59 $\pm$ 0.03 \\
& \multirow{2}{*}[0em]{\multrow{RoBERTa\\sentiment + HMM}} & AUC & 0.47 $\pm$ 0.06 & 0.64$\ast$ $\pm$ 0.02 & 0.69$\ast$ $\pm$ 0.01 & 0.54$\ast$ $\pm$ 0.04 & 0.58$\ast$ $\pm$ 0.05 \\
& & Accuracy &  0.49 $\pm$ 0.06 & 0.60 $\pm$ 0.02 & 0.61 $\pm$ 0.02 & 0.54 $\pm$ 0.04 & 0.57 $\pm$ 0.06 \\
\hline
\end{tabular}\label{table:uni_psych}
\end{adjustbox}
 \end{removedtable}

\begin{removedtable}
\centering
\caption{Binary classification performance for multimodal interview analysis pipeline for dichotomized social network and psychological well-being.
\textbf{Bold} number refers to the best performance for classifying each assessment.
"$\ast$" number refers to the model performance better than the random chance (\textit{AUROC} $>$ 0.5). 5 samples were excluded from the experiment for NIH toolbox scores (Negative Affect, Social Satisfaction, and Psychological well-being) because of missing data. 
}
\begin{adjustbox}{width=\textwidth}
\begin{tabular}{|c|c|c|c|c|c|c|c|}
\hline
\textbf{Modality} &
\textbf{Fusion Method} & \textbf{Metric} & \textbf{LSNS<=12} & \textbf{Neuro<16} & \textbf{Neg.aff.<44.10} & \textbf{Soc.sat.<48.66} & \textbf{psych.wb<53.70}\\
\hline
\multirow{6}{*}[0em]{\multrow{Face \&\\Cardiovascular\\ \& Demographics}} 
& \multirow{2}{*}[0em]{\multrow{Majority voting}} & AUC & 0.52$\ast$ $\pm$ 0.07 & 0.63$\ast$ $\pm$ 0.06 & 0.78$\ast$ $\pm$ 0.05 & 0.68$\ast$ $\pm$ 0.03 & \textbf{0.62$\ast$ $\pm$ 0.06} \\
& & Accuracy & 0.52 $\pm$ 0.06 & 0.60 $\pm$ 0.06 & 0.75 $\pm$ 0.06 & 0.63 $\pm$ 0.05 & 0.61 $\pm$ 0.08 \\
\cline{2-8}
& \multirow{2}{*}[0em]{\multrow{Average score}} & AUC & 0.53$\ast$ $\pm$ 0.07 & 0.62$\ast$ $\pm$ 0.06 & \textbf{0.79$\ast$ $\pm$ 0.06} & \textbf{0.68$\ast$ $\pm$ 0.04} & 0.59$\ast$ $\pm$ 0.07 \\
& & Accuracy& 0.55 $\pm$ 0.07 & 0.61 $\pm$ 0.07 & 0.75 $\pm$ 0.08 & 0.63 $\pm$ 0.06 & 0.57 $\pm$ 0.09 \\
\cline{2-8}
& \multirow{2}{*}[0em]{\multrow{Selected vote}} & AUC & 0.48 $\pm$ 0.08 & 0.61$\ast$ $\pm$ 0.06 & 0.72$\ast$ $\pm$ 0.05 & 0.64$\ast$ $\pm$ 0.07 & 0.57$\ast$ $\pm$ 0.07 \\
& & Accuracy & 0.47 $\pm$ 0.07 & 0.59 $\pm$ 0.06 & 0.66 $\pm$ 0.07 & 0.60 $\pm$ 0.05 & 0.57 $\pm$ 0.08 \\
\hline
\multirow{6}{*}[0em]{\multrow{Audio \& \\ Language \&\\ Demographics}} 
& \multirow{2}{*}[0em]{\multrow{Majority vote}} & AUC & 0.58$\ast$ $\pm$ 0.08 & 0.63$\ast$ $\pm$ 0.07 & 0.66$\ast$ $\pm$ 0.04 & 0.38 $\pm$ 0.07 & 0.46 $\pm$ 0.07 \\
&   & Accuracy & 0.53 $\pm$ 0.06 & 0.56 $\pm$ 0.06 & 0.58 $\pm$ 0.04 & 0.42 $\pm$ 0.06 & 0.46 $\pm$ 0.06 \\
\cline{2-8}
& \multirow{2}{*}[0em]{\multrow{Average score}} & AUC & 0.50 $\pm$ 0.09 & 0.64$\ast$ $\pm$ 0.07 & 0.67$\ast$ $\pm$ 0.04 & 0.43 $\pm$ 0.08 & 0.39 $\pm$ 0.07 \\
& & Accuracy & 0.49 $\pm$ 0.07 & 0.62 $\pm$ 0.07 & 0.62 $\pm$ 0.05 & 0.48 $\pm$ 0.07 & 0.44 $\pm$ 0.06 \\
\cline{2-8}
& \multirow{2}{*}[0em]{\multrow{Selected vote}} & AUC & 0.57$\ast$ $\pm$ 0.08 & 0.58$\ast$ $\pm$ 0.09 & 0.63$\ast$ $\pm$ 0.07 & 0.42 $\pm$ 0.10 & 0.45 $\pm$ 0.08 \\
& & Accuracy & 0.55 $\pm$ 0.08 & 0.54 $\pm$ 0.07 & 0.57 $\pm$ 0.06 & 0.47 $\pm$ 0.08 & 0.47 $\pm$ 0.08 \\
\hline
\multirow{6}{*}[0em]{\multrow{Face \&\\Cardiovascular \\ \& Audio \& \\ Language \&\\ Demographics}} 
& \multirow{2}{*}[0em]{\multrow{Majority vote}} & AUC & \textbf{0.60$\ast$ $\pm$ 0.05} & 0.64$\ast$ $\pm$ 0.06 & 0.73$\ast$ $\pm$ 0.04 & 0.51$\ast$ $\pm$ 0.04 & 0.57$\ast$ $\pm$ 0.07 \\
& & Accuracy & 0.56 $\pm$ 0.06 & 0.58 $\pm$ 0.06 & 0.68 $\pm$ 0.07 & 0.53 $\pm$ 0.04 & 0.56 $\pm$ 0.08 \\
\cline{2-8}
& \multirow{2}{*}[0em]{\multrow{Average score}} & AUC & 0.49 $\pm$ 0.06 & \textbf{0.71$\ast$ $\pm$ 0.06} & 0.77$\ast$ $\pm$ 0.05 & 0.61$\ast$ $\pm$ 0.06 & 0.57$\ast$ $\pm$ 0.08 \\
& & Accuracy & 0.48 $\pm$ 0.08 & 0.65 $\pm$ 0.06  & 0.70 $\pm$ 0.05 & 0.57 $\pm$ 0.05 & 0.55 $\pm$ 0.06 \\
\cline{2-8}
& \multirow{2}{*}[0em]{\multrow{Selected vote}} & AUC & 0.59$\ast$ $\pm$ 0.07 & 0.62 $\ast$ $\pm$ 0.06 & 0.70$\ast$ $\pm$ 0.05 & 0.54$\ast$ $\pm$ 0.08 & 0.55$\ast$ $\pm$ 0.08 \\
& & Accuracy & 0.55 $\pm$ 0.07 & 0.57 $\pm$ 0.06 & 0.64 $\pm$ 0.04 & 0.54 $\pm$ 0.07 & 0.53 $\pm$ 0.08 \\
\hline  
\end{tabular}\label{table:multi_psych}
\end{adjustbox}
\end{removedtable}

\end{document}